\pdfoutput=1
\documentclass[floats,floatfix,showpacs,amssymb,prd,twocolumn,superscriptaddress,nofootinbib,longbibliography,reprint]{revtex4-2}
\usepackage{amssymb,amsmath,verbatim,mathtools,needspace,enumitem,etoolbox,graphicx,physics,microtype,afterpage,xspace,tabularx,lmodern,multirow}
\usepackage{gensymb}
\usepackage{placeins}
\usepackage[dvipsnames]{xcolor}
\definecolor{linkcolor}{rgb}{0.0,0.3,0.5}
\usepackage[unicode, colorlinks=true, linkcolor=linkcolor, citecolor=linkcolor, filecolor=linkcolor, urlcolor=linkcolor, linktocpage, breaklinks]{hyperref}
\usepackage[all]{hypcap}
\usepackage[T1]{fontenc}
\usepackage[utf8]{inputenc}
\usepackage{mathrsfs}
\usepackage[dvipsnames]{xcolor}

\usepackage[normalem]{ulem}
\hypersetup{colorlinks=true,citecolor=romared,linkcolor=romared,urlcolor=romared}
\definecolor{romared}{RGB}{142,0,28}
\usepackage{aas_macros}
\usepackage{makecell}
\usepackage{soul}
\usepackage[nolist,nohyperlinks]{acronym}
\usepackage{orcidlink}
\usepackage{academicons}
\usepackage{lipsum}

\definecolor{orcidlogocol}{HTML}{A6CE39}
\usepackage{mathtools}

\def\prd{PRD}
\def\apj{ApJ}                 %
\def\apjl{ApJL}               %
\def\apjs{ApJS}               %
\def\mnras{MNRAS}             %
\def\aap{A\&A}                %
\def\araa{ARA\&A}             %
\def\icarus{Icarus}           %
\def\jcap{JCAP}

\newcommand{\orcid}[1]{\href{https://orcid.org/#1}{\includegraphics[width=10pt]{orcid.pdf}}}
\renewcommand{\vec}[1]{\boldsymbol{#1}}

\newcommand{\be}{\begin{equation}}
\newcommand{\ee}{\end{equation}}
\newcommand{\beq}{\begin{eqnarray}}
\newcommand{\eeq}{\end{eqnarray}}

\allowdisplaybreaks

\allowdisplaybreaks

\newcommand{\eden}{\mathfrak{e}}

\newcommand{\prj}{\mathrm{q}}
\newcommand{\asp}{\eta}

\usepackage{color}

\definecolor{darkpurple}{RGB}{102,0,153}

\definecolor{mypurple}{RGB}{130, 0, 130} %

 \hypersetup{
    colorlinks=true,
    linkcolor=mypurple, %
     citecolor=mypurple, %
    urlcolor=mypurple   %
 }

\begin{document}

\pagenumbering{arabic}

\title{Spiral Density Waves and Torque Balance in the Kerr Geometry}
\author{Conor Dyson}
\email{conor.dyson@nbi.ku.dk}
\affiliation{ Center of Gravity, Niels Bohr Institute, Blegdamsvej 17, 2100 Copenhagen, Denmark} 
\author{Daniel J. D'Orazio}
\email{dorazio@stsci.edu}
\affiliation{Space Telescope Science Institute, 3700 San Martin Drive, Baltimore, MD 21218, USA 3}
\affiliation{Department of Physics and Astronomy, Johns Hopkins University, 3400 North Charles Street, Baltimore, Maryland 21218, USA}
\affiliation{Niels Bohr Institute, Blegdamsvej 17, 2100 Copenhagen, Denmark} 
\date{\today}

\begin{abstract}
Extreme mass-ratio inspirals (EMRIs) in relativistic accretion discs are a key science target for the upcoming LISA mission. Existing models of disc–EMRI interactions typically rely on crude dynamical friction or Newtonian planetary migration prescriptions, which fail to capture the relativistic fluid response induced by the binary potential. In this work we address this gap by providing the relativistic calculation. We apply standard methods from self-force theory, black hole perturbation theory, and relativistic stellar perturbation theory to compute the relativistic analogue of the spiral density wave and torque-balance results underlying planetary migration. Specifically, we calculate the response of a perfect fluid disc to the perturbing potential of an EMRI, consistently incorporating pressure effects. Using a master enthalpy-like variable and linearised fluid theory, we present a set of proof-of-principle calculations that reconstruct the fluid perturbations and relativistic spiral arm structure for a range of spin values in the Kerr geometry. We derive a relativistic torque-balance equation that enables us to compute and compare local torques with the angular momentum advected through the disc, finding striking agreement when applied to our relativistic disc–EMRI solutions. This opens a promising route towards establishing torque-balance relations between integrated disc torques arising from fluid perturbations and the forces acting on EMRIs embedded in matter.
\end{abstract}

\maketitle

\section{Introduction}

Since the first detection of a binary black hole merger in 2015~\cite{LIGOScientific:2016aoc}, gravitational wave astronomy has progressed rapidly, with over 200 events published to date~\cite{LIGOScientific:2025slb,LIGOScientific:2018mvr,LIGOScientific:2020ibl,KAGRA:2021vkt}. This progress will continue with next generation ground based detectors such as the Einstein Telescope~\cite{Maggiore:2019uih} and Cosmic Explorer~\cite{Evans:2021gyd}. Looking beyond ground based observatories, the coming decade will also see the launch of space-based detectors operating in the milli-hertz band, including the Laser Interferometer Space Antenna (LISA)~\cite{Colpi:2024xhw}, TianQin~\cite{TianQin:2020hid,Li:2024rnk}, and Taiji~\cite{Gong:2021any}. In this frequency range, intermediate and extreme mass ratio inspirals are among the most exciting sources~\cite{LISA:2022yao,LISACosmologyWorkingGroup:2022jok,LISA:2022kgy}. These systems consist of a supermassive black hole with mass $\gtrsim 10^{4} M_{\odot}$ orbited by a much less massive companion. Due to the large mass ratio, the secondary can remain in band for years, providing a probe of the geometry and matter content near the central object~\cite{LISAConsortiumWaveformWorkingGroup:2023arg,Barack:2018yvs,Wardell:2021fyy,Chua:2020stf,Katz:2021yft,Hughes:2021exa}. In this vein, studies of accretion discs and active galactic nuclei suggest that gravitational wave signals will be influenced by their surrounding environments~\cite{1973A&A....24..337S,Novikov:1973kta,2002ApJ...565.1257T, YLevin:2003, YLevin:2007, Barausse:2007dy,Abramowicz:2011xu,Yunes:2011ws,Kocsis:2011dr,Barausse:2014tra,Derdzinski:2018qzv,Duffell:2019uuk,Derdzinski:2020wlw, Tagawa:2020qll,Pan:2021ksp, Pan:2021oob,Zwick:2021dlg, Derdzinski:2022ltb,Cole:2022yzw,Speri:2022upm,Garg:2022nko,Morton:2023wxg,Tiede:2023cje,Zwick:2024yzh,Garg:2024oeu,Garg:2024yrs,Garg:2024zku,Khalvati:2024tzz,Duque:2024mfw, Spieksma:2025wex, Zwick_eccenv+2025, Zwick:2026}. Incorporating such effects into waveform models is therefore important both for signal detection and for controlling biases in parameter estimation and tests of general relativity~\cite{Cole:2022yzw,Zwick:2022dih,Garg:2024qxq,Khalvati:2024tzz}.

State of the art vacuum EMRI waveform modelling relies on the gravitational self-force formalism, which exploits the disparate length scales and adiabatic nature of EMRI systems to construct precision waveform models \cite{Pound:2012dk,Chapman-Bird:2025xtd,Speri:2023jte,Pound:2017psq,vandeMeent:2017bcc,Lynch:2024ohd,Upton:2023tcv,Lynch:2021ogr,Leather:2024mls,Bourg:2024vre,Miller:2020bft,Nasipak:2023kuf,Bourg:2024cgh,Upton:2025bja}. Within this framework, the separation of dissipative and orbital timescales leads to a controlled phasing expansion,
\begin{equation}
    \Phi = \frac{1}{\epsilon}\,\phi^{(0)} + \phi^{(1)} + \mathcal{O}(\epsilon).
\end{equation}
where $\epsilon = m_p/M$ is the mass-ratio in the system and $\Phi$ is the total waveform phase. The $\phi^{(0)}$ term is the zeroth post adiabatic (0PA) contribution which drives the adiabatic evolution of the system, while $\phi^{(1)}$ represents the first post adiabatic (1PA) correction. Obtaining the full 1PA dynamics for generic vacuum orbits in Kerr spacetimes remains a major long term goal of the second-order self-force programme. 

Looking beyond the vacuum model, toward systems in the presence of environments, this picture is modified. Provided the additional field is conservative at leading order, the phasing expansion becomes
\begin{equation}
    \Phi = \frac{1}{\epsilon}\,\left(\phi^{(0,0)} + \lambda\,\phi^{(0,1)}\right) + \phi^{(1,0)} + \mathcal{O}(\epsilon,\lambda),
\end{equation}
where $\lambda$ characterises the strength of the matter field and now corrects the leading order 0PA phasing, motivating accurate calculations of their contribution. Historically, such corrections have often been treated using approximations based on rectilinear motion or constant background Newtonian flows. More recent work has begun to calculate these effects in a more consistent manner \cite{Vicente:2025gsg,Brito:2014wla,Rahman:2025mip,Li:2025ffh,Dyson:2025dlj,Polcar:2025yto,Duque:2025yfm,Duque:2023seg, HegadeKR:2025dur,HegadeKR:2025rpr}.

Modelling the orbital evolution of body moving through a surrounding environment has a long history, with notable early contributions from \citet{chandrasekhar1943dynamical} and \citet{Ostriker_1999}, developing the theory of dynamical friction for rectilinear motion through homogeneous, isotropic collisionless and collisional media, respectively. Focusing on the latter, i.e., fluids, much work has been done in the ensuing decades to extend linear-theory results to include, more general orbits \citep[e.g.,][and references therein]{O’Neill_2025}, non-linear effects through numerical calculations \citep[e.g.,][]{Kim_NLGDF:2010}, relativistic velocities \citep{Barausse:2007ph}, and general relativity, first for non-spinning black holes~\cite{Vicente:2022ivh,Traykova:2021dua,Traykova:2023qyv} and subsequently including spin~\cite{Dyson:2023fws,Wang23}. 
More recently, self consistent calculations of EMRIs in environments have been carried out using exact solutions of the Einstein equations~\cite{Cardoso:2022whc,Duque:2023seg}. While promising, these approaches rely on spherical symmetry of both the geometry and matter distribution, making it difficult to extend to more general configurations.

The assumption of a homogeneous medium with a uniform background flow breaks down for a perturber orbiting a central body in a thin accretion disc. Where the shearing Keplerian flow of the disc generates a significantly different fluid response. Seminal works such as \citet{LinPapa:1979}, and \citet{Goldreich79, Goldreich80} showed that angular momentum exchange in such discs is dominated by resonant interactions between the secondary and the disc. These interactions excite spiral density waves, underpinning a large body of work on Newtonian ``disc-satellite'' interactions ~\cite[e.g.,][]{Ward:1986, MeyerSicardy:1987, Lubow:1991, AL:1994, Ward:1997, Tanaka2002, Tanaka2004, Tsang14, Miranda20, Fairbairn22, Tanaka2024, FairbairnDittmann:2025}, and inducing ``migration'' of the perturbing body through the disc. The results of these calculations have been widely applied to the formation of planetary system architectures \citep[e.g.,][]{KleyNelson:2012}, as well as a wide range of other applications, including EMRI dynamics 
\citep[e.g.,][]{Yunes:2011ws,Kocsis:2011dr, Copparoni+2025}.
Hirata \citep{Hirata1, Hirata2} extended some of these approaches to relativistic EMRIs using a Hamiltonian based torque balance formalism. Such Hamiltonian approaches have been recently applied to study torque exchange in collisionless discs \cite{Duque:2025yfm,HegadeKR:2025dur,HegadeKR:2025rpr}. Here we present the first relativistic calculation of the linear response of a pressure-supported disc to an EMRI.

The goal of this work is to calculate the fluid perturbations and associated torque densities in an EMRI system in the spacetime of a spinning black hole. We achieve this by combining the two parameter expansion framework \cite{Brito:2023pyl,Dyson:2025dlj,Datta:2025ruh,Rahman:2025mip,Polcar:2025yto,Li:2025ffh} with self-force theory, Newtonian planetary migration methods and the master enthalpy formalism of Ipser and Lindblom \cite{Ipser91,Ipser92}. 

We consider an adiabatic disc with a radial temperature gradient around a Kerr black hole, in the presence a much less massive companion at strong relativistic
separations of order $10$ gravitational radii. 
We restrict to the regime where the secondary mass is smaller than the thermal mass of the disc \cite{Goodman_2001} and the linearised fluid prescription should hold. The structure of the paper is as follows. In Section~\ref{sec:nonlinearsetup} we present the relativistic nonlinear fluid equations. In Section~\ref{sec:pertscheme} we derive the linearised perturbation equations. In Section~\ref{sec:masterenthalpy} we construct the master enthalpy equation. In Section~\ref{sec:backgrouninpuot} we derive the background disc solution. In Section~\ref{sec:numerical} we describe the numerical methods and operator analysis used to solve the master equation. In Section~\ref{sec:Solutions} we explore solutions: \S\ref{sec:spiral_morph} considers spiral density wave morphology and \S\ref{sec:Torque} provides a new form of angular momentum balance equation extending the Newtonian prescription of advection-torque balance to the relativistic regime. We summarise and consider future directions in Section~\ref{sec:Summary}.

We use the mostly plus metric signature and geometric units with $G=c=1$. Spacetime components of tensors are denoted by Greek indices.

\section{Set-Up and Non-linear equations of motion}\label{sec:nonlinearsetup}
We begin with the Einstein equations coupled to matter,
\begin{align}\label{eq:einsteineq}
G_{\mu\nu} &= 8 \pi T^{fl}_{\mu\nu}
\end{align}
modelled as a perfect fluid,
\begin{equation}\label{eq:eommain}
T_{fl}^{\mu\nu} = (\eden + p)U^\mu U^\nu + p g^{\mu\nu} .
\end{equation}
With $U^{\mu}$ describing the four flow of the fluid, $p$ is the pressure, and $\eden = \rho(c^2 + \epsilon)$ is the combination of rest-mass and internal energy density. Through conservation of the stress energy tensor we obtain the fluid equations of motion in the following form,
\begin{equation}\label{eq:eom}
\nabla_{\mu}T_{fl}^{\mu\nu} = 0.     
\end{equation}
This equation takes the explicit form,
\begin{equation}\label{eq:explciiteEOM}
\left(\eden + p\right)A^{\nu} = - \nabla^{\nu}p - U^{\nu}\nabla_{\mu}\left(\left[\eden+p\right]U^{\mu}\right),
\end{equation}
where $U^{\mu}\nabla_{\mu}U^{\nu} = A^{\nu}$ is the four acceleration of the fluid element. Due to the normalisation of the four-velocity $U^{\mu}U_{\mu}=-1$, we must satisfy the condition $A^{\mu}U_{\mu}=0$ and define the projection tensor
\begin{equation}
\prj^{\mu\nu} = g^{\mu\nu} + U^{\mu}U^{\nu}.
\end{equation}
Contraction with this tensor picks out elements of a four-vector orthogonal to the fluid flow and satisfies the conditions $\prj^{\mu\nu} = \prj^{\mu\alpha}\prj_{\alpha}^{\;\nu}$, $\prj^{\mu}_{\;\alpha}A^{\alpha} = A^{\mu}$, and $\prj^{\mu}_{\;\alpha}U^{\alpha} = 0$. Beginning from Eq. \eqref{eq:explciiteEOM}, we can contract with $U_{\mu}$ and $\prj^{\alpha}_{\;\mu}$ respectively to obtain the relativistic energy and momentum conservation equations,
\begin{align}
\label{eq:eulerequations}
\left(\eden+p\right)\nabla_{\mu}U^{\mu} + U^{\mu}\nabla_{\mu}\eden&=0,\\
\label{eq:eulerequations2}
\left(\eden+p\right) U^{\mu}\nabla_{\mu}U^{\nu} +\prj^{\mu\nu}\nabla_{\mu}p&=0.
\end{align}

\section{Perturbative Scheme}\label{sec:pertscheme}
\subsection{Linearising the Metric}
In order to gain a handle on the problem, we elect to take a perturbative approach. In many astrophysical contexts, it is possible to perform expansions around some background geometry. Take a metric parametrically close to the Kerr spacetime \cite{StewartWalker}, then we can expand our full metric in the form,
\begin{equation}
g_{\mu\nu} = g^\mathrm{kerr}_{\mu\nu} + h_{\mu\nu}.
\end{equation}
In the context we are considering here, the implicit assumption is that the effect on the geometry due to the small body and the matter field is perturbatively small in comparison to the background. Here, $g_{\mu\nu}^{kerr}$ is the Kerr metric and, $h_{\mu\nu}$ contains the collection of all perturbative content. In order for this form of expansion to be valid, we must specify that the stress energy tensor of the isolated fluid in the absence of the secondary is also subleading and controlled by a small parameter \cite{Brito:2023pyl,Dyson:2025dlj,Datta:2025ruh,Rahman:2025mip,Polcar:2025yto}.
Unlike in \cite{Dyson:2025dlj}, we leave this parameter unspecified as an $\mathcal{O}(1)$ counting parameter $\lambda$. Next, we wish to introduce the effect of the secondary black hole. The incorporation of which can be developed formally through matched asymptotic expansions and the self-consistent formalism of~\citet{Pound:2017psq}. For the purposes of this work we take the skeletonised approach, which re-expresses the vacuum content of the smaller black hole as a point particle sourcing term to give, 
\begin{align}\label{eq:einsteineqII}
G_{\mu\nu} &= 8 \pi \left(\epsilon T^{pp}_{\mu\nu}+\lambda T^{fl}_{\mu\nu} \right),
\end{align}
where $\epsilon$ is now a $\mathcal{O}(1)$ counting parameter associated with the mass ratio between the secondary and central black hole, $ m_p/M \ll 1$, and
\begin{equation}\label{eq:ppstressenergy}
T_{pp}^{\mu\nu} = m_p \int \dot{x}^{\mu}(\tau)\dot{x}^{\nu}(\tau)\frac{\delta\big(x^{\mu}-x^{\mu}(\tau)\big)}{\sqrt{-g}}d\tau.
\end{equation}
Due to the nature of the fluid stress-energy tensor in Eq.~\eqref{eq:eommain}, at leading order the fluid variables have the scalings, $\eden\sim\mathcal{O}(\lambda)$, $p\sim\mathcal{O}(\lambda)$, and $U^{\mu}\sim\mathcal{O}(1)$.
A perturbative expansion provides the form of the metric perturbation,
\begin{equation}
h_{\mu\nu} = \sum_{|n|+|m|\geq1} \epsilon^m \lambda^n h^{(m,n)}_{\mu\nu},\\
\;\;\;\notag
\end{equation}
which allows us to expand Eqs.~\eqref{eq:eulerequations} and \eqref{eq:eulerequations2} and collect the $\mathcal{O}(\epsilon\lambda )$ expressions to give,
\vspace{0.1cm}

\begin{align}\label{eq:eulerequationsmetricexpand}
&(\eden+p)\nabla_{\mu}U^{\mu} + U^{\mu}\nabla_{\mu}\eden=-\frac{1}{2}(\eden+p)U^{\mu}\nabla_{\mu}h^{(1,0)},\\
& (\eden+p)U^{\mu}\nabla_{\mu}U^{\nu} +\prj^{\mu\nu}\nabla_{\mu}p=\prj^{\nu}_{\;\mu}f^{\mu}_{(1,1)}.
\end{align}

where,
\begin{align}
\label{eq:Effective force}
f_{\mu}^{(1,1)}=(\eden+p)U^{\alpha} U^{\beta}\left( \frac{h^{(1,0)}_{\alpha\beta;\mu}}{2}- h_{\mu\alpha;\beta}^{(1,0)} \right) \\+ h_{\mu\nu}^{(1,0)}\partial^{\nu}p.
\notag
\end{align}
Here, all gradients and projectors are now defined with respect to the background Kerr metric and non-linear fluid flow. After the expansion we have reprojected the second equation with the projector tensor to retain the orthogonal nature of the four-acceleration to the four-flow. 

At this point, we have defined the non-linear fluid equations experiencing the effective force of a secondary black hole on a fixed background. In principle, these equations could be incorporated into a numerical relativistic hydrodynamics solver on a static metric \citep[e.g., AthenaK][]{AthenaK:2024}. For the purposes of this work, however, we now turn to a simplified linearised fluid prescription.
\subsection{Linearising the Fluid}
In line with \cite{Datta:2025ruh} and our previous motivation we expand the fluid variables as
\begin{align*}
\eden &= \sum_{n\geq1,m\geq0}\epsilon^m \lambda^{n}\eden^{(m,n)},\\
p &= \sum_{n\geq1,m\geq0}\epsilon^m \lambda^{n} p^{(m,n)},\\
U^{\mu} &= \sum_{n\geq0,m\geq0}\epsilon^m \lambda^{n} U^{\mu}_{(m,n)}.
\end{align*}
Where here, $\eden^{(0,1)},p^{(0,1)}$, and $U^{\mu}_{(0,0)}$ constitute the background fluid quantities on the Kerr geometry in the absence of the secondary BH and self-gravity. Having also introduced the expansion of our fluid variables, we are now in a position to linearise the fluid elements in Eq.~\eqref{eq:eulerequationsmetricexpand} and obtain the $\mathcal{O}(\epsilon\lambda )$ component of the linearised Euler equations,

\begin{widetext}
\begin{align}\label{eq:eomEnergyPert}
\left(\eden^{(1,1)} +p^{(1,1)}\right)\nabla_{\mu}U^{\mu}+\left(\eden +p\right)\nabla_{\mu}U^{\mu}_{(1,0)} + U^{\mu}\partial_{\mu}\eden^{(1,1)}+U^{\mu}_{(1,0)}\partial_{\mu}\eden = -\frac{1}{2}(\eden +p)U^{\mu}\nabla_{\mu}h^{(1,0)},
\end{align}

\begin{align}\label{eq:eomMomPert}
\left(\eden^{(1,1)}+p^{(1,1)}\right) U^{\mu}\nabla_{\mu}U^{\nu}+
\left(\eden+p\right)&\left(U^{\mu}\nabla_{\mu}U^{\nu}_{(1,0)}+U_{(1,0)}^{\mu}\nabla_{\mu}U^{\nu}\right) +\prj^{\mu\nu}\nabla_{\mu}p^{(1,1)} + U^{\nu}U^{\mu}_{(1,0)}\nabla_{\mu}p= \prj^{\nu}_{\; \mu}f^\mu_{(1,1)}.
\end{align}
\end{widetext}
For simplicity of notation, any occurrence of a background fluid quantity, $\eden^{(0,1)}, p^{(0,1)},$ and $U^{\mu}_{(0,0)}$ has had the counting indices removed, all occurrences of the covariant derivatives and projectors are  now related to the Kerr metric and the background fluid flow, and $h^{(1,0)} = g^{\mu \nu} h_{\mu \nu}^{(1,0)}$ is the trace of our metric perturbation.

\section{Master Enthalpy Equation}\label{sec:masterenthalpy}
The use of a master enthalpy equation in studying disc dynamics in the presence of a secondary perturber has seen wide use in the Newtonian community. From the seminal works of Goldreich and Tremaine \cite{Goldreich79,Goldreich80} to more recent works understanding the outflow conditions and extended density structure of the disc 
\cite{Fairbairn22,Fairbairn25,Ogilvie01,Ogilvie02,Miranda19,Miranda20,Tsang14,Zhang2006}. In the relativistic community,
seminal works, such as Ipser and Lindblom \cite{Ipser91,Ipser92}, have derived relativistic forms of the master enthalpy equation for, non-barotropic, self-gravitating perfect fluid flows. In the following section, we connect these approaches by deriving the fully relativistic master enthalpy equation describing perturbations of a perfect fluid induced by an independently sourced metric perturbation (i.e. due to a small secondary perturber) in the Kerr spacetime. The derivation follows closely that of Ipser and Lindblom \cite{Ipser91} with some minor deviations; due to this and the novelty of its usage in the context of a secondary perturber, we detail our version of the derivation in full.

\subsection{Killing Symmetries of Background Fluid Flow} 

As is often crucial in the derivation of such a master equation we will require some level of symmetry in our background flow in order for our linearised fluid equations to sufficiently simplify and become amenable to re-expressing as a single scalar equation. In particular, in the case of a relativistic disc, we will require that our background flow follows the isometries of the Kerr geometry, i.e, that it is stationary and axi-symmetric and can be written as,
\begin{equation}
U_{(0,0)}^{\mu}  = \alpha\left(t^{\mu}+\Omega \phi^{\mu}\right).
\end{equation}
Where in Boyer-Lindquist (BL) coordinates, $t^{\mu}$ and $\phi^{\mu}$ represent the coordinate Killing vectors of the Kerr geometry and $ \alpha$ and $\Omega$ are generic functions of the radial and polar ($r$ and $\theta$) BL  coordinates. With these restrictions on the background flow, at leading order the momentum equation simplifies to, 
\begin{align}
U_{(0,0)}^{\mu}\nabla_{\mu}U_{(0,0)}^{\nu} =-\frac{\nabla^{\mu}p}{\eden+p}\equiv -R^{\mu}.
\end{align}

\subsection{Fluid Perturbation Prescription}
To this point, our equations of motion are still underdetermined. In order to remedy this issue we must choose an equation of state in the co-moving Lagrangian frame of the fluid. We begin by denoting the leading order Lagrangian perturbations by $\Delta p$, $\Delta \eden$. 
One can then ``close'' the system by enforcing that local adiabatic perturbations to the fluid take the form, 
\begin{equation}
    \Delta p  = c^2_s\Delta\eden
\end{equation}
By allowing $c_s$ to be an arbitrary axi-symmetric, stationary function, and identifying it with the sound-speed given by the background relation, $p = c_s^2 \mathfrak{e}$, we fix our fluid to be in a $\gamma\rightarrow 1$ adiabatic state which as discussed in  \cite{Miranda20,Miranda19ALMA} is distinct from the standard local isothermal approximation. To relate the Lagrangian and Eulerian perturbations of our fluid we write $\Delta p = p^{(1,1)} + \mathcal{L}_{\xi}p$ and $\Delta \eden = \eden^{(1,1)} + \mathcal{L}_{\xi}\eden$, where $\mathcal{L}_{\xi}$ is the Lie derivative along the Lagrangian displacement vector field $\xi$.
For adiabatic perturbations, we can define the displacement vector between the perturbed and background flow through,
\begin{equation}
    U^{\nu}_{(1,0)} = -\left(\mathcal{L}_{\xi}U\right)^{\nu}  =U^{\mu}\partial_{\mu}\xi_{(1,0)}^{\nu}-\xi_{(1,0)}^{\mu}\partial_{\mu}U^{\nu}.
\end{equation}
Using that perturbations are assumed to take the form $ \sim e^{i m\phi-i\omega t}$ and $\tilde{\omega} = m \Omega - \omega $, this can be reduced to the form,
\begin{equation}
    U^{\mu}_{(1,0)} = \xi_{(1,0)}^{\nu}\left(i \tilde{\omega} \alpha \delta^{\mu}_{\nu} - \partial_{\nu}U^{\mu}\right).
\end{equation}
Due to the assumed symmetries of the background, the matrix $\partial_\nu U^\mu$ is nilpotent, $\partial_\nu U^\mu\,\partial_\rho U^\nu = 0$. Hence, given $M^\mu{}_\nu = \delta^\mu{}_\nu - \partial_\nu U^\mu/(i\tilde{\omega}\alpha)$, we have $(M^{-1})^\mu{}_\nu = \delta^\mu{}_\nu + \partial_\nu U^\mu/(i\tilde{\omega}\alpha)$.
Specifying the form of $U^{\mu}$ we then obtain the equations,
\begin{equation}\label{eq:dispalcementvec}
    \xi^{\mu}_{(1,0)} = \frac{1}{i \tilde{\omega} \alpha }\left(\delta^{\mu}_{\nu} +\frac{\partial_{\nu}U^{\mu}}{i \tilde{\omega}\alpha}\right)U_{(1,0)}^{\nu}.
\end{equation}
Now that we have determined our displacement vector in terms of the fluid perturbations, we are primed to relate our density perturbation to our pressure perturbation in the Eulerian frame:
\begin{align}
    \eden^{(1,1)} &=  \Delta \eden - \xi^{\mu}\partial_{\mu}\mathfrak{e} \\
    &= \frac{1}{c_s^2} \left[p^{(1,1)} + \xi^{\mu}_{(1,0)}\partial_{\mu}p \right] - \xi^{\mu}_{(1,0)}\partial_{\mu}\eden. 
\end{align}
Inputting the explicit form of Eq.~\eqref{eq:dispalcementvec}, one finds the expression,
\begin{equation}
    \eden^{(1,1)} = (\eden+p)\left[ \frac{p^{(1,1)}}{(\eden+p)c_s^2}- \frac{ B_{\nu}U_{(1,0)}^{\nu} }{i\tilde\omega\alpha }\right],
\end{equation}
where,
\begin{equation}
   B_{\mu}  = \frac{1}{\eden+p}\left(\partial_{\mu}\eden-\frac{\partial_{\mu}p}{c_s^2} \right)  = - \frac{\partial_{\mu}\log c_s^2}{1 + c_s^2},
\end{equation}
corresponds precisely to terms like Eq.~(18) in \cite{Fairbairn2022}, describing the thermal structure of the disc. 
\subsection{Reducing to the Single Master Variable}
Having closed our system of equations, we are equipped to express our fluid perturbations in terms of a single master variable, an enthalpy like perturbation, which we define by $\mathfrak{h}^{(1,0)}$. %
We define this perturbation through the identification, 
\begin{align}\label{eq:PresEnthrecon}
   p^{(1,1)}&= (\eden+p) \mathfrak{h}^{(1,0)},
\end{align}
leading to the expression,
\begin{align}\label{eq:denEnthecon}
   \eden^{(1,1)}& = (\eden+p)\left[ \frac{\mathfrak{h}^{(1,0)}}{c_s^2}- \frac{ B_{\nu}U_{(1,0)}^{\nu} }{i \tilde\omega\alpha }\right],
\end{align}
for the energy density perturbation. 
Our final task is to re-express the velocity perturbations in terms of the enthalpy scalar.
We will achieve this by substituting the expressions in Eqs.~(\ref{eq:PresEnthrecon}, \ref{eq:denEnthecon}) into Eq.~\eqref{eq:eomMomPert}. In doing so it becomes apparent that the assumed symmetries in our background flow allow the resulting equation to become purely algebraic in $U^{\mu}_{(1,0)}$, taking the form,
\begin{equation}\label{eq:VeolcityReconstructionecon}
\left(\mathcal{Q}^{-1}\right)^{\nu}_{\;\mu} U_{(1,0)}^{\mu} =  \frac{\prj^{\nu}_{\mu}f^{\mu}_{(1,1)}}{\eden+p}- \prj^{\nu\mu}\nabla_{\mu}\mathfrak{h}^{(1,0)}-\mathfrak{h}^{(1,0)}B^{\nu},
\end{equation}
where we have defined the linear operator,

\begin{align}
\left(\mathcal{Q}^{-1}\right)^{\nu}_{\;\;\mu}  = i \tilde{\omega} \alpha \delta^{\nu}_{\;\mu}&+\Gamma^{\nu}_{\alpha\mu}U^{\alpha} 
    + \nabla_{\mu}U^{\nu}\\&+ U^{\nu}R_{\mu} + \frac{R^{\nu} B_{\mu}  }{i \alpha\tilde\omega}.\notag
\end{align}
Inverting $\mathcal{Q}^{-1}$, we express all fluid quantities in terms of known expressions and the master enthalpy like perturbation. Substituting the expressions from equations~\eqref{eq:PresEnthrecon},\eqref{eq:denEnthecon}, and \eqref{eq:VeolcityReconstructionecon} into the energy equation~\eqref{eq:eomEnergyPert} yields a scalar equation expressing all perturbed fluid quantities purely in terms of a master variable $\mathfrak{h}^{(1,0)}$, and lower order quantities, as in \cite{Brito:2023pyl, Dyson:2025dlj}. 
The explicit form of the master equation, is then given by,
\begin{widetext}
\begin{align}\label{eq:MasterEquations}
c_s^2\nabla_{\mu}\left(\mathcal{Q}^{\mu}_{\;\alpha}q^{\alpha\nu}\nabla_{\nu}\mathfrak{h}^{(1,0)}\right)+\left[R_{\mu}\mathcal{Q}^{\mu}_{\;\alpha}q^{\alpha\nu} + c_s^2 \mathcal{Q}^{\nu}_{\;\alpha}B^{\alpha}\right]\nabla_{\nu}\mathfrak{h}^{(1,0)} -\left[i\alpha\tilde{\omega} - (c_s^2\nabla_{\mu} + R_{\mu})\mathcal{Q}^{\mu}_{\;\alpha}B^{\alpha} \right]\mathfrak{h}^{(1,0)}=\mathfrak{F}.
\end{align}
with source,
\begin{equation}\label{eq:MasterEquationsSource}
    \mathfrak{F} = \frac{i \alpha \tilde{\omega}c_s^2}{2 }h^{(1,0)}+ \left[c_s^2\nabla_{\mu}+R_{\mu}\right]
\left(\frac{\mathcal{Q}^{\mu}_{\;\nu}\prj^{\nu}_{\;\alpha}f^{\alpha}_{(1,1)}}{\eden+p}\right),
\end{equation}
\end{widetext}
and where, $\prj^{\mu}_{\;\nu}f^{\nu}_{(1,1)}$ is the effective force of the metric perturbation onto the fluid given by Eq.~\eqref{eq:Effective force}. Unlike in \cite{Ipser91,Ipser92}, we have derived our equations in terms of the full fluid perturbation as opposed to its projected form $\hat{U}_{(1,0)}^{\mu} = q^{\mu}_{\;\nu}U_{(1,0)}^{\nu}$, allowing for explicit inversion of $\mathcal{Q}^{-1}$ provided $m\neq0$. This choice has the subsequent effect of not automatically conserving the normalization of our four flows ($U_{\mu} U^{\mu}=-1$) to order $\mathcal{O}(\epsilon^2)$. We remedy this by introducing a perturbative non-affine re-parameterisation of our non-linear four flow,
\begin{align}
    \tilde{U}^{\mu}(\tilde\chi) &= \frac{d\chi}{d\tilde\chi}\,U^{\mu}(\chi) \nonumber \\
    &= U^{\mu}(\chi) - \epsilon f'(\chi)\,U^{\mu}(\chi) + \mathcal{O}(\epsilon^2),
\end{align}
with $\tilde\chi = \chi + \epsilon f(\chi)$,where $\chi$ is the original parameter along the flow lines, $\epsilon$ is the same small mass-ratio parameter as before, $f$ is a free function fixed below, and $\tilde{U}$ is the re-parameterised four-flow. We are always free to make such a re-parameterisation, and we use this freedom to impose $\tilde{U}^{\mu}\tilde{U}_{\mu} = -1 + \mathcal{O}(\epsilon^2)$, so that $\tilde{\chi}$ is the proper time along the perturbed flow to this order.

By setting the transformation to take the form,
\begin{equation}
    f' = -\left(U_{\mu}^{(0,0)}U^{\mu}_{(1,0)} + \frac{1}{2}h^{(1,0)}_{\mu\nu}U^{\mu}_{(0,0)}U^{\nu}_{(0,0)}\right),
\end{equation}
we fix the normalisation. and obtain an altered description of our four-velocity given by,
\begin{equation}
    \tilde{U}^{\mu} = U^{\mu}_{(0,0)} + \epsilon \left( U^{\mu}_{(1,0)} - f' U^{\mu}_{(0,0)}\right)+\mathcal{O}(\epsilon^2 \lambda).
\end{equation}
In our case, such a re-parameterisation does not affect the definition of our scalar fluid variables, $\mathfrak{e}$ and $p$, due to their invariance under propagation along the background flow (i.e $U^{\mu}_{(0,0)}\nabla_{\mu}p = 0$). Hence, having reconstructed our perturbed four-flow through the master enthalpy formalism, we simply make the replacement, 
\begin{equation}
  U^{\mu}_{(1,0)} \rightarrow  \tilde{U}^{\mu}_{(1,0)} = U^{\mu}_{(1,0)} - f' U^{\mu}_{(0,0)},
\end{equation}
in all reconstructed quantities. An example of how this reparameterisation fixes the normalization in practice can be found in appendix ~\ref{app:fourflownorm}.

\section{Background disc \& Secondary Metric Perturbation} \label{sec:backgrouninpuot}
Before solving the master enthalpy equation, we first define the background disc state and compute the metric perturbations induced by the secondary black hole. Our background disc-setup closely follows the three-dimensional set-up of~\citet{Tanaka2024} promoted to a fully relativistic formulation in the Kerr geometry. We consider a three-dimensional, locally isothermal thin disc that is poloidally isothermal but allows for radial temperature gradients and constant aspect ratio in the disc.
The metric perturbation induced by the presence of the small secondary BH are then related to a point source on a circular equatorial geodesic in the Kerr geometry. Where,
\begin{equation}\label{eq:Kerr_Metric}
     \begin{aligned}
        ds^2 =& - \frac{\Delta}{\Sigma}\Bigl(dt - a\sin^2\theta d\phi\Bigr)^2 + \frac{\Sigma}{\Delta}dr^2 \\
        &+ \Sigma d\theta^2 + \frac{\sin^2\theta}{\Sigma}\Bigl(adt - (r^2+a^2)d\phi\Bigr)^2 \, ,
    \end{aligned}
\end{equation}
is the line element associated with the Kerr metric in Boyer-Lindquist co-ordinates. With,
\begin{equation}
    \Sigma = r^2 + a^2\cos^2\theta  \,  , \quad \Delta = r^2-2r+a^2    \,  ,
\end{equation}
$a$ being the Kerr spin parameter and we have set $M=1$.
\subsection{Background Thin Disc}
We build our background disc to be a stationary, axi-symmetric, locally isothermal thin disc on the Kerr geometry. Further, we fix our disc to be of constant (small) aspect ratio which we denote $\eta\ll1$.

The introduction of a small aspect ratio introduces a new perturbative parameter in which we project and expand our solutions around the equatorial plane. We will describe this additional expansion in \S\ref{sec:numerical} as the projected modified local approximation (PMLA), in line with previous works by Potter \cite{Potter_2021} and by Tanaka \textit{et al.} \cite{Tanaka2002,Tanaka2004,Tanaka2024}. 

All calculations, unless otherwise stated, are expanded consistently up to and including order $\cos^2\theta=z^2 \sim \mathcal{O}(\eta^2)$ contributions. Our setup begins with the background fluid equation, Eq. (\ref{eq:eulerequations2}),
\begin{align}\label{eq:backgroundfluidEOM}
\left(\eden + p\right) U^{\mu}\nabla_{\mu} U^{\nu} + \prj^{\mu\nu} \nabla_{\mu} p &= 0.
\end{align}
We assume our background flow follows the isometries of the Kerr spacetime, i.e., it is stationary and axi-symmetric, $U^{\mu} = \alpha(r,\theta)\{1, 0,0,\Omega(r,\theta) \}$ in Boyer-Lindquist coordinates and we impose that, $p(r,\theta) = c^2_s(r)\eden(r, \theta)$ fixing the disc to be poloidally isothermal but radially non-isothermal. We assume that the disc is in hydrostatic equilibrium, which reduces Eq.~\eqref{eq:backgroundfluidEOM} to the form,
\begin{equation}\label{eq:backgroundfluidEOMstatic}
    \Gamma^{\mu}_{\alpha\nu}U^{\nu}U_{\mu} = \frac{c_s^2}{1+c_s^2}\frac{\partial_{\alpha}p}{p}.
\end{equation}

Within the thin disc framework, we expect scalings of the fluid variables with the disc aspect ratio to arise analogously to the Newtonian case \cite{2020ApJ...892...65M,2014ApJ...782..112T}, namely, $p\sim \eta^2, c_s \sim \eta,\delta\omega \sim \eta^2,$ and $\delta\alpha \sim \eta^2,$  where $\Omega = \Omega^{k}(1 + \delta\omega)$ is the angular frequency of a fluid element, with angular frequency perturbed by $\delta\omega$ away from the Kerr value ($\Omega^{k}$), and similarly, $\alpha = \alpha^{k}(1 + \delta\alpha)$.
Here it is useful to differentiate the off-equatorial contributions to the angular frequency in the Kerr geometry from the corrections arising due to pressure. The purely geometric terms are 
\begin{align}\label{eq:backgroundVel}
    \Omega^{k} &= \frac{1}{r^{3/2}+a}\left(1 + \frac{r^2+ 2 a \sqrt{r} - 3 a^2}{2\sqrt{r}(a+r^{3/2})}z^2\right),\\
   \alpha^{k} &= \frac{\left(a+r^{3/2}\right)}{\sqrt{2 a r^{3/2}+(r-3) r^2}}\times\\ \nonumber
   &\left(1-\frac{\left(3 a^4-6 a^3 \sqrt{r}+4 a^2 r (r+1)-r^4\right)}{2 r^2 \left(2 a+(r-3) \sqrt{r}\right) \left(a+r^{3/2}\right )} z^2\right),
\end{align}
and the two pressure correction terms are related through, 
\begin{equation}
    \delta\alpha = \frac{\left(a^2-2 a \sqrt{r}+r^2\right) }{2 a r^{3/2}+(r-3) r^2}\delta\omega.
\end{equation}
The angular frequency perturbation $\delta\omega$ is fixed below upon choice of thermodynamic profile and hence pressure gradient in the disc. Next taking Eq.~\eqref{eq:backgroundfluidEOMstatic}, we self-consistently expand the polar equation to $\mathcal{O}(z^2)$ and find a solution for pressure field of the form, 
\begin{equation}\label{eq:pressureprolfe1}
    p(r,z) = p_0(r)e^{- \frac{z^2}{2\eta^2}},
\end{equation}
where we have defined,
\begin{equation}\label{eq:hdefinition}
\eta(r) \equiv c_s\sqrt{\frac{2 a r^{3/2} - (3-r)r^2}{(1+c_s^2)(3 a^2-4a\sqrt{r} + r^2)}},
\end{equation}
as the Boyer-Lindquist analogue of the dimensionless disc aspect ratio.

Fixing $\eta$ to be a constant now fixes the thermodynamic profile of our disc. Inverting the above relation for the sound speed gives,
\begin{equation}
    c_s = \eta \sqrt{\frac{4 a \sqrt{r} -3 a^2-r^2}{3a^2\eta^2+(3+\eta^2-r)r^2-2a\sqrt{r}(2\eta^2+r)}}.
\end{equation}
Substituting this result into the radial component of Eq.~\eqref{eq:backgroundfluidEOMstatic} then fixes the angular frequency perturbation,
\begin{equation}
   \delta\omega = \frac{\eta^2 \left(3 a^2 - 4 a \sqrt{r} + r^2\right) \partial_r p _0}{\left(2 a r^{3/2} + (-3 + r) r^2\right) p_0}.
\end{equation}
The only freedom remaining in our disc it to fix the equatorial density profile. For simplicity and ease of connection to the Newtonian literature, we choose,
\begin{equation}
    \label{eq:background_eofr}
    \eden_{(0,1)} = \eden_I \left(\frac{r}{r_{I}}\right)^{-2},
\end{equation}
where subscript $I$ denotes the value of a variable at the innermost stable circular orbit (ISCO). We remind the reader that in the relativistic case $\mathfrak{e}_{(0,1)}$ comprises both mass and internal energy density. We have now self-consistently fixed our background disc profile with pressure and constant scale height in the Kerr geometry \footnote{In order to obtain an asymptotically flat solution one would require a density profile which decays at least as fast as  $\sim r^{-3}$. This is not a problem for or purposes however as we will only be concerned with the fluid dynamics within a finite extent of the central SMBH.}.

\subsection{Point Particle Metric Data}\label{app:Lorenz} 

Having formulated our background disc profile we now turn to the construction of the metric perturbation due to the small body. The self-force formalism provides a first-principles framework for describing asymmetric binaries in a systematic expansion in their mass ratio \cite{Pound:2017psq, Mino_1997,Pound:2012dk,Pound_2021}. Through this formalism, one finds that to leading order in the global spacetime, the presence of the smaller body can be represented as a point particle, the metric perturbation for which is governed by,
\begin{equation}
    E_{\mu\nu}[h^{(1,0)}] = T^{pp}_{\mu\nu}.
\end{equation}
Where $E_{\mu\nu}$ is the linearised Einstein operator and $T^{pp}_{\mu\nu}$ is the stress energy tensor for a point particle on a fixed orbit (Eq.~\eqref{eq:ppstressenergy}).

\subsubsection{Lorenz Gauge Solution}
General relativity naturally leaves a wide class of gauge freedom in the calculation of such metric perturbations. Some gauges, such as radiation gauges \cite{Toomani:2021jlo}, lead to extended singular structures that are not localised at the position of the small body, making them difficult to use as sources in higher-order equations of motion. Among the available gauges, the most practical in the Kerr geometry has been the Lorenz gauge, defined by the condition
\begin{equation}
\nabla^\mu \bar{h}_{\mu \nu}^{(1,0)} = 0,
\end{equation}
where $\bar{h}_{\mu \nu} = h_{\mu \nu} - \frac{1}{2} g_{\mu \nu} h$ is the trace-reversed metric perturbation. Within this gauge, metric perturbations take the familiar singular structure of $h^{(1,0)}\sim 1/|\vec{r} - \vec{r} _{\rm p}|$ near the object, where $r_p$ is the position of the secondary, keeping in close connection with the Newtonian analogue. Recent breakthroughs have been made in calculating such perturbations in the Kerr geometry. Dolan et al.~\cite{Dolan_2022,Dolan_2024, Wardell_2025} have developed a prescription for constructing the Lorenz-gauge metric perturbation on the Kerr background from scalar variables that satisfy decoupled, separable equations \cite{Teukolsky73}. In this paper, we make use of the implementation in Ref.~\cite{Dolan_2024} to compute the metric perturbation of a pointlike body on a circular equatorial orbit of a rotating BH in Lorenz gauge.

Notably, within the class of point-particle metric solutions, the Lorenz gauge is inherently asymptotically irregular. In order to obtain a solution that is regular in this gauge, one must remove the static monopole and dipole perturbations of the system, which arise due to the presence of the small body. This class of modified Lorenz-gauge solutions is commonly referred to as being within the Berndtson class of solutions \cite{berndtson2009harmonicgaugeperturbationsschwarzschild,Dolan_2024,Miller:2020bft}. We work within this class of solutions.

\subsubsection{Metric Data Construction}

To numerically produce the required metric data, we follow the approach taken in \cite{Dyson:2025dlj}. Our goal is to construct the Fourier $m$-modes of the point particle metric perturbations denoted, $h^m_{i j}$, whereby,
\begin{equation}
h_{i j} = \sum_m h^m_{i j} e^{i m \phi-i \omega_m t}
\end{equation}
and, in this context tetrad components are denoted by Latin indices. To do so, we first construct the Kinnersley tetrad components of the Lorenz gauge solutions projected onto a spin-weighted spherical harmonic basis, as in ~\cite{Dolan_2024}. Having obtained the spherical harmonic modes, we then construct the $m$-modes through,
\begin{equation}
    h^m_{i j}(r,\theta) = \sum_{\ell = \max(m,s)}^{\ell_{\rm max}} h^{\ell m}_{i j}(r)\;_sY_{\ell m}(\theta,\phi=0),
\end{equation}
typically taking $\ell_{\rm max} = 40$. The spin weight $s$ is determined by the particular tetrad component being constructed. Having built the $m$-mode tetrad components, we numerically transform from the Kinnersley tetrad to Boyer–Lindquist coordinate components through the change of basis operation, 
\begin{equation}
h_{\mu\nu}^{(m)} = \sum_{i j}\tilde{Z}^{i}_{\mu}\tilde{Z}^{j}_{\nu}h_{ij}^{(m)},
\end{equation}
with $\tilde{Z}^{i}_{\mu} = \{ -n_{\mu}, -l_{\mu}, \bar{m}_{\mu}, m_{\mu} \}$, and where $l, n$, and $m$ are the standard principal null directions of the Kinnersley tetrad in Kerr,
 \begin{align*}
    l^{\mu} &= \left(\frac{r^2+a^2}{\Delta},1,0,\frac{a}{\Delta} \right),\\
    n^{\mu} &= \frac{1}{2 \Sigma}\left(r^2+a^2,-\Delta,0,a \right),\\
        m^{\mu} &= \frac{1 }{\sqrt{2}(r+ i a \cos \theta)} \left( i a \sin \theta,0,1,\frac{i}{\sin\theta} \right).\\
 \end{align*}

\begin{figure}[!htbp]
    \centering
    \includegraphics[width=1\columnwidth]{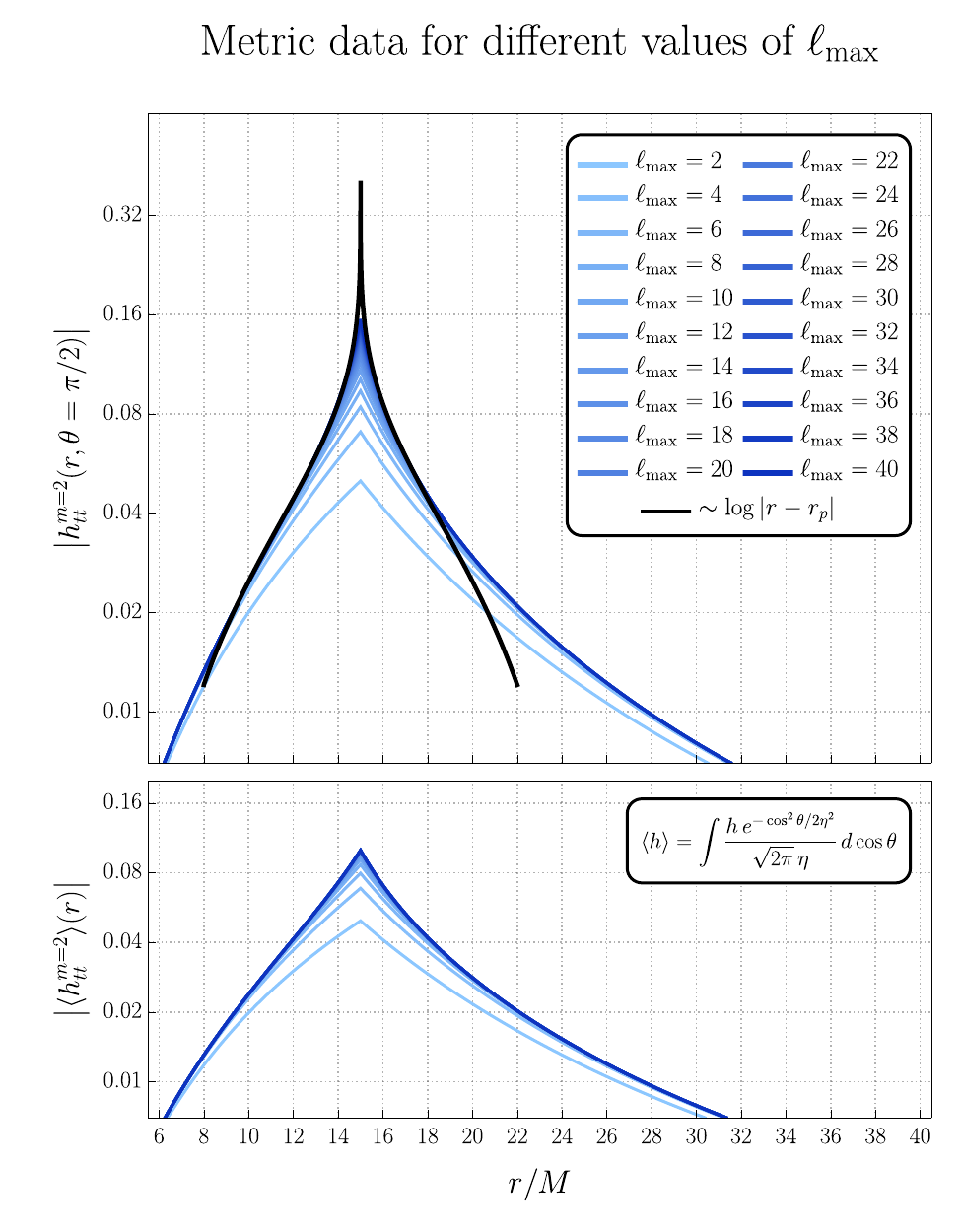}
       \caption{Convergence of the $m = 2$ metric perturbation $|h^{m=2}_{tt}|$ with the number of spherical-harmonic modes, for $a = 0.6M$ and $r_p = 15M$. The lightest line corresponds to $\ell_{\rm max} = 2$, with successively darker lines representing $\ell_{\rm max} = 4, 6, \dots$ up to $\ell_{\rm max} = 40$. \textit{Top panel:} the equatorial slice $\theta = \pi/2$, with the expected logarithmic divergence at the particle, $\sim \log|r - r_p|$, overlaid in black; near $r_p$ the partial sums approach it only slowly. \textit{Bottom panel:} the same partial sums after the height average which we will introduce in Eq.~\eqref{eq:heightprojector} with $\eta = 0.1$, the form in which the metric enters our equations. The average removes the divergence, and the sums converge rapidly in $\ell_{\rm max}$.}
    \label{fig:metricdata}
\end{figure}

\subsubsection{Singular Structure}

As stated previously, the Lorenz gauge solutions have a special singularity structure near the secondary, making them particularly suitable for use in the source function of our equations. Due to the act of projection in the angular decomposition, Lorenz gauge solutions have different levels of regularity depending on whether one considers the four-dimensional, Fourier $m$-mode, or spherical harmonic $\ell$-mode solutions. The general singularity structure in each case is given by,
\begin{align*}
h_{\mu\nu}^{4d}|_{\vec{r}\rightarrow \vec{r}_p} &\sim \frac{1}{|\vec{r}-\vec{r}_p|},\\
h_{\mu\nu}^{m}|_{\vec{r}\rightarrow \vec{r}_p} &\sim \log{|\vec{r}-\vec{r}_p|},\\
h_{\mu\nu}^{\ell m}|_{r\rightarrow r_p} &\sim |r-r_p|.\
\end{align*}
In our approach we construct our $m$-modes from a finite sum of spherical-harmonic $l$-modes, resulting in $m$-mode metric data that exhibits sharp but continuous behaviour across the particle’s location, rather than the true logarithmic behaviour of the exact solution as can be seen in the top panel of Fig.~\ref{fig:metricdata}. Therefore the $m$-mode metric data remains poorly converged near the particle, even at these high values of $\ell_{max} = 40$. In practice, for our application to thin discs, the metric will only enter through its height averaged form, leading to an input that is well converged as can be see in the bottom panel of Fig.~\ref{fig:metricdata}.

\section{Solution Method and Implementation}\label{sec:numerical}
Having obtained the master enthalpy equation \eqref{eq:MasterEquations}, and all necessary background and metric contributions, we are poised to begin solving our equation. In its current form, it is a second-order linear partial differential equation. Rather than solving the equation in this form explicitly, we will make use of the thin disc profile of our background solution to project to a system of ordinary differential equations. To define the working equations, we perform a vertical (height) averaging as in Potter \cite{Potter_2021} and adopt a modified local approximation (as described in \citet{Tanaka2024}) of the master enthalpy equation, Eq.~(\ref{eq:MasterEquations}). We denote this combination of approaches the Projected Modified Local Approximation (PMLA). This procedure reduces the problem to a system of de-coupled radial ordinary differential equations and corresponds to solving an equation akin to the fully relativistic analogue of the dominant $n=0$ mode identified in \cite{Tanaka2024}. In this sense we solve for the dominant cylindrical contribution to the three-dimensional fluid configuration rather than taking a two-dimensional disc set-up a priori. We detail this set-up in the following sub-sections.

\subsection{Height Averaging the Master Equation}
We follow closely the approaches in previous Newtonian works, where we consistently height-average our equations under the operator,
\begin{equation}\label{eq:heightprojector}
  \left<\cdot\right>= \int\frac{(\cdot )e^{ -\frac{z^2}{2\asp^2}}}{\sqrt{2 \pi}\asp}dz,
\end{equation}
where $z=\cos\theta$ and $\asp$ is fixed to be a constant and defined through Eq.~\eqref{eq:hdefinition}. 

We simultaneously re-expand in the aspect ratio and retain all terms up to and including $\asp^2$ in the modified local approximation, i.e., the PMLA. The resulting equations can be seen as governing the solution to the zeroth order ($n=0$) term in a Hermite polynomial expansion of the polar degree of freedom. The notebooks providing the explicit form of these projected equations will be provided publicly in forthcoming work.

In order to obtain a decoupled system of equations at this level we also take advantage of the up-down symmetry inherent in our background disc and equatorial orbits. This property prevents the existence of any terms of the form, $\asp$ or $z$ in our master equation, problematic terms of the form $z \partial_z\mathfrak{h}$ or $ \partial^2_z\mathfrak{h}$ still arise however. We overcome the existence of such terms by integrating by parts until no further poloidal derivatives act on the master variable within the inner product defined by Eq.~\eqref{eq:heightprojector}.

After averaging Eq.~\eqref{eq:MasterEquations}, appropriately reducing the derivatives through integration by parts, and re-expanding our equation to $\mathcal{O}(\asp^2)$, we arrive at a system of decoupled ordinary differential equations, 
\begin{equation}
    \mathcal{L}_m[\mathfrak{h}(r)] = \mathfrak{F}_m(r),
\end{equation}
within the PMLA. Where $\mathfrak{h}(r)$ is defined by,
\begin{equation}
    \mathfrak{h}(r):=  \int\frac{\mathfrak{h}(r,z)e^{ -\frac{z^2}{2\asp^2}}}{\sqrt{2 \pi}\asp}dz
\end{equation}
and is hence formally given as the zeroth order term in a Hermite polynomial expansion of $z$ in the field,
\begin{align}
    \mathfrak{h}(r,z) &= \mathfrak{h}_0(r) + \mathcal{O}(z^2) + \cdots,
\end{align}
where the coefficient of the $\mathcal{O}(z)$ term is necessarily zero due to the up-down symmetry of our problem. Hence, for the fiducial scale height parameter we will employ in this work ($\eta = 0.1$), we expect an error of $\sim \mathcal{O}(z^2)\sim \mathcal{O}(\eta^2)\sim 1\%$ between our projected solution and the full 3D solution evaluated on the equatorial plane (See Tanaka \cite{Tanaka2002} for a direct comparison between the dominant $n=0$ mode and the subdominant vertical modes).

We also reduce  $\mathcal{L}_m[\mathfrak{h}(r)]$ such that the coefficient of the principal piece is moved into the sourcing function, leading to a form,
\begin{equation}\label{eq:reduceeq}
   \mathfrak{h}_m''(r)+ A_m(r)\mathfrak{h}_m'(r)+ B_m(r)\mathfrak{h}_m(r) = \mathfrak{F}_m(r).
\end{equation}
At the level of the operator, we retain the leading two orders in $\asp$, given by $A_m \sim A^{(0)}(r) + \mathcal{O}(\asp^2)$ and $B_m = B^{(-2)}(r)/\asp^2 + B^{(0)}(r) + \mathcal{O}(\asp^2)$.

We have now dropped the perturbative indices for readability. The term $B^{(0)}(r)$ contains all information regarding the pressure and thermodynamic structure of the background disc; all other terms are purely geometric, arising from the geodesic structure of the Kerr spacetime.

We solve this system of equations using the standard variation-of-parameters approach: we first build the Green's function of the system from its homogeneous solutions, and then integrate it directly against the source. The source function is constructed with the same PMLA approximations. The point-particle metric perturbation enters through its height-averaged form using Eq.~\eqref{eq:heightprojector}. This leads to an integrable potential, which does not include any sense of an ad hoc softening length scale that are common in such calculations, instead arriving at a relativistic form similar to the the $m$-mode decomposed potential in Eq.~(83) of \cite{brown2024horseshoesspiralwavescapturing} .
\subsection{Operator Analysis \& Resonant Locations}
In constructing solutions to Eq.~\eqref{eq:reduceeq}, it is useful to first analyse the structure of the differential operator and its associated Wronskian. The Wronskian can be written as, 
\begin{equation}\label{eq:wronsdef}
    \mathcal{W}_{m}(r) = \mathcal{W}^{p}_{m} e^{-\int^r A_{m}(s)\, ds}.
\end{equation}
Here $A_m(s)$ is as defined in Eq.~(\ref{eq:reduceeq}),
and $\mathcal{W}_p$ is an unspecified constant that will be fixed by boundary conditions. Notably, even given the complex structure of our master enthalpy equation, it is possible to find an analytic solution for the Wronskian, given by 
\begin{align}\label{eq:wronskain}
    \mathcal{W} &= \mathcal{W}_p
    \frac{m^2 \left(\Omega^{k}_{r}-\Omega^{k}_{r_p}\right)^2-\kappa_r^2}
    {\left(2 a +(r-3)\sqrt{r}\right) (\Omega^{k}_{r})^2 (\Omega^{k}_{r_p})^2 \Delta_r}
    \sqrt{r}.
\end{align}
Here we have defined the radial epicycle frequency in Kerr by \cite{Torok:2005ct},
\begin{equation}
     \kappa_r^2 = \frac{-3 a^2 \sqrt{r} + 8 a r + (r-6) r^{3/2}}
     {r^{5/2} \left(a+r^{3/2}\right)^2}.
\end{equation}
The circular geodesic frequency in Kerr is now given by it purely equatorial form, $\Omega^k = 1/(a+r^{3/2})$, and the pressure and height corrections present have been re-expanded and now arise as subleading terms, beyond second order in the expansion of $A_m(s)$.

We find that the Wronskian vanishes precisely at points where the difference in angular frequency between the secondary and a fluid element coincides with the fluid’s radial epicycle frequency, leading to potential degeneracies in the solution space at these locations. These points are commonly known as Lindblad resonance locations and have been extensively studied in the literature \cite{Goldreich79,Goldreich80}. In nearly Keplerian discs, these resonances are well known to be the primary mechanism for the transfer of energy and angular momentum between the secondary and the fluid. For a given $m$-mode, each resonance occurs as a pair: one at an orbital radius interior to the secondary’s orbit and one at an orbital radius exterior to it. 

\begin{figure}[htbp]
    \centering
    \includegraphics[width=1.0\columnwidth]{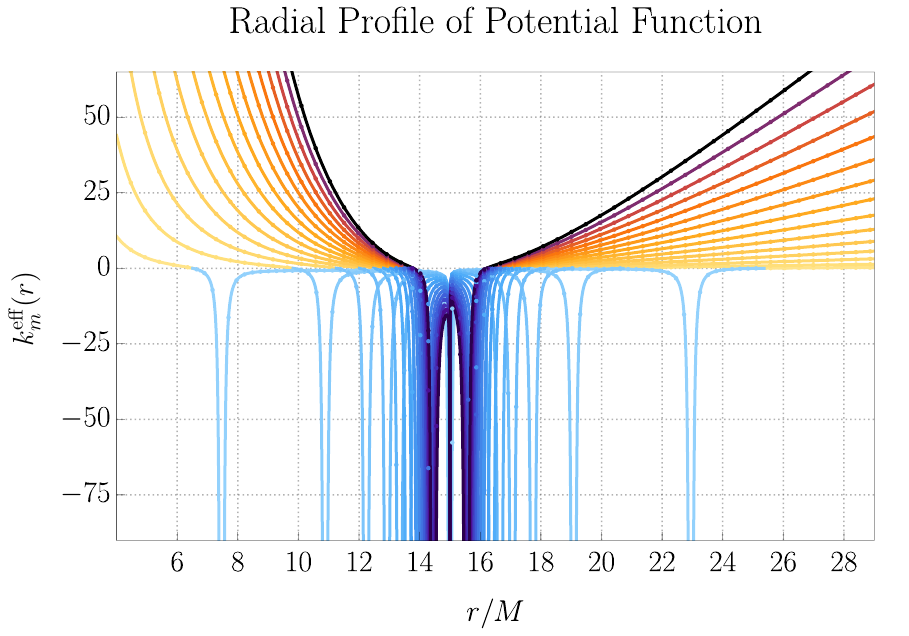}
    \caption{Plot exemplifying the structure of the effective wave-number as given in Eq.~(\ref{eq:BoundaryConditions}), of the enthalpy wave equation for the fiducial parameter set, $r_p = 15$, $\asp = 0.1$, and $a = 0.6$, for a range of $m$-modes. Key to this plot is the identification of wave regions, i.e., when the potential is positive, as opposed to excitation regions where the potential is negative. Naturally, for higher $m$-modes, the excitation region contracts towards the location of the secondary. Here the lightest lines, with divergences further from the particle location, correspond to the $m=1$ potential while subsequent darker lines, which diverge closer to the secondary, represent the potential of the $m=2$, $m=3 \cdots,m=15$ potential functions. The differential in colour schemes is simply to make clear the positions at which each equation transitions from a wavelike to an exponentially growing character.}
    \label{fig:OperatorPlot}
\end{figure}

As shown in Fig.~\ref{fig:OperatorPlot}, for a given $m$-mode the region between these two locations (which act as turning points) contains a negative effective potential in Eq.~\eqref{eq:reduceeq}. In this region, fluid perturbations exhibit exponential growth where waves are ``excited''. Outside these turning points, the potential becomes positive, yielding wavelike solutions that propagate through the disc.

Importantly, the PMLA does not eliminate the effect of pressure, which would normally shift the resonance locations; instead, it effectively re-expands these shifts and places them in the potential describing the wave zone, so that mode propagation can always occur at a finite distance from the secondary, resulting in convergent torques and mode sums. This can be seen in Fig.~\ref{fig:OperatorPlot} through the mismatch in the positions of the Lindblad resonances (which now occur at their pressure-less positions) and the turn over from positive to negative signature in the effective potential.

\begin{figure}[htbp]
    \centering
    \includegraphics[width=1\columnwidth]{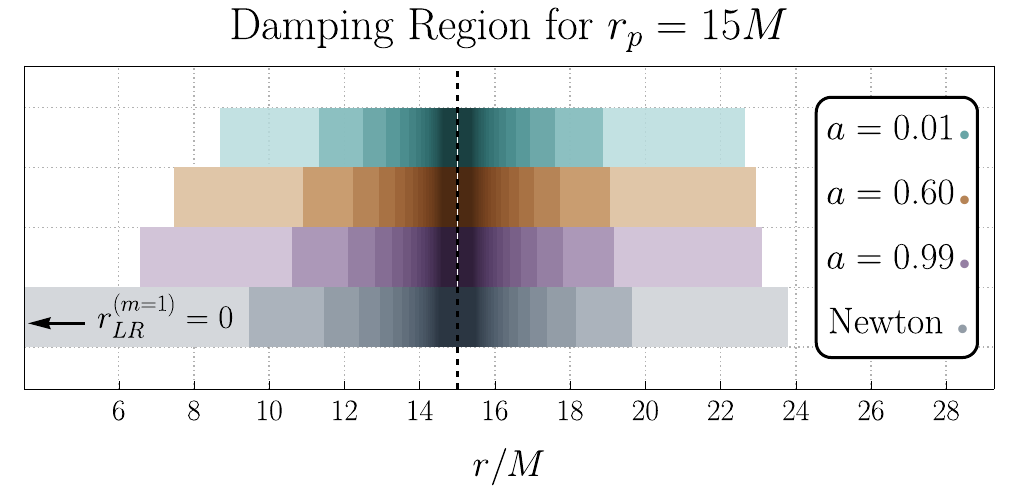}
    \caption{Illustration of the damping regions arising between Lindblad resonances for a range of black hole spins at $r_p=15M$, and for a reference Newtonian analogue. The outermost region, shown as the lightest (non-white) colour on each bar, corresponds to the area between the $m=1$ inner and outer Lindblad resonances. Moving inward from the edges toward the secondary location, each successive darkening of the colour indicates crossing an additional resonant point. This colour progression provides a visual representation of how the resonance structure and excitation regions evolve with black hole spin compared to the Newtonian case.  }
    \label{fig:DampedRegions}
\end{figure}

 As illustrated in Fig.~\ref{fig:DampedRegions}, the positions of relativistic Lindblad resonances differ significantly from their Newtonian counterparts where notably, in the Newtonian case, the $m=1$ resonance does not have an inner counterpart, since its location is fixed at $r = 0$. In the relativistic case, all spin values produce an inner resonance for that mode which can subsequently contribute to the system’s dynamics. Moving from the Schwarzschild to Kerr, spin generally widens the separation between the resonances, which may contribute to the decrease in total torque with spin found in \cite{Duque:2025yfm,HegadeKR:2025dur} and in Sec.~\ref{sec:Torque}. In addition, increasing spin draws the inner resonance closer to the central black hole at a faster rate than it pushes the outer resonance outward.

An important but often overlooked aspect in relativistic disc–EMRI studies is the role of transfer of angular momentum between the orbiting body and fluid elements at these resonant location. In systems with EMRIs in near co-rotation with the fluid it is these interactions, rather than classical dynamical friction that dominate the system’s evolution. Consequently, strong-field EMRI calculations that model disc effects solely through a dynamical-friction force are likely to miss the dominant contribution.

\subsection{Boundary Conditions}

In order to build the solutions of Eq.~\eqref{eq:reduceeq} we must first define the physical boundary conditions for our master equation $\mathfrak{L}[\mathfrak{h}_m] = 0$. 
We fix ingoing and outgoing wave conditions at $r_{\min} = r_I + 0.1M$, just outside the ISCO, and at $r_{\mathrm{max}} = 80M$, respectively,
thus enforcing that we do not have reflections from waves at the boundaries of our domain. Similarly to \cite{Tsang_2014, Miranda19,Miranda20}, we perform a second-order WKB approximation of the field at the boundaries, to give
\begin{align}\label{eq:BoundaryConditions}
    \frac{d\mathfrak{h}^{\pm}_m}{d r} &=\pm i k_m^{\mathrm{eff}}\mathfrak{h}_m^{\pm}\\ \nonumber
    &:= \pm i(k_m \pm \frac{i}{2k_m}\frac{d k_m}{dr}\pm  \frac{i}{2} A_m){\mathfrak{h}^{\pm}_m}.
\end{align}
corresponding to outgoing ($+$) solutions at $r_{\mathrm{max}}$ and ingoing solutions ($-$) near the ISCO. Wherein we have defined the wavenumber as, 
\begin{equation}
\label{eq:WKBwavenumber}
k^2_m(r) = B_m - \frac{1}{2}\frac{d A_m}{dr}- \frac{1}{4}A_m^2.
\end{equation} 
Having specified the ingoing and outgoing boundary conditions, we are now primed to solve for the homogeneous solutions required for our variation-of-parameters approach. 
The homogeneous solutions however exhibit integrable divergences at co-rotation $(r=r_p)$, which must be treated before proceeding.

\subsection{Divergence at Co-rotation}

The co-rotation resonance is another point of extensive interest in the planetary migration community, arising at the point where the orbital frequency of fluid elements match the frequency of the secondary perturber. Performing a local expansion of our height-averaged equations, we find that our master equation at that point takes the following leading-order form,
\begin{equation}
    \mathcal{L}[\mathfrak{h}]|_{r\rightarrow r_p} = \frac{d^2\mathfrak{h}}{dr^2} - \frac{4 \kappa_{r_p}^2}{9 m^2 r_p \Delta_{r_p}(\Omega^k_{r_p})^4(r-r_p)^2} \mathfrak{h},
\end{equation}
where $r_p$ is the location of the secondary. We find that our operator develops a regular-singular point at this location, making it ill-suited for numerical implementation. Notably we do not find that an irregular-singular point arises at co-rotation as in \citet{Tanaka2024}. We attribute this to the fact that our set-up essentially restricts to the $n=0$ mode of \cite{Tanaka2024} and we expect future works focused on solving for the full three-dimensional structure would arrive at similar difficulties. Although the singularity is a regular-singular point, numerical integration through this point is still troublesome.  
We circumvent this issue by removing its singular behaviour analytically, solving numerically for the residual regularised field, and reinstating the singularity after the fact. Performing a local analysis of the operator $\mathcal{L}_m[\mathfrak{h}_m]=0$ around the point $r=r_p$, we find two independent solutions, one regular and one singular. The singular solution takes the form,
\begin{equation}\label{eq:divergentfact}
   S^p_m =  (r-r_p)^{\frac{1}{2} \left(1-\sqrt{1+\frac{16 \kappa_{r_p}^2}{9 m^2 r_p \Delta_{r_p}(\Omega^k_{r_p})^4}}\right)}.
\end{equation}
 For fiducial parameters $a = 0.6M$, $r_p = 15M$, and $m = 2$, we find $S^p_m \sim (r - r_p)^{-0.08}$ giving rise to an integrable divergence. In order to remove this singularity, we substitute $\mathfrak{h}_m = S^p_m \psi_m$ into Eq.~(\ref{eq:reduceeq}) and obtaining a new equation $\hat{\mathfrak{L}}[\psi_m] = 0$, which is now amenable to simple inbuilt ODE solvers in \texttt{Mathematica}.

\subsection{Homogenous Solutions}
\begin{figure}[htbp]
    \centering
    \includegraphics[width=1\columnwidth]{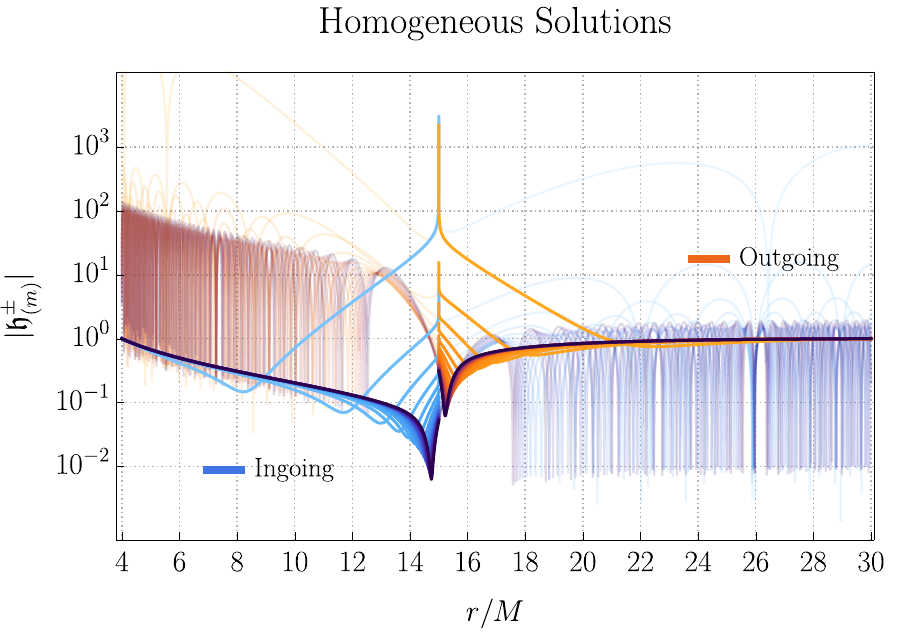}
    \caption{The independent $m$-mode solutions to the homogeneous enthalpy equations using the same lighter-to-darker colour scheme employed in previous figures to indicate increasing $m$-mode contributions. This example corresponds to the fiducial parameter set $r_p = 15$, $\asp = 0.1$, and $a = 0.6$. Solutions satisfying outgoing boundary conditions at the outer boundary are shown in {\it orange}, while those satisfying ingoing boundary conditions at the inner boundary are shown in {\it blue}. The ingoing solutions exhibit smooth behaviour within the domain interior to the particle (opaque blue) and rapidly oscillatory behaviour in the exterior domain (translucent blue). The use of differing opacities allows for easier identification of the uniform solution behaviour across the respective domains. }
    \label{fig:homogenoussolutons}
\end{figure}

Having appropriately rescaled the master equation through the introduction of the divergent factor in Eq.~\eqref{eq:divergentfact} and similarly rescaled the boundary conditions, we compute the homogeneous solutions for $\psi_m$ corresponding to both ingoing and outgoing boundary conditions on a domain running from $r_{\min} = r_I + 0.1M$ to $r_{\max} = 80M$. We then analytically reinstate the correct singularity structure at co-rotation, yielding the full homogeneous ingoing ($\mathfrak{h}^-_m(r)$) and outgoing ($\mathfrak{h}^+_m(r)$) solutions across the computational domain. In Fig.~\ref{fig:homogenoussolutons}, we present these solutions, with ingoing modes shown in {\it blue} and outgoing modes in {\it orange}. The absolute value of the ingoing solutions displays smooth behaviour within the domain interior to the particle (opaque blue), while it shows rapidly oscillatory behaviour in the exterior domain (translucent blue). The opposite pattern is observed for the outgoing solutions. As in Fig.~\ref{fig:OperatorPlot}, the lightest lines in each colour scheme correspond to the $m=1$ mode, with successively darker lines representing $m=2, m=3, \ldots$, up to $m_{\rm max}=15$.

\subsection{Particular Solutions}
Having obtained our homogeneous solutions we build the full particular solution through the method of variation of parameters. Whereby our solution is written in the form,
\begin{equation}
    \mathfrak{h}_m(r) = C^+_m(r)\mathfrak{h}^+_m(r)+ C^-_m(r)\mathfrak{h}^-_m(r),
\end{equation}
with, 
\begin{align}\label{eq:coefficeints}
    C^+_m(r) &= \int_{r_{\mathrm{min}}}^{r}\frac{\mathfrak{h}_m^-(s)\mathfrak{F}_m(s)}{\mathcal{W}^{+}_m(s)}ds, \\
    C^-_m(r) &= \int_{r}^{r_{\mathrm{max}}}\frac{\mathfrak{h}_m^+(s)\mathfrak{F}_m(s)}{\mathcal{W}^{-}_m(s)}ds. \nonumber
\end{align}

The Wronskian is defined as in Eq.~\eqref{eq:wronsdef},
and the associated Wronskian coefficient, $\mathcal{W}p$, is fixed via

\begin{equation}\label{eq:wronskcoefs}
\mathcal{W}p^{\pm}
\equiv
\mathfrak{h}^- \frac{d\mathfrak{h}^+}{dr}-\mathfrak{h}^+ \frac{d\mathfrak{h}^-}{dr}
\bigg|_{r = r_{\mathrm{max}}, \; r = r_{min}}.
\end{equation}

In practice, we compute two distinct Wronskian coefficients: $\mathcal{W}p^{+}$ evaluated at the outer boundary $r = r_{\mathrm{max}}$, and $\mathcal{W}_p^{-}$ evaluated at the inner boundary $r = r_{min}$. The introduction of separate Wronskian coefficients for the $C^{+}$ and $C^{-}$ normalisation constants is motivated purely by numerical considerations. In particular, we find that the delicate cancellations required to reconstruct the fluid variables are significantly more stable when the Wronskian is fixed independently at each boundary. 

We find that the integrals in Eqs.~\eqref{eq:coefficeints} diverge due to contributions at the inner and outer Lindblad resonances, which corresponds to the zeros of the Wronskian in Eq.~\eqref{eq:wronskain}, where our solution space becomes degenerate.

\begin{figure}[htbp]
    \centering
    \includegraphics[width=1\columnwidth]{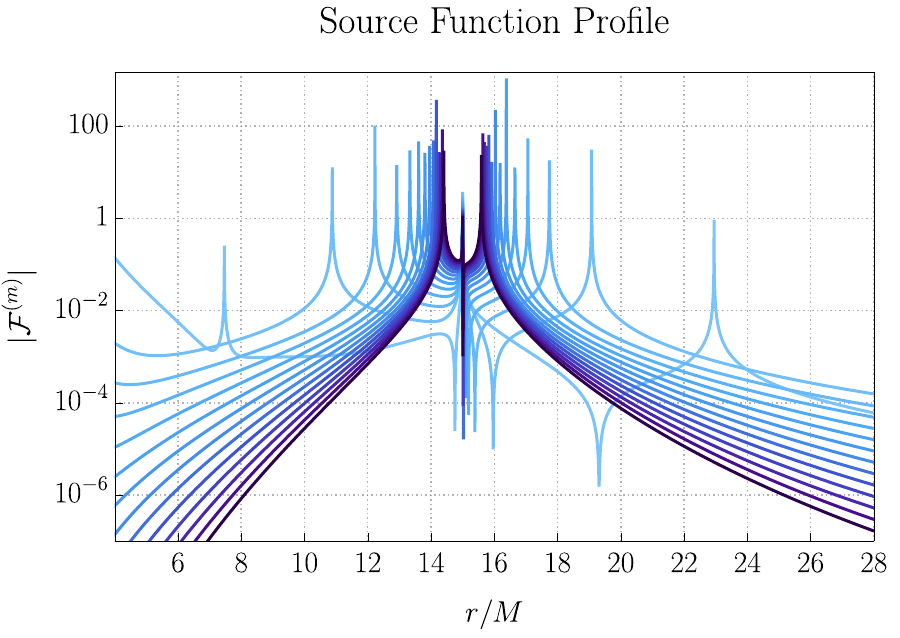}
    \caption{Plot exemplifying the sourcing function, Eq.~(\ref{eq:MasterEquationsSource}), of the master enthalpy equation with contributions from a high-accuracy interpolant of the effective forcing function, Eq.~(\ref{eq:Effective force}), built using the height averaged metric data contracted with the inversion matrix $\mathcal{Q}$ in the PMLA. The example is given for the fiducial parameter set $r_p = 15$, $\asp = 0.1$, and $a = 0.6$ for a range of $m$-modes. Notably, this plot explicitly shows the Lindblad resonant locations.}
    \label{fig:sourceplot}
\end{figure}

Unlike in numerical schemes normally employed in the Newtonian literature \cite{Miranda19, Miranda20, Fairbairn22, Fairbairn_2025}, the variation of parameters scheme causes these resonances to arise as $1/\delta r^2$ divergences in the integrands of Eqs.~\eqref{eq:coefficeints} (where $\delta r$ is the distance from the divergent location). This arises because $\mathcal{W} \sim \delta r$ and $\mathcal{F} \sim 1/\delta r$ near the resonance, leading to a total divergence of the form $1/\delta r^2$. Application of our method to the Newtonian analogue in \cite{Miranda20} reveals precisely the same singularity structure. This combination of divergences in the variation of parameters framework leads to non-integrable contributions in our final solution. Examples of the $1/\delta r$ divergences which partially give rise to the non-integrable contributions can be seen explicitly in Fig.~\ref{fig:sourceplot}.

We alleviate these non-integrable contributions 
through Hadamard regularisation \cite{Blanchet_2000}. The principle behind this regularisation procedure is basic: we remove the minimal content of the integrand needed to make it finite. In practice, the approach is carried out as follows: Given an integral, 
\begin{equation}
I(r) = \int_{r_{a}}^r\frac{F(x)}{(r_s-x)^2}dx,
\end{equation}
with a singular part at a unspecified location $r_s\in(r_{a},r)$ and $F'(r_s)=0$ we define its partie-finie contribution as
\begin{equation}
    \text{p.f.}[I](r) =F(r_s)\left[\frac{1}{r_a-r_s} - \frac{1}{r-r_s}\right]+\int_{r_a}^r\frac{F(x)-F(r_s)}{(r_s-x)^2}dx.
\end{equation}
The condition of $F'(r_s)=0$ arises naturally in this problem due to the structure of the enthalpy equation near Lindblad resonances. For a more general construction applicable for a generic smooth function $F(x)$ one would apply the same construction with additional Logarithmic terms. This approach explicitly removes the divergent contribution from the integrand with the characteristic form,
\begin{equation}
\int_{r_s-\epsilon}^{r_s+\epsilon}\frac{F(x)}{(r_s-r)^2}dx \sim\frac{F(r_{s})}{\epsilon},
\end{equation}
leading to a robust method of numerically applying Hadamard regularisation. The key step is to add and subtract the term $F(r_s)\int (x-r_s)^{-2}\,dx$. Then one of the terms is kept inside the integral, so the integrand becomes $F(x)-F(r_s)$, which is integrable. The other is evaluated using the fundamental theorem of calculus (FToC), which gives only the boundary terms. Strictly, this integral is ill-defined across $r_s$, and the FToC cannot formally be applied to it. Applying it nonetheless discards the $2F(r_s)/\epsilon$ divergence, so the added and subtracted terms no longer cancel. Their combination instead carries a $1/\epsilon$ divergence that is designed to exactly cancel the divergence in the original integral, leaving only its finite part.\footnote{By construction, the scheme is therefore equivalent to Hadamard regularisation, and in practice we find it to be numerically stable.}
Performing the same regularisation on the full solution at both Lindblad locations for every $m$-mode gives the full particular solutions to our master equation, 
\begin{equation}
    \mathfrak{h}_m(r) = \text{p.f.}[C^+_m](r)\mathfrak{h}^+_m(r)+ \text{p.f.}[C^-_m](r)\mathfrak{h}^-_m(r).\\
\end{equation}
We calculate the particular solutions for $1 \leq m \leq m_{\mathrm{max}} = 15$, but exclude $m = 1$ from the flux and torque sums that follow. Unlike the $m \geq 2$ modes, the resulting inner $m = 1$ flux behaves irregularly, we attribute to the WKB condition of Eq.~\eqref{eq:BoundaryConditions} not being an appropriate inner boundary condition for this mode.

\subsection{Reconstruction of Fluid Variables}
We next turn to reconstructing our fluid variables. We do so by first applying the PMLA to the velocity reconstruction operator given in Eq.~\eqref{eq:VeolcityReconstructionecon}. In performing the reconstruction we find a delicate cancellation between terms linear in the enthalpy and terms linear in the metric perturbation. Schematically, we find that the operator $\mathcal{Q}$ diverges as $\mathcal{Q} \sim 1/\delta r$ in the vicinity of Lindblad resonances. In contrast, the terms involving the enthalpy and metric perturbations remain finite. However, their combined contribution on the right-hand side of Eq.~\eqref{eq:VeolcityReconstructionecon} vanishes linearly, scaling as $\sim \delta r$, at the Lindblad resonances. This delicate cancellation serves to recover the correct smooth velocity profile of the fluid away from co-rotation. We find that this cancellation can be made numerically stable by fixing two separate Wronskian coefficients, one at each domain boundary, as in Eq.~\eqref{eq:wronskcoefs}. Since this cancellation occurs exactly at the positions where Hadamard regularisation is applied, it further suggests that the need for regularisation arises only as a divide-by-zero-type divergence in the enthalpy construction, and is not a physical divergence associated with any of the true fluid properties. Having calculated the four-velocity of our fluid perturbations, it is then straightforward to calculate the pressure and density perturbations through Eqs.~\eqref{eq:PresEnthrecon} and \eqref{eq:denEnthecon}.

\section{Solution Properties}
\label{sec:Solutions}

Having derived the master equation for non-barotropic linear perturbations to a perfect fluid in the Kerr geometry (\S\ref{sec:pertscheme}, \S\ref{sec:masterenthalpy}), and having developed a solution method for a thin, background disc perturbed by an orbiting secondary (\S\ref{sec:backgrouninpuot}, \S\ref{sec:numerical}), we now explore the character of solutions, focusing on density wave morphology and angular momentum balance in the disc.

\subsection{Spiral Morphology} 
\label{sec:spiral_morph}

\begin{figure}[htbp]
    \centering
\includegraphics[width=1.0\columnwidth]{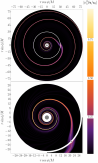}
   \caption{Extended structure of spiral density waves in the Kerr geometry for parameters $\asp = 0.1$, $a = 0.6M$, and $r_p = 30M$.
   {\it Top panel:} Large-scale structure of the disc with the analytic Newtonian spiral-arm prediction \citep{Ogilvie02} overlaid. {\it Bottom panel:} Zoomed-in view of the inner disc, highlighting the detailed arm structure near the secondary and the emergence of both primary and secondary spiral arms near the central black hole. 
   The discrepancy between the Kerr spiral arms and the analytic Newtonian prediction is clearly visible in the inner regions. The plotted quantity is chosen because it exhibits constant perturbation amplitude at large radius.
   We denote $\delta\mathfrak{e} = \mathfrak{e}_{(1,1)}$ and $\mathfrak{e}_{(0,1)} = \mathfrak{e}_0$ in the legends for clarity. 
   }
\label{fig:densities2d}
\end{figure}

In Fig.~\ref{fig:densities2d} we plot the perturbed disc energy density (analogue of Newtonian mass density), showing clear inner and outer spiral arms connecting at the location of the secondary, as expected from the Newtonian case. 

While our solutions span the entire domain, for the purpose of plotting we have excised a region of width $\sim 1M$ around the secondary orbit, as the reconstruction of the 2D solution leads to divergences on the circle that intersects the co-rotation singularity. The inner white circle denotes $r_{min} = r_{ISCO} +0.1M$. To compare the phase structure of the relativistic and Newtonian cases, we overlay the analytical approximation (for $|r-r_p|/r_p\gtrsim h$) for the Newtonian spiral wavefront from \citep{Ogilvie02} (white dashed line).

In the top panel, the agreement appears remarkable, with the outer Kerr spiral slightly more tightly wound (shifted to smaller radii) than the Newtonian spiral. 
Focusing on the inner disc, in the lower panel, reveals the detailed structure of the inner spiral arms and a rapid divergence between the Kerr and Newtonian predictions, with the Kerr arms again more tightly wound.
The bottom panel also clearly demonstrates the emergence of a secondary spiral arm, arising at $r\lesssim0.5r_p$\footnote{Features in the higher-order spirals may not be fully resolved given the maximum $m$-mode considered here.}. \citet{Miranda20} have shown that the emergence of multiple arms in the inner disc is a natural outcome of density wave propagation in differentially rotating discs, making their appearance in the Kerr geometry intriguing but unsurprising.

\begin{figure*}[htbp]  
    \centering

\includegraphics[width=1\textwidth]{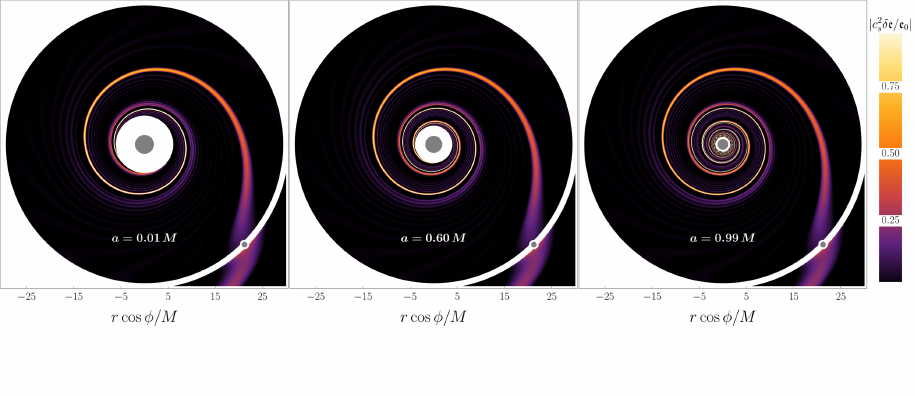}
   \caption{Equatorial-plane density profiles for a fixed disc aspect ratio $\asp = 0.1$ and secondary orbital radius $r_p = 30M$, shown for increasing values of the black hole spin parameter, $a = 0.01M$, $0.6M$, and $0.99M$ from left to right. 
   These panels illustrate the systematic modification of the inner spiral arm structure with increasing black hole spin, highlighting the growing deviation from the low-spin morphology in the relativistic regime. The progression also demonstrates the robustness of our method at high spin values, enabling the resolution of spiral density waves deep in the Kerr potential where relativistic effects are strongest. The legends are the same as described in Fig.~\ref{fig:densities2d}.}
    \label{fig:densitiesspindep}
\end{figure*}

Fig.~\ref{fig:densitiesspindep} explores the change in inner-spiral structure for different values of the central BH spin, at the same binary separation, $r_p=30M$. Notably, the primary effect of the spin is to allow the spirals to propagate more closely to the central black hole due to the contraction of the ISCO. This allows higher order spirals to become even more pronounced in the highest spin, right panel. Beyond this, at fixed radius the inner spiral arms wind slightly more tightly for higher spins. The strength of the spiral also decreases with increasing spin, implying that for a fixed orbital radius the integrated torques may weaken for higher spins. This is consistent with recent findings \cite{Duque:2025yfm,HegadeKR:2025rpr} as well as ours below.

Finally, we note that the close agreement with the Newtonian spiral structure is expected for an equatorial disc with weak pressure gradients, given the similarity in disc rotational frequencies in the Newtonian and Kerr cases $\Omega_{\mathrm{N}} = r^{-3/2}$ while $\Omega_{\mathrm{k}} = (r^{3/2}+a)^{-1}$. Further insight into how spiral wave morphologies depend on spin and disc thickness can be gained by analysing the spiral wavefronts. These can be accessed from the radial wave number in Eq. (\ref{eq:WKBwavenumber}) \citep[][]{Ogilvie02, Rafikov_NLTidWave:2002, Miranda20},
however, we leave further study of the rich spiral structure in the Kerr geometry for a future analysis, and now turn to angular momentum and torque balance in the disc.

\subsection{Torque Density and Advected Angular Momentum Balance}\label{sec:Torque}

In black hole perturbation theory, a common approach to understanding the evolution of a secondary BH in the presence of matter is to first solve for the response of the matter fields and then treat the resulting perturbations as test fields on the background spacetime, exploiting the spacetime's Killing symmetries to define energy and angular momentum fluxes through the horizon and to infinity \citep{Cardoso:2021wlq, Brito:2023pyl,Dyson:2025dlj, Rahman:2023sof, Datta:2025ruh}. These fluxes of energy and angular momentum are then often related directly back to the evolution of the secondary's orbit. 

This approach will encounter significant difficulties when applied to discs with viscous and heat dissipation and, more generally, to matter configurations where the interplay between the matter and secondary do not imprint directly on the fluxes at infinity. In many situations, the matter field and the secondary will exchange energy and angular momentum locally, and the corresponding fluxes measured at the horizon or at infinity will not be representative of the actual exchange governing the orbital evolution of the secondary. 

By contrast, in Newtonian theory and in the hydrodynamical simulation literature, disc–secondary interactions are commonly quantified by computing the torque as a local quantity which is integrated to obtain the total torque exerted by the perturber on the disc and vice versa by Newton's third law.\footnote{See for example \citep[][]{LISA_IMRI:2025}, targeted towards intermediate mass ratio inspirals.} In this section, we outline a pathway for describing an analogous construction of local torque densities within the relativistic framework of black hole perturbation theory, which we hope in future works may allow for a formal connection to the motion of EMRIs in matter. Although presented in the context of accretion discs, the approach we describe applies more generally to any relativistic prescription of matter in a perturbative binary black hole spacetime. Related ideas have appeared previously in the literature for describing the excitation of conserved quantities locally, for example in work associated with the \textsc{GR-Chombo} numerical relativity collaboration \cite{2021JOSS....6.3703A,Clough:2021qlv}; however, while these methods move in a similar direction, they do not directly extend to an analogue of a local momentum–torque balance that we seek to construct.
Our goal here is therefore to formulate a balance equation that relates the local flux of momentum through the matter to the locally applied force the EMRI applies to the matter. The outlook is that integrated matter torques are the physical quantities that may be formally connected to the torques on the EMRI orbit in future studies.

\subsubsection{General Torque Balance Equation}
We motivate a representation of the disc torques that connects directly to the Newtonian literature and takes the representative form described in App.~A of \cite{Miranda_2016}. To this end, we return to the fully non-linear Einstein equations. For circular orbits the primary means for understanding orbital evolution is through angular momentum exchange between the disc and the secondary. In this vein we begin with the equation,
\begin{equation}\label{eq:balanceeqnonoin}
\nabla_{\mu}\left(\xi^{\nu}_{\phi} T^{\mu}_{\nu}\right)
= T^{\mu}_{\nu}\nabla_{\mu}\xi^{\nu}_{\phi}.
\end{equation}
Here, $T^{\mu\nu}$ denotes the full fluid stress–energy tensor on the full spacetime manifold, assumed to satisfy the usual conservation equations, and $\xi^{\mu}_{\phi} = \{0,0,0,1\}$ is the coordinate basis vector associated with the azimuthal direction. At this stage, $\xi^{\mu}_{\phi}$ is not assumed to be a Killing vector.

This expression may be written equivalently as
\begin{equation}\label{eq:torquebalnonlinear}
\partial_{\mu}\left(\sqrt{-g}T^{\mu}_{\phi}\right)
= \sqrt{-g}T^{\mu}_{\nu}\Gamma^{\nu}_{\mu\phi},
\end{equation}
where, $\sqrt{-g}$ is the volume element of the full nonlinear spacetime and all other quantities also remain fully non-linear. We next expand this equation to $\mathcal{O}(\lambda \epsilon^{2})$, the order at which we expect energy and angular momentum transfer to arise.

Before performing this expansion, however, we note that our objective is to obtain an expression for the orbit-averaged torque. Accordingly, we work with the orbit-averaged form of Eq.~\eqref{eq:torquebalnonlinear}. The averaging operator for circular orbits is defined by, 
\begin{equation*}
 \langle X \rangle
= \frac{\omega}{2\pi}\int_{0}^{2\pi/\omega} Xdt  .
\end{equation*}
Here, $\omega$ denotes the orbital frequency of the perturber and may be generalised beyond circular motion to include inclined, eccentric, and fully generic orbits as in \cite{Mathews:2025nyb}. At $\mathcal{O}(\lambda \epsilon^{2})$, we are concerned only with quantities that are either quadratic in first-order perturbations or associated with the static ($m=0$ for circular) modes of the second-order perturbations. Since our solutions are constructed in terms of azimuthal $m$-modes and for circular orbits, quadratic combinations of first-order quantities may be readily obtained by,
\begin{equation}
\langle X Y \rangle = \Re\!\left[\sum_{m} x_{m}\,\bar{y}_{m}\right].
\end{equation}
Where bar is the conjugation operator, $X=\sum x_m e^{im \phi} e^{- i \omega_m t}$ is a generic perturbative term, and similarly for $Y$. In the present work, we solve the field equations only to first order and therefore do not have access to the static second-order contributions that enter the leading-order angular momentum balance. We collectively denote these missing terms by $S^{(2)}_{m=0}$.

At this stage, we must also clarify a simplification employed throughout our prescription. Specifically, we have treated the background disc solution as time independent. Such an assumption is, in general, inconsistent for realistic disc configurations, particularly for adiabatic fluids or discs with heat dissipation or viscosity, where one must account for the secular evolution of the disc. A more appropriate assumption is that the time derivatives of the background disc quantities are small. For example, if the energy density admits the expansion $\mathfrak{e} = \mathfrak{e}_{(0,1)} + \mathfrak{e}_{(1,1)}$, then we require
\begin{equation}
   \frac{\partial \mathfrak{e}_{(0,1)}}{\partial t} \sim \mathcal{O}(\lambda \epsilon^{2}) \, .
\end{equation}

Alternatively, one may appeal to a multiscale expansion and reinstate the independence of the background quantities on the \emph{fast} time ($t$) by introducing a new, independent \emph{slow} time ($T_D$) variable governing the secular evolution of the disc, defined as $T_D = \epsilon^{2} t$. In this framework,
\begin{equation}
   \partial_t \mathfrak{e}_{(0,1)} \;\Rightarrow\;
   \left(\partial_t + \epsilon^{2}\partial_{T_D}\right)\mathfrak{e}_{(0,1)}
   = \epsilon^{2}\frac{\partial \mathfrak{e}_{(0,1)}}{\partial T_D} \, .
\end{equation}
Such slow-time evolution of the background disc parameters is expected even in the example considered here, where the initial thermodynamic profile of our adiabatic fluid must evolve according to the energy equation, as discussed in \cite{Miranda20}.

Then, taking Eq.~\eqref{eq:torquebalnonlinear} , including the new slow timescale, expanding to order $\mathcal{O}(\lambda \epsilon^2)$, and averaging over circular orbits we obtain the new balance equation given by,
\begin{widetext}
\begin{align}
    \label{eq:TorqueEq}
   \partial_{T_D}\left(\sqrt{-g} T^{t}_{\phi}\right) + &\partial_i\left[ \sqrt{-g} \left( \left<\delta^2T^{i}_{\phi}\right> +\left<h \delta T^{i}_{\phi}\right>+ \frac{1}{4}T^{i}_{\phi}\left(\left<h^2\right> - 2 \left<h^{\alpha}_{\beta}h^{\beta}_{\alpha}\right>\right) \right)\right]\\ \notag
   = &\sqrt{-g}\left[ \left<\delta T^{\alpha\beta}\nabla_{\phi}h_{\alpha\beta}\right>
   - 2 T^{\alpha\beta} \left( \left<h^{\;\gamma}_{\alpha} \nabla_{[\gamma}h_{\beta]\phi}\right>
   - \frac{1}{4} \left<h \nabla_{\phi}h_{\alpha \beta}\right> \right) \right]+ S^{(2)}_{m=0},
\end{align}
\end{widetext}
where all gradients and determinants are with respect to the background Kerr geometry, all occurrences of $h$ refer to the metric perturbation and its trace, and $ i \in \{r,\theta\}$. If we now impose in Eq.~\eqref{eq:TorqueEq} our restrictions on the background flow, $U^r = U^{\theta} = 0$, then the explicit contributions to the balance equation become, 
\begin{align}
    T^{t}_{\;\;\phi} = &(\eden+p)U^{t}_{(0,0)}U_{\phi}^{(0,0)},  
\end{align}

\vspace{0.1cm}

\begin{align}
  \delta T^{i}_{\;\;\phi} = &(\eden+p)U^{i}_{(1,0)}U_{\phi}^{(0,0)},
\end{align}

\vspace{0.1cm}

\begin{align}
  \delta T^{\mu\nu} = &2(\eden+p)U^{(\mu}_{(1,0)}U^{\nu)}  \\ \notag
  +&(\eden^{(1,1)}+p^{(1,1)})U^{\mu}U^{\nu}\\ \notag
  + &p^{(1,1)}g^{\mu\nu} - p h^{\mu\nu}_{(1,0)}, 
\end{align}

\vspace{0.1cm}

\begin{align}
   \delta^2T^{i}_{\;\;\phi} = &(\eden+p)U^{i}_{(1,0)}\left( U_{\phi}^{(1,0)} + h^{(1,0)}_{\phi \mu}U^{\mu} \right)\\  \notag
   &+ \left( \eden^{(1,1)}+p^{(1,1)} \right) U^{i}_{(1,0)}U_{\phi}.
\end{align}
By replacing $\phi$ with $t$ in all of the above, one also recovers the energy balance equations and although we derive this equation enforcing a Killing background flow and circular orbit frequencies the result can be naturally extended to generic configurations. 

In connection with the Newtonian literature and previous work in black hole perturbation theory, we define the advected angular momentum density as a subset of the terms quadratic in the variation of the stress energy tensor, given by,
\begin{align}
    \dot{J}^{i}_{\mathrm{Adv}} &= \sqrt{-g} \left<\delta^2T\right>^{i}_{\;\;\phi}  - \dot{J}^{i}_{\mathrm{\dot{M}}}.
\end{align}
with 
\begin{align}
\label{eq:Jdot_mass}
    \dot{J}^{i}_{\mathrm{\dot{M}}} &= \sqrt{-g} \left<\left( \eden^{(1,1)} \right) U^{i}_{(1,0)}\right>U_{\phi}.
\end{align}
We choose this separation explicitly such that in the Newtonian limit, 
\begin{align}
\dot{J}^{i}_{\mathrm{Adv}} &\rightarrow r^2 \rho_0\left< \delta U^r \delta U^{\phi} \right>,\\ 
\dot{J}^{i}_{\mathrm{\dot{M}}} &\rightarrow r^2 \left< \delta\rho \; \delta U^r\right>U^{\phi} .
\end{align}
Directly identifying, $\dot{J}^{i}_{\mathrm{Adv}}$ with the relativistic analogue of the advected angular momentum flux through the disc \citep[e.g.,][]{Goldreich79}, and $\dot{J}^{i}_{\dot{M}}$ with the angular momentum transport associated with the radial motion of perturbed fluid elements. Crucially, as emphasised in previous Newtonian analyses \cite{1990ApJ...362..395L,2010ApJ...724..448M}, the term $\dot{J}^{i}_{\dot{M}}$ is precisely the contribution expected to balance the otherwise unknown static second-order terms, which we collectively denote by $S^{(2)}_{m=0}$.\footnote{We thank Callum Fairbairn for drawing our attention to these references.}

Having accounted for the second-order stress--energy contributions, we now perform a canonical decomposition of the remaining terms in the torque balance equation. We separate the residual contributions into those that are linear in a single variation of the stress--energy tensor and a single metric perturbation, which we identify as the derivatives of the \emph{matter torques}, $\partial T^{\phi}_{\mathrm{Mat}}$, and those that depend only on the background fluid flow and are quadratic in the metric perturbations sourced by the secondary, which we interpret as derivatives of the
\emph{geometric torques}, $\partial T^{\phi}_{\mathrm{Geo}}$. This decomposition yields,
\begin{align}\label{eq:gravTorque}
    \partial T^{\phi}_{\mathrm{Mat}} = &\sqrt{-g}\left<\delta T^{\alpha\beta}\nabla_{\phi}h_{\alpha\beta}\right>\\ \notag
    &- \partial_{i} \left( \sqrt{-g} \left<h \delta T^{i}_{\phi}\right>  \right),
\end{align}
and 
\begin{align}\label{eq:geoTorque}
    \partial T^{\phi}_{\mathrm{Geo}} =&- 2 \sqrt{-g} T^{\alpha\beta} \left<h^{\;\gamma}_{\alpha} \nabla_{[\gamma}h_{\beta]\phi}\right>\\ \notag
   &
   -\frac{1}{2} \sqrt{-g} T^{\alpha\beta}  \left<h \nabla_{\phi}h_{\alpha \beta}\right> \\ \notag
   &+\frac{1}{4}\partial_i\left[ \sqrt{-g}T^i_{\phi} \left(\left<h^2\right> - 2 \left<h^{\mu}_{\nu}h^{\nu}_{\mu}\right>\right) \right]. 
\end{align}

This canonical separation of our equations gives rise to its collected form,
\begin{equation}
\label{eq:balanceequationreduced}
\begin{aligned}
\partial_{T_D}\left(\sqrt{-g} T^{i}_{\phi}\right)
+& \partial_{i}\left(\dot{J}^{i}_{\mathrm{Adv}}+\dot{J}^{i}_{\dot{M}} \right)
=\\ 
&\;\;\;\;\;\;\;\;\;\;\;\;\partial T^{\phi}_{\mathrm{Mat}} + \partial T^{\phi}_{\mathrm{Geo}} + S^{(2)}_{m=0}.
\end{aligned}
\end{equation}
Similarly to our analysis of the angular momentum fluxes we now find that in the Newtonian limit,
\begin{align}
\partial T^{\phi}_{\mathrm{Mat}} &\rightarrow r^2 \left<\delta\rho \partial_{\phi} \Phi_{p} \right>,\\ 
\partial T^{\phi}_{\mathrm{Geo}} &\rightarrow 0,
\end{align}
where $ \Phi_{p}$ is the Newtonian potential of the smaller body. Making a direct connection to previous works invoking torque-advection balance \cite[e.g.,][]{Miranda_2016}.
This along with the identification of the advected angular momentum gives a direct relativistic analogue of the full advected angular momentum -- torque density balance equation. 
\subsubsection{Application to the Relativistic Disc}
\begin{figure}[htbp]
    \centering
    \includegraphics[width=0.99\columnwidth]{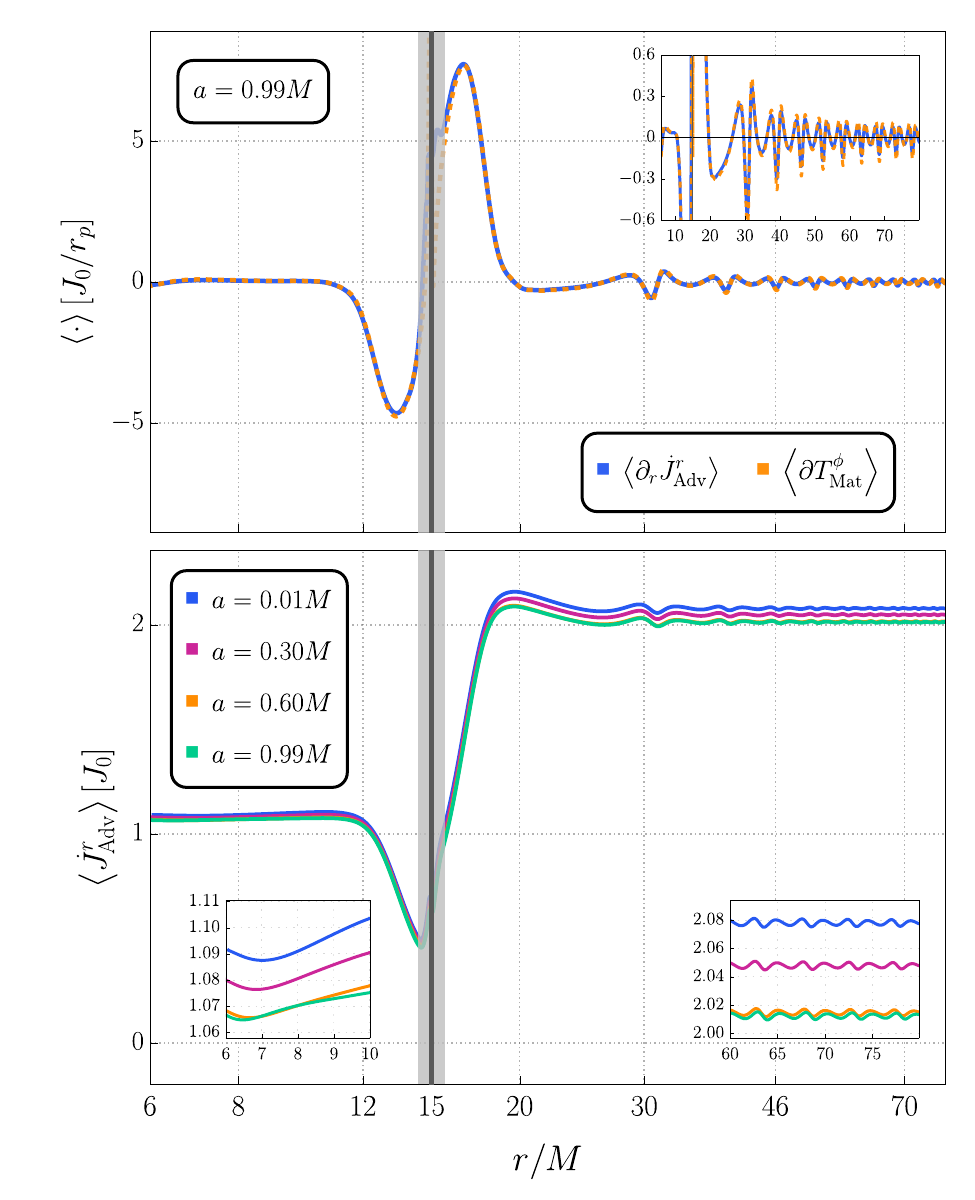}
    \caption{Torque--flux balance and asymptotic fluxes for $\eta = 0.1$ and $r_p = 15M$, summed over $m = 2$--$15$ (the $m = 1$ mode is excluded; see Sec.~\ref{sec:numerical}). The dark vertical line marks the orbital radius, and the surrounding grey band ($|r - r_p| < 0.66M$) the region close to the secondary around which the torque balance in our solutions is poor. 
    \textit{Top panel:} contributions to the torque-density balance equation for $a = 0.99M$, constructed from products of the first-order metric and fluid perturbations, normalised by $J_0/r_p$. This shows global balance between the advected angular-momentum and matter torque densities and exhibits the characteristic trough and peak on either side of co-rotation. \textit{Inset:} zoomed-in view illustrating the agreement away from $r_p$. \textit{Bottom panel:} total advected angular-momentum flux, normalised by $J_0$, for $a = 0.01M$, $0.3M$, $0.6M$ and $0.99M$; the insets zoom on the inner ($6$--$10M$) and outer ($60$--$80M$) flux. Modes $m > 15$, not included here, are estimated to add a further $\approx 25\%$ to the outer and $\approx 30\%$ to the inner flux (App.~\ref{app:fellconver}). }
    \label{fig:torquebalance}
\end{figure}

Having obtained our solutions for the fluid variables, along with the input metric data and a prescribed background fluid, we apply the PMLA to our torque balance Eq.~\eqref{eq:TorqueEq} to leading $\mathcal{O}(1)$ in the scale height expansion. This results in a purely radial, orbit-averaged equation. We present the results of applying the torque balance equation to our fluid solutions in  Fig.~\ref{fig:torquebalance}.

In Fig.~\ref{fig:torquebalance} and hereafter, we normalise our torques and torque density analogously to the Newtonian literature \citep[e.g.,][]{Goldreich80, rafikov2025indirectforcesdiscplanetinteraction}, with the exception that our torque is per unit volume, not area, and we here adjust for our initial choice of disc normalisation at the ISCO (Eq. (\ref{eq:background_eofr})) by multiplying by a factor of $\mathfrak{e}_I (r_p/r_I)^{-2}$, to give,
\begin{equation}
\label{eq:J0norm}
    J_0 = \left(\frac{m_p}{M}\right)^2 \eta^{-3} r_p^2 \mathfrak{e}_I r_I^2 \Omega_p^2 .
\end{equation}
Figure~\ref{fig:torquebalance} shows that, even with the PMLA, there is remarkable agreement between the radial profile of the radial derivative of the orbit-averaged advected angular momentum flux ($\partial_r\dot{J}^{r}_{\mathrm{Adv}}$) and our relativistic analogue of the disc excitation torque density ($\partial T^{\phi}_{\mathrm{Mat}}$): integrated away from the secondary, the two agree to $\approx 1\%$ inside and $\approx 3\%$ outside its orbit. Furthermore, the extended structure of the torque clearly exhibits the same qualitative behaviour as excitation torque densities previously calculated in the Newtonian literature, including the primary trough and peak just inside and outside the secondary's radius, as well as the so-called negative torque density and torque wiggles at larger distances from the secondary (see \cite{TrqWiggles:2024, rafikov2025indirectforcesdiscplanetinteraction} for recent examples).

The $\partial_r\dot{J}^{r}_{\mathrm{Adv}}$ contribution depends entirely on perturbed fluid quantities, while $\partial T^{\phi}_{\mathrm{Mat}}$ depends on a cross product between the metric perturbation and perturbed fluid quantities. This comparison serves as a strong check on both the validity of the approximations used throughout this work, the correctness of the numerical implementation, including the Hadamard regularisation procedure employed to remove the non-integrable divergences arising at Lindblad resonances, and the reconstruction of our fluid variables.

We also find that the new class of torques, which we denoted above as {\it geometric torques}, is strongly subdominant: for all four spins its magnitude peaks at $\sim 10^{-5}$ of the {\it matter torque}, so we do not show it in Fig.~\ref{fig:torquebalance}. The contribution of the $\dot{J}^{r}_{\dot{M}}$ term, by contrast, grows steadily away from the secondary. As noted previously in the literature \cite{1990ApJ...362..395L,2010ApJ...724..448M}, and as can be seen by taking the orbit-averaged energy equation to second order, this growth arises from the neglect of static second-order contributions; specifically, we expect terms of the form $\mathfrak{e}_{(0,1)} U_{\phi}^{(0,0)} U^{r}_{(2,0), m=0}$ to balance it. Without a careful calculation of these second-order static pieces, or of the slow-time evolution of our background variables (which also arises at second order), we cannot perform a full balance check that includes this term. Nevertheless, the strong agreement between $\partial_r \dot{J}^{r}_{\mathrm{Adv}}$ and $\partial T^{\phi}_{\mathrm{Mat}}$ suggests a clear route toward a torque-balance framework for relativistic adiabatic discs within black hole perturbation theory calculations.

In the linear, adiabatic, Newtonian case, 
the torque density excited into the disc by the perturber is completely carried away by the advected angular momentum flux, hence, allowing an association of the asymptotic angular momentum flux with the excitation torque \citep[e.g.,][]{Miranda19ALMA, Miranda20}. Here we have found that a similar matching holds in the linear, adiabatic, relativistic case (Fig.~\ref{fig:torquebalance}), motivating that our angular momentum flux is an indicator of the total relativistic torque, which includes full disc contributions from Linblad, co-rotation, and non-resonant torques.
In Fig.~\ref{fig:torquebalance} we provide an example calculation of the advected fluxes for a range of spin values at fixed $r_p = 15M$. We emphasis that due to the low number of $m$-modes present in our current calculation, these disc torque and flux values should only be considered to have converged to around $75\%$ of their total value as we discuss further in Fig.~\ref{fig:FluxConvmmode} of appendix~\ref{app:fellconver}. Along these lines we also include a convergence analysis of our flux values for each $m$-mode contribution as a function of the $\ell_{max}$ value considered in constructing the height averaged metric perturbation data for the source input. We find this sum is well converged in relation to the other approximations already present in this work and show the exponential decay in Fig.~\ref{fig:FluxConvL}.   Finally we stress that relating these to the torque on the secondary requires a torque-balance law that has not yet been derived within self-force theory (see below); with that caveat, if such a law holds, the negative of their difference gives the net torque on the secondary.

We identify two key results from Fig.~\ref{fig:torquebalance}. First, for all spin values considered, the outer angular momentum flux exceeds the inner flux. By analogy with the Newtonian case, this would imply a negative net torque and hence inward migration of the secondary at $r_p = 15M$. This is consistent with recent relativistic studies of pressure-less discs, which find a torque reversal only much closer to the black hole, near the ISCO, at a location that depends on spin and on the disc profile \cite{HegadeKR:2025dur, Duque:2025yfm}.

Second, and in agreement with recent relativistic studies, we find a decrease in the total torque with increasing black hole spin. Upcoming follow-up work will explore a wider region of parameter space in order to develop a more complete understanding of how our model differs from other recent relativistic treatments which treat the disc as a pressure-less collection of particles \citep{HegadeKR:2025dur, Duque:2025yfm}.

Furthermore, this result points toward a path forward for the potential identification of a Newton’s third-law--type balance, from which one could hope to identify torques, $T_{\text{secondary}} = -\,T_{\text{Mat}} $. If such a connection exists, a fundamental advance in perturbative matter-field binary modelling could be achieved. Specifically, rather than calculating higher-order metric perturbations to determine the forces acting on a small body for orbital evolution, one could instead solve for the appropriate matter-field configuration of the binary, construct $\partial T^{\phi}$ for a given matter-field theory, and integrate over the volume to determine the binary evolution. We expect such a balance would take the schematic form,

\begin{equation}
    \dot{J}_{\mathrm{secondary}}\sim - \int \partial T^{\phi}_{\mathrm{Mat}}\;dV.
\end{equation}
Such an exploration will be of critical importance in future studies. A concrete connection between the force applied to the matter field and the force applied to the secondary will require a derivation of a first principles torque-balance derivation within self-force theory. However, even at the level of a heuristic implementation, the application of such a torque balance in this manner, we expect, would already represent a substantial improvement over the approximations presently employed in the literature.

\section{Conclusion}
\label{sec:Summary}
In this work, we have presented the first fully relativistic calculation of the linear response of a pressure-supported disc to the gravitational potential of an extreme mass-ratio inspiral. Our approach leverages results from multiple fields, adapting solutions developed for self-gravitating relativistic stars and accretion discs and combining these with state-of-the-art self-force calculations whilst guiding our calculations from current results in the planetary migration literature. Using these results, we derive a new form of master equation for the perturbations to a perfect fluid in the Kerr geometry. This equation incorporates the effective force of an arbitrary, independently sourced metric perturbation; here we specialise to metric perturbations arising from a small secondary black hole on a circular orbit in the Kerr black hole, and to the case of an adiabatic fluid with adiabatic index approaching unity.

After constructing the relativistic background flow, we construct the $m$-mode content of the EMRI metric perturbation from its $\ell$-modes and show that, once height-averaged, it converges exponentially in $\ell$ without any softening length. We then describe in detail the procedure for obtaining our height-averaged master equations, highlighting the subtleties encountered at different stages of calculating the enthalpy solutions. With the fluid solutions reconstructed, we analyse the resulting spiral-arm structures and extended flow patterns, illustrating the deviations between the Kerr and Newtonian spiral-arm structures. We also compute the extended profiles for a range of spin values, demonstrating the contraction of the ISCO and the subtle increase in the winding rate of the inner arm with increasing spin.

We then focus on the angular momentum balance, examining the interplay between advected angular momentum through the disc and the matter torques excited by the secondary potential. We derive a new, fully relativistic form of the torque balance equation, showing remarkable agreement between our definition of the advected flux densities and matter torque densities across the domain.

Our torque balance equation provides a crucial step toward understanding the interaction between orbital mechanics and disc physics. Instead of relying on traditional calculations of advected fluxes through the disc (as typically done in black hole perturbation theory), our approach relates these effects directly to local torques. In future work, this method could allow one to compute forces on the orbiting body without needing to perform complicated second order self-force type calculations.

There are a number of immediate future directions in which to take this initial study. Firstly, we have computed solutions up to a maximum azimuthal Fourier mode of $m_{\mathrm{max}}=15$. This captures the disc morphology and the local torque balance, but the $m > 15$ modes still carry $\approx 20$--$25\%$ of the one-sided fluxes  as we show in appendix.~\ref{app:fellconver}; extending the mode sum is a priority for future work. Secondly, we do not calculate the contribution of the $m=0$ mode to the fluid perturbations. Calculation of this term will be necessary in order to source the static second order equations needed to perform a full torque balance.

Future developments of this framework should also include the explicit incorporation of the baryon number conservation equation, allowing a proper separation of rest-mass energy from internal energy. Such a split is essential for accurately tracking the thermodynamic evolution of the system over time, particularly when additional cooling effects are included.  

Extensions to ideal magnetohydrodynamics (MHD) also appear feasible and promising both at linear order, within the framework laid out here, and also numerically. Indeed future extensions to linear theory could utilise the full non-linear fluid equations, or their MHD extensions, coupled with the linear metric perturbations from the secondary, as derived in Eqs.~(\ref{eq:eulerequationsmetricexpand}-\ref{eq:Effective force}). 

Incorporating dissipative effects, such as viscosity, could be handled self-consistently through the incorporation of causal first order fluid theories such as those by  Bemfica, Disconzi, Noronha, and Kovtun (BDNK) \cite{PhysRevX.12.021044,Kovtun:2019hdm,PhysRevD.98.104064,PhysRevD.100.104020}. Such models, however, would involve background accretion and radial inflow, which would likely need to be addressed at the level of solving the full system of coupled fluid equations rather than through a reduced master equation. Progress in calculating master equations with BDNK has been made recently however, in the context of spherically symmetric and low-spin viscous stars \cite{Boyanov:2024jge,Redondo-Yuste:2024vdb,Redondo-Yuste:2025ktt,Caballero:2025omv}.

In this work, we have primarily focused on the formulation and methodology, providing several proof-of-principle calculations to demonstrate the robustness and potential of these results. Follow-up studies will explore the underlying physics and parameter space in greater detail, examining how relativistic effects modify the dynamics compared to the wealth of Newtonian studies already present in the literature. These results represent the first steps toward fully relativistic fluid modelling around EMRIs and pave the way for more complex and astrophysically relevant models of strong-field EMRIs in active galactic nuclei.

\section{Acknowledgments}
CD and DJD thank Callum Fairbairn for pointing us to invaluable references and for providing helpful clarifications during the course of this work. DJD thanks Roman Rafikov for useful discussions and for spotting multiple spiral arms in preliminarily plots from an impressive distance. CD thanks Francisco Duque, Philip Kirkeberg, Maarten van de Meent, David O’Neil, Adam Pound, Jaime Redondo-Yuste, Laura Sberna, Andrew Spiers, and Chris Tiede for valuable discussions while carrying out this research. CD is grateful to Scott Hughes and the MIT Kavli institute for hospitality during the early stages of development of this work. This work also makes use of a modified form of the Kerr-Lorenz-Circ \texttt{Mathematica} code developed by Sam Dolan \cite{Dolan_2022,Dolan_2024}. This work makes use of the Black Hole Perturbation Toolkit \cite{BHPToolkit} and the xPert sub-package of the xAct codebase \cite{Brizuela:2008ra}. The Tycho supercomputer hosted at the SCIENCE HPC center at the University of Copenhagen was used for this work. The Center of Gravity is a Center of Excellence funded by the Danish National Research Foundation under grant No. 184. CD acknowledges support by VILLUM Foundation (grant no.VIL37766 and no. VIL53101) and the DNRF Chair program (grant no. DNRF162) by the Danish National Research Foundation. 
DJD acknowledges support from the Danish Independent Research Fund through Sapere Aude Starting Grant No. 121587 and the STScI Director's Discretionary Fund.

\appendix

\section{Numerical Calculations and Checks}
In this appendix we provide complementary numerical checks and calculations performed throughout the process of our example calculations for the benefit of the reader.

\subsection{Four-Flow Normalisation}\label{app:fourflownorm}
\begin{figure}[!htbp]
    \centering
    \includegraphics[width=0.99\columnwidth]{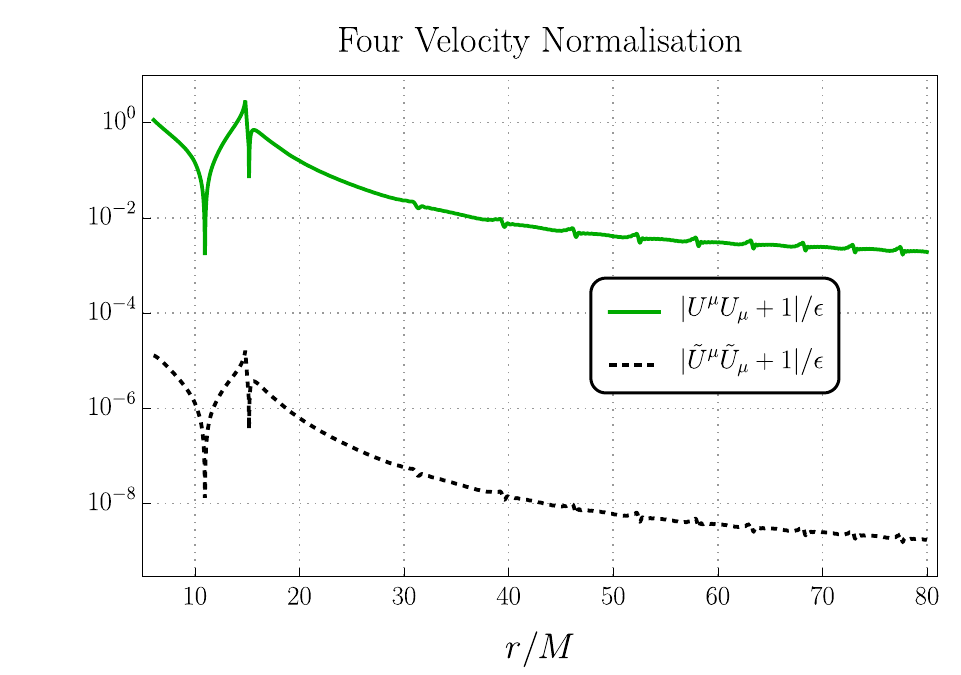}
    \caption{Plot of the four-flow along the slice $\phi=0$ as a function of radius, including the background plus perturbed flow. We show an example plot for $a = 0.01M$, $\eta = 0.1$, $r_p = 15M$, with the tilded flow representing the quantity including the non-affine reparameterisation.
}
    \label{fig:fourflow}
\end{figure}

In Sec.~\ref{sec:masterenthalpy} we describe how, through the introduction of a perturbative non-affine reparameterization, one can explicitly fix the flow to retain unity normalisation. We show this for a specific example in Fig.~\ref{fig:fourflow} where we see that after performing the perturbative non-affine re-parameterisation the inconsistency in four flow normalisation drops form being $\mathcal{O}(1)$ to well below the $\mathcal{O}(\eta^2)$ error floor introduced by the PMLA.

\subsection{Mode Convergence}\label{app:fellconver}

\begin{figure}[!htbp]
    \centering
    \includegraphics[width=1\columnwidth]{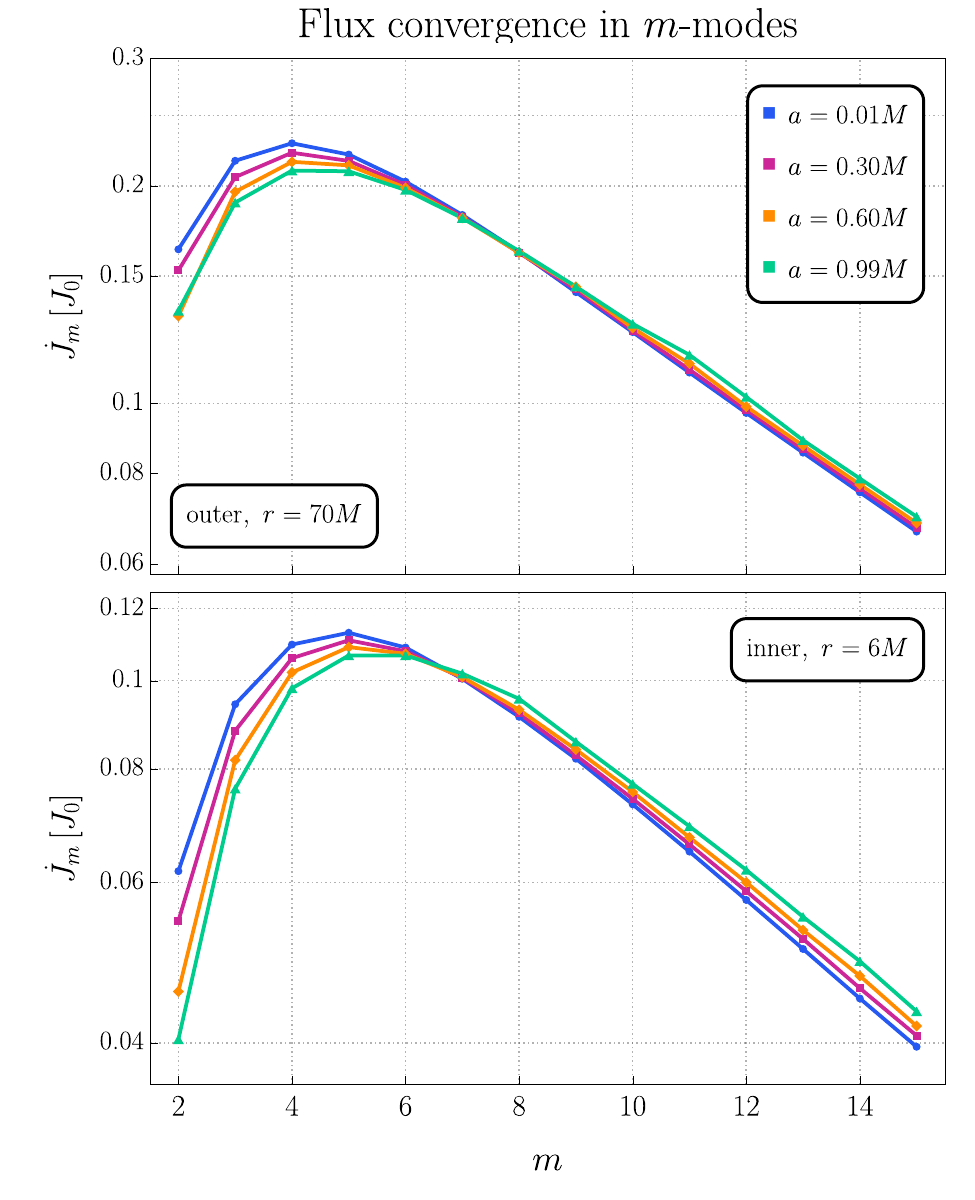}
    \caption{Advected angular momentum flux carried by each azimuthal mode, $m = 2$--$15$, for $r_p = 15M$, $\eta = 0.1$ and spins $a = 0.01M$, $0.3M$, $0.6M$ and $0.99M$, in units of $J_0$. \textit{Top panel:} outer flux at $r = 70M$. \textit{Bottom panel:} inner flux at $r = 6M$ ($r_{\min} = 6.07M$ for $a = 0.01M$). Beyond the peak at $m \approx 4$--$6$, the flux decays exponentially in $m$. We see that increasing the spin generally reduces the flux in the low modes ($m \lesssim 6$) and raises it in the higher modes ($m \gtrsim 8$).}
    \label{fig:FluxConvmmode}
\end{figure}

The flux carried by each azimuthal mode decays exponentially in $m$ beyond $m \approx 4$--$6$ as can be seen in Fig.~\ref{fig:FluxConvmmode}. The decay is however quite slow, so the $m > 15$ modes still carry a significant fraction of the total flux. Fitting an exponential to $m = 12$--$15$ and summing the remaining series, we estimate that these modes carry $19$--$22\%$ of the full outer flux and $22$--$26\%$ of the full inner flux, for all the spins we compute at $r_p = 15M$. Including them raises the one-sided fluxes in Fig.~\ref{fig:torquebalance} by $\approx 25$--$30\%$ hence the flux values in Fig.~\ref{fig:torquebalance} should be read as lower bounds with the primary purpose of these results to exemplify the torque balance and the similarity of our results with the Newtonian literature.

\begin{figure}[!htbp]
    \centering
    \includegraphics[width=1\columnwidth]{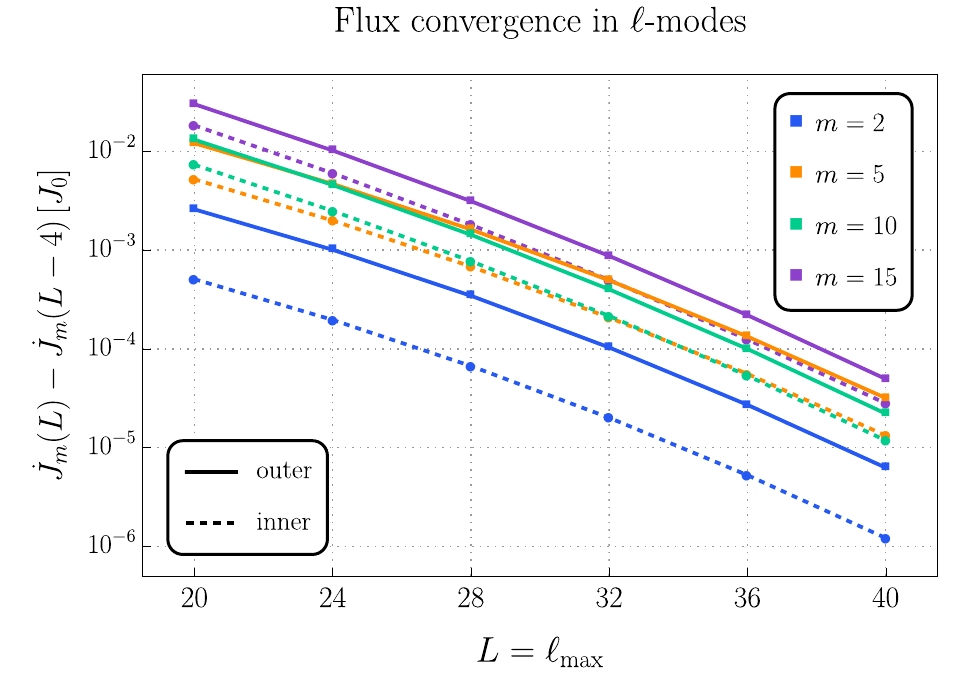}
    \caption{Convergence of the advected angular momentum flux with the number of spherical-harmonic $\ell$-modes used to construct the height-averaged metric perturbation, for $a = 0.6M$, $r_p = 15M$ and $\eta = 0.1$. We show the flux added to individual azimuthal modes ($m = 2, 5, 10, 15$) by each successive block of four $\ell$-modes, $\dot{J}_m(L) - \dot{J}_m(L-4)$, in units of $J_0$. Solid lines show the outer flux (averaged over $60$--$79M$) and dashed lines the inner flux (at $6M$); both decay exponentially in $\ell$.}
    \label{fig:FluxConvL}
\end{figure}

Finally our approach uses the full height-averaged $m$-mode retarded metric perturbation data as input for the source. The initial $m$-mode data is constructed from a finite number of $\ell$-modes, whose sum converges slowly near the particle. Under the PMLA however, the effect of this slow convergence on the height-averaged equations is small, and the torques converge exponentially with $\ell_{max}$. To show this, we vary the maximum $\ell$-mode, $L$, included in the sum. As Fig.~\ref{fig:FluxConvL} shows, each successive block of $\ell$-modes adds an exponentially smaller contribution to the flux, so that every individual $m$-mode used in our solutions are well converged in $\ell$.

\newpage

\bibliography{ref}

\begin{thebibliography}{160}%
\makeatletter
\providecommand \@ifxundefined [1]{%
 \@ifx{#1\undefined}
}%
\providecommand \@ifnum [1]{%
 \ifnum #1\expandafter \@firstoftwo
 \else \expandafter \@secondoftwo
 \fi
}%
\providecommand \@ifx [1]{%
 \ifx #1\expandafter \@firstoftwo
 \else \expandafter \@secondoftwo
 \fi
}%
\providecommand \natexlab [1]{#1}%
\providecommand \enquote  [1]{``#1''}%
\providecommand \bibnamefont  [1]{#1}%
\providecommand \bibfnamefont [1]{#1}%
\providecommand \citenamefont [1]{#1}%
\providecommand \href@noop [0]{\@secondoftwo}%
\providecommand \href [0]{\begingroup \@sanitize@url \@href}%
\providecommand \@href[1]{\@@startlink{#1}\@@href}%
\providecommand \@@href[1]{\endgroup#1\@@endlink}%
\providecommand \@sanitize@url [0]{\catcode `\\12\catcode `\$12\catcode
  `\&12\catcode `\#12\catcode `\^12\catcode `\_12\catcode `\%12\relax}%
\providecommand \@@startlink[1]{}%
\providecommand \@@endlink[0]{}%
\providecommand \url  [0]{\begingroup\@sanitize@url \@url }%
\providecommand \@url [1]{\endgroup\@href {#1}{\urlprefix }}%
\providecommand \urlprefix  [0]{URL }%
\providecommand \Eprint [0]{\href }%
\providecommand \doibase [0]{https://doi.org/}%
\providecommand \selectlanguage [0]{\@gobble}%
\providecommand \bibinfo  [0]{\@secondoftwo}%
\providecommand \bibfield  [0]{\@secondoftwo}%
\providecommand \translation [1]{[#1]}%
\providecommand \BibitemOpen [0]{}%
\providecommand \bibitemStop [0]{}%
\providecommand \bibitemNoStop [0]{.\EOS\space}%
\providecommand \EOS [0]{\spacefactor3000\relax}%
\providecommand \BibitemShut  [1]{\csname bibitem#1\endcsname}%
\let\auto@bib@innerbib\@empty
\bibitem [{\citenamefont {Abbott}\ \emph {et~al.}(2016)\citenamefont {Abbott}
  \emph {et~al.}}]{LIGOScientific:2016aoc}%
  \BibitemOpen
  \bibfield  {author} {\bibinfo {author} {\bibfnamefont {B.~P.}\ \bibnamefont
  {Abbott}} \emph {et~al.} (\bibinfo {collaboration} {LIGO Scientific,
  Virgo}),\ }\bibfield  {title} {\bibinfo {title} {{Observation of
  Gravitational Waves from a Binary Black Hole Merger}},\ }\href
  {https://doi.org/10.1103/PhysRevLett.116.061102} {\bibfield  {journal}
  {\bibinfo  {journal} {Phys. Rev. Lett.}\ }\textbf {\bibinfo {volume} {116}},\
  \bibinfo {pages} {061102} (\bibinfo {year} {2016})},\ \Eprint
  {https://arxiv.org/abs/1602.03837} {arXiv:1602.03837 [gr-qc]} \BibitemShut
  {NoStop}%
\bibitem [{\citenamefont {Abac}\ \emph {et~al.}(2025)\citenamefont {Abac} \emph
  {et~al.}}]{LIGOScientific:2025slb}%
  \BibitemOpen
  \bibfield  {author} {\bibinfo {author} {\bibfnamefont {A.~G.}\ \bibnamefont
  {Abac}} \emph {et~al.} (\bibinfo {collaboration} {LIGO Scientific, VIRGO,
  KAGRA}),\ }\bibfield  {title} {\bibinfo {title} {{GWTC-4.0: Updating the
  Gravitational-Wave Transient Catalog with Observations from the First Part of
  the Fourth LIGO-Virgo-KAGRA Observing Run}},\ }\href
  {https://doi.org/10.48550/arXiv.2508.18082} {\bibfield  {journal} {\bibinfo
  {journal} {arXiv e-prints}\ ,\ \bibinfo {eid} {arXiv:2508.18082}} (\bibinfo
  {year} {2025})},\ \Eprint {https://arxiv.org/abs/2508.18082}
  {arXiv:2508.18082 [gr-qc]} \BibitemShut {NoStop}%
\bibitem [{\citenamefont {Abbott}\ \emph {et~al.}(2019)\citenamefont {Abbott}
  \emph {et~al.}}]{LIGOScientific:2018mvr}%
  \BibitemOpen
  \bibfield  {author} {\bibinfo {author} {\bibfnamefont {B.~P.}\ \bibnamefont
  {Abbott}} \emph {et~al.} (\bibinfo {collaboration} {LIGO Scientific,
  Virgo}),\ }\bibfield  {title} {\bibinfo {title} {{GWTC-1: A
  Gravitational-Wave Transient Catalog of Compact Binary Mergers Observed by
  LIGO and Virgo during the First and Second Observing Runs}},\ }\href
  {https://doi.org/10.1103/PhysRevX.9.031040} {\bibfield  {journal} {\bibinfo
  {journal} {Phys. Rev. X}\ }\textbf {\bibinfo {volume} {9}},\ \bibinfo {pages}
  {031040} (\bibinfo {year} {2019})},\ \Eprint
  {https://arxiv.org/abs/1811.12907} {arXiv:1811.12907 [astro-ph.HE]}
  \BibitemShut {NoStop}%
\bibitem [{\citenamefont {Abbott}\ \emph {et~al.}(2021)\citenamefont {Abbott}
  \emph {et~al.}}]{LIGOScientific:2020ibl}%
  \BibitemOpen
  \bibfield  {author} {\bibinfo {author} {\bibfnamefont {R.}~\bibnamefont
  {Abbott}} \emph {et~al.} (\bibinfo {collaboration} {LIGO Scientific,
  Virgo}),\ }\bibfield  {title} {\bibinfo {title} {{GWTC-2: Compact Binary
  Coalescences Observed by LIGO and Virgo During the First Half of the Third
  Observing Run}},\ }\href {https://doi.org/10.1103/PhysRevX.11.021053}
  {\bibfield  {journal} {\bibinfo  {journal} {Phys. Rev. X}\ }\textbf {\bibinfo
  {volume} {11}},\ \bibinfo {pages} {021053} (\bibinfo {year} {2021})},\
  \Eprint {https://arxiv.org/abs/2010.14527} {arXiv:2010.14527 [gr-qc]}
  \BibitemShut {NoStop}%
\bibitem [{\citenamefont {Abbott}\ \emph {et~al.}(2023)\citenamefont {Abbott}
  \emph {et~al.}}]{KAGRA:2021vkt}%
  \BibitemOpen
  \bibfield  {author} {\bibinfo {author} {\bibfnamefont {R.}~\bibnamefont
  {Abbott}} \emph {et~al.} (\bibinfo {collaboration} {KAGRA, VIRGO, LIGO
  Scientific}),\ }\bibfield  {title} {\bibinfo {title} {{GWTC-3: Compact Binary
  Coalescences Observed by LIGO and Virgo during the Second Part of the Third
  Observing Run}},\ }\href {https://doi.org/10.1103/PhysRevX.13.041039}
  {\bibfield  {journal} {\bibinfo  {journal} {Phys. Rev. X}\ }\textbf {\bibinfo
  {volume} {13}},\ \bibinfo {pages} {041039} (\bibinfo {year} {2023})},\
  \Eprint {https://arxiv.org/abs/2111.03606} {arXiv:2111.03606 [gr-qc]}
  \BibitemShut {NoStop}%
\bibitem [{\citenamefont {Maggiore}\ \emph {et~al.}(2020)\citenamefont
  {Maggiore} \emph {et~al.}}]{Maggiore:2019uih}%
  \BibitemOpen
  \bibfield  {author} {\bibinfo {author} {\bibfnamefont {M.}~\bibnamefont
  {Maggiore}} \emph {et~al.},\ }\bibfield  {title} {\bibinfo {title} {{Science
  case for the Einstein telescope}},\ }\href
  {https://doi.org/10.1088/1475-7516/2020/03/050} {\bibfield  {journal}
  {\bibinfo  {journal} {\jcap}\ }\textbf {\bibinfo {volume} {2020}},\ \bibinfo
  {eid} {050} (\bibinfo {year} {2020})},\ \Eprint
  {https://arxiv.org/abs/1912.02622} {arXiv:1912.02622 [astro-ph.CO]}
  \BibitemShut {NoStop}%
\bibitem [{\citenamefont {Evans}\ \emph {et~al.}(2021)\citenamefont {Evans}
  \emph {et~al.}}]{Evans:2021gyd}%
  \BibitemOpen
  \bibfield  {author} {\bibinfo {author} {\bibfnamefont {M.}~\bibnamefont
  {Evans}} \emph {et~al.},\ }\bibfield  {title} {\bibinfo {title} {{A Horizon
  Study for Cosmic Explorer: Science, Observatories, and Community}},\ }\href
  {https://doi.org/10.48550/arXiv.2109.09882} {\bibfield  {journal} {\bibinfo
  {journal} {arXiv e-prints}\ ,\ \bibinfo {eid} {arXiv:2109.09882}} (\bibinfo
  {year} {2021})},\ \Eprint {https://arxiv.org/abs/2109.09882}
  {arXiv:2109.09882 [astro-ph.IM]} \BibitemShut {NoStop}%
\bibitem [{\citenamefont {Colpi}\ \emph {et~al.}(2024)\citenamefont {Colpi}
  \emph {et~al.}}]{Colpi:2024xhw}%
  \BibitemOpen
  \bibfield  {author} {\bibinfo {author} {\bibfnamefont {M.}~\bibnamefont
  {Colpi}} \emph {et~al.},\ }\bibfield  {title} {\bibinfo {title} {{LISA
  Definition Study Report}},\ }\href
  {https://doi.org/10.48550/arXiv.2402.07571} {\bibfield  {journal} {\bibinfo
  {journal} {arXiv e-prints}\ ,\ \bibinfo {eid} {arXiv:2402.07571}} (\bibinfo
  {year} {2024})},\ \Eprint {https://arxiv.org/abs/2402.07571}
  {arXiv:2402.07571 [astro-ph.CO]} \BibitemShut {NoStop}%
\bibitem [{\citenamefont {Mei}\ \emph {et~al.}(2021)\citenamefont {Mei} \emph
  {et~al.}}]{TianQin:2020hid}%
  \BibitemOpen
  \bibfield  {author} {\bibinfo {author} {\bibfnamefont {J.}~\bibnamefont
  {Mei}} \emph {et~al.} (\bibinfo {collaboration} {TianQin}),\ }\bibfield
  {title} {\bibinfo {title} {{The TianQin project: current progress on science
  and technology}},\ }\href {https://doi.org/10.1093/ptep/ptaa114} {\bibfield
  {journal} {\bibinfo  {journal} {PTEP}\ }\textbf {\bibinfo {volume} {2021}},\
  \bibinfo {pages} {05A107} (\bibinfo {year} {2021})},\ \Eprint
  {https://arxiv.org/abs/2008.10332} {arXiv:2008.10332 [gr-qc]} \BibitemShut
  {NoStop}%
\bibitem [{\citenamefont {Li}\ \emph {et~al.}(2025{\natexlab{a}})\citenamefont
  {Li} \emph {et~al.}}]{Li:2024rnk}%
  \BibitemOpen
  \bibfield  {author} {\bibinfo {author} {\bibfnamefont {E.-K.}\ \bibnamefont
  {Li}} \emph {et~al.},\ }\bibfield  {title} {\bibinfo {title} {{Gravitational
  wave astronomy with TianQin}},\ }\href
  {https://doi.org/10.1088/1361-6633/adc9be} {\bibfield  {journal} {\bibinfo
  {journal} {Reports on Progress in Physics}\ }\textbf {\bibinfo {volume}
  {88}},\ \bibinfo {eid} {056901} (\bibinfo {year} {2025}{\natexlab{a}})},\
  \Eprint {https://arxiv.org/abs/2409.19665} {arXiv:2409.19665 [astro-ph.GA]}
  \BibitemShut {NoStop}%
\bibitem [{\citenamefont {{Gong}}\ \emph {et~al.}(2021)\citenamefont {{Gong}},
  \citenamefont {{Xu}}, \citenamefont {{Gui}}, \citenamefont {{Huang}},\ and\
  \citenamefont {{Lau}}}]{Gong:2021any}%
  \BibitemOpen
  \bibfield  {author} {\bibinfo {author} {\bibfnamefont {X.}~\bibnamefont
  {{Gong}}}, \bibinfo {author} {\bibfnamefont {S.}~\bibnamefont {{Xu}}},
  \bibinfo {author} {\bibfnamefont {S.}~\bibnamefont {{Gui}}}, \bibinfo
  {author} {\bibfnamefont {S.}~\bibnamefont {{Huang}}},\ and\ \bibinfo {author}
  {\bibfnamefont {Y.-K.}\ \bibnamefont {{Lau}}},\ }\bibinfo {title} {{Mission
  Design for the TAIJI Mission and Structure Formation in Early Universe}},\
  in\ \href {https://doi.org/10.1007/978-981-15-4702-7_24-1} {\emph {\bibinfo
  {booktitle} {Handbook of Gravitational Wave Astronomy}}},\ \bibinfo {editor}
  {edited by\ \bibinfo {editor} {\bibfnamefont {C.}~\bibnamefont {{Bambi}}},
  \bibinfo {editor} {\bibfnamefont {S.}~\bibnamefont {{Katsanevas}}},\ and\
  \bibinfo {editor} {\bibfnamefont {K.~D.}\ \bibnamefont {{Kokkotas}}}}\
  (\bibinfo {year} {2021})\ p.~\bibinfo {pages} {24}\BibitemShut {NoStop}%
\bibitem [{\citenamefont {Seoane}\ \emph {et~al.}(2023)\citenamefont {Seoane}
  \emph {et~al.}}]{LISA:2022yao}%
  \BibitemOpen
  \bibfield  {author} {\bibinfo {author} {\bibfnamefont {P.~A.}\ \bibnamefont
  {Seoane}} \emph {et~al.} (\bibinfo {collaboration} {LISA}),\ }\bibfield
  {title} {\bibinfo {title} {{Astrophysics with the Laser Interferometer Space
  Antenna}},\ }\href {https://doi.org/10.1007/s41114-022-00041-y} {\bibfield
  {journal} {\bibinfo  {journal} {Living Rev. Rel.}\ }\textbf {\bibinfo
  {volume} {26}},\ \bibinfo {pages} {2} (\bibinfo {year} {2023})},\ \Eprint
  {https://arxiv.org/abs/2203.06016} {arXiv:2203.06016 [gr-qc]} \BibitemShut
  {NoStop}%
\bibitem [{\citenamefont {Auclair}\ \emph {et~al.}(2023)\citenamefont {Auclair}
  \emph {et~al.}}]{LISACosmologyWorkingGroup:2022jok}%
  \BibitemOpen
  \bibfield  {author} {\bibinfo {author} {\bibfnamefont {P.}~\bibnamefont
  {Auclair}} \emph {et~al.} (\bibinfo {collaboration} {LISA Cosmology Working
  Group}),\ }\bibfield  {title} {\bibinfo {title} {{Cosmology with the Laser
  Interferometer Space Antenna}},\ }\href
  {https://doi.org/10.1007/s41114-023-00045-2} {\bibfield  {journal} {\bibinfo
  {journal} {Living Rev. Rel.}\ }\textbf {\bibinfo {volume} {26}},\ \bibinfo
  {pages} {5} (\bibinfo {year} {2023})},\ \Eprint
  {https://arxiv.org/abs/2204.05434} {arXiv:2204.05434 [astro-ph.CO]}
  \BibitemShut {NoStop}%
\bibitem [{\citenamefont {Arun}\ \emph {et~al.}(2022)\citenamefont {Arun} \emph
  {et~al.}}]{LISA:2022kgy}%
  \BibitemOpen
  \bibfield  {author} {\bibinfo {author} {\bibfnamefont {K.~G.}\ \bibnamefont
  {Arun}} \emph {et~al.} (\bibinfo {collaboration} {LISA}),\ }\bibfield
  {title} {\bibinfo {title} {{New horizons for fundamental physics with
  LISA}},\ }\href {https://doi.org/10.1007/s41114-022-00036-9} {\bibfield
  {journal} {\bibinfo  {journal} {Living Rev. Rel.}\ }\textbf {\bibinfo
  {volume} {25}},\ \bibinfo {pages} {4} (\bibinfo {year} {2022})},\ \Eprint
  {https://arxiv.org/abs/2205.01597} {arXiv:2205.01597 [gr-qc]} \BibitemShut
  {NoStop}%
\bibitem [{\citenamefont {Afshordi}\ \emph {et~al.}(2023)\citenamefont
  {Afshordi} \emph {et~al.}}]{LISAConsortiumWaveformWorkingGroup:2023arg}%
  \BibitemOpen
  \bibfield  {author} {\bibinfo {author} {\bibfnamefont {N.}~\bibnamefont
  {Afshordi}} \emph {et~al.} (\bibinfo {collaboration} {LISA Consortium
  Waveform Working Group}),\ }\bibfield  {title} {\bibinfo {title} {{Waveform
  Modelling for the Laser Interferometer Space Antenna}},\ }\href
  {https://doi.org/10.48550/arXiv.2311.01300} {\bibfield  {journal} {\bibinfo
  {journal} {arXiv e-prints}\ ,\ \bibinfo {eid} {arXiv:2311.01300}} (\bibinfo
  {year} {2023})},\ \Eprint {https://arxiv.org/abs/2311.01300}
  {arXiv:2311.01300 [gr-qc]} \BibitemShut {NoStop}%
\bibitem [{\citenamefont {Barack}\ and\ \citenamefont
  {Pound}(2019)}]{Barack:2018yvs}%
  \BibitemOpen
  \bibfield  {author} {\bibinfo {author} {\bibfnamefont {L.}~\bibnamefont
  {Barack}}\ and\ \bibinfo {author} {\bibfnamefont {A.}~\bibnamefont {Pound}},\
  }\bibfield  {title} {\bibinfo {title} {{Self-force and radiation reaction in
  general relativity}},\ }\href {https://doi.org/10.1088/1361-6633/aae552}
  {\bibfield  {journal} {\bibinfo  {journal} {Rept. Prog. Phys.}\ }\textbf
  {\bibinfo {volume} {82}},\ \bibinfo {pages} {016904} (\bibinfo {year}
  {2019})},\ \Eprint {https://arxiv.org/abs/1805.10385} {arXiv:1805.10385
  [gr-qc]} \BibitemShut {NoStop}%
\bibitem [{\citenamefont {Wardell}\ \emph {et~al.}(2023)\citenamefont
  {Wardell}, \citenamefont {Pound}, \citenamefont {Warburton}, \citenamefont
  {Miller}, \citenamefont {Durkan},\ and\ \citenamefont
  {Le~Tiec}}]{Wardell:2021fyy}%
  \BibitemOpen
  \bibfield  {author} {\bibinfo {author} {\bibfnamefont {B.}~\bibnamefont
  {Wardell}}, \bibinfo {author} {\bibfnamefont {A.}~\bibnamefont {Pound}},
  \bibinfo {author} {\bibfnamefont {N.}~\bibnamefont {Warburton}}, \bibinfo
  {author} {\bibfnamefont {J.}~\bibnamefont {Miller}}, \bibinfo {author}
  {\bibfnamefont {L.}~\bibnamefont {Durkan}},\ and\ \bibinfo {author}
  {\bibfnamefont {A.}~\bibnamefont {Le~Tiec}},\ }\bibfield  {title} {\bibinfo
  {title} {{Gravitational Waveforms for Compact Binaries from Second-Order
  Self-Force Theory}},\ }\href {https://doi.org/10.1103/PhysRevLett.130.241402}
  {\bibfield  {journal} {\bibinfo  {journal} {Phys. Rev. Lett.}\ }\textbf
  {\bibinfo {volume} {130}},\ \bibinfo {pages} {241402} (\bibinfo {year}
  {2023})},\ \Eprint {https://arxiv.org/abs/2112.12265} {arXiv:2112.12265
  [gr-qc]} \BibitemShut {NoStop}%
\bibitem [{\citenamefont {Chua}\ \emph {et~al.}(2021)\citenamefont {Chua},
  \citenamefont {Katz}, \citenamefont {Warburton},\ and\ \citenamefont
  {Hughes}}]{Chua:2020stf}%
  \BibitemOpen
  \bibfield  {author} {\bibinfo {author} {\bibfnamefont {A.~J.~K.}\
  \bibnamefont {Chua}}, \bibinfo {author} {\bibfnamefont {M.~L.}\ \bibnamefont
  {Katz}}, \bibinfo {author} {\bibfnamefont {N.}~\bibnamefont {Warburton}},\
  and\ \bibinfo {author} {\bibfnamefont {S.~A.}\ \bibnamefont {Hughes}},\
  }\bibfield  {title} {\bibinfo {title} {{Rapid generation of fully
  relativistic extreme-mass-ratio-inspiral waveform templates for LISA data
  analysis}},\ }\href {https://doi.org/10.1103/PhysRevLett.126.051102}
  {\bibfield  {journal} {\bibinfo  {journal} {Phys. Rev. Lett.}\ }\textbf
  {\bibinfo {volume} {126}},\ \bibinfo {pages} {051102} (\bibinfo {year}
  {2021})},\ \Eprint {https://arxiv.org/abs/2008.06071} {arXiv:2008.06071
  [gr-qc]} \BibitemShut {NoStop}%
\bibitem [{\citenamefont {Katz}\ \emph {et~al.}(2021)\citenamefont {Katz},
  \citenamefont {Chua}, \citenamefont {Speri}, \citenamefont {Warburton},\ and\
  \citenamefont {Hughes}}]{Katz:2021yft}%
  \BibitemOpen
  \bibfield  {author} {\bibinfo {author} {\bibfnamefont {M.~L.}\ \bibnamefont
  {Katz}}, \bibinfo {author} {\bibfnamefont {A.~J.~K.}\ \bibnamefont {Chua}},
  \bibinfo {author} {\bibfnamefont {L.}~\bibnamefont {Speri}}, \bibinfo
  {author} {\bibfnamefont {N.}~\bibnamefont {Warburton}},\ and\ \bibinfo
  {author} {\bibfnamefont {S.~A.}\ \bibnamefont {Hughes}},\ }\bibfield  {title}
  {\bibinfo {title} {{Fast extreme-mass-ratio-inspiral waveforms: New tools for
  millihertz gravitational-wave data analysis}},\ }\href
  {https://doi.org/10.1103/PhysRevD.104.064047} {\bibfield  {journal} {\bibinfo
   {journal} {Phys. Rev. D}\ }\textbf {\bibinfo {volume} {104}},\ \bibinfo
  {pages} {064047} (\bibinfo {year} {2021})},\ \Eprint
  {https://arxiv.org/abs/2104.04582} {arXiv:2104.04582 [gr-qc]} \BibitemShut
  {NoStop}%
\bibitem [{\citenamefont {Hughes}\ \emph {et~al.}(2021)\citenamefont {Hughes},
  \citenamefont {Warburton}, \citenamefont {Khanna}, \citenamefont {Chua},\
  and\ \citenamefont {Katz}}]{Hughes:2021exa}%
  \BibitemOpen
  \bibfield  {author} {\bibinfo {author} {\bibfnamefont {S.~A.}\ \bibnamefont
  {Hughes}}, \bibinfo {author} {\bibfnamefont {N.}~\bibnamefont {Warburton}},
  \bibinfo {author} {\bibfnamefont {G.}~\bibnamefont {Khanna}}, \bibinfo
  {author} {\bibfnamefont {A.~J.~K.}\ \bibnamefont {Chua}},\ and\ \bibinfo
  {author} {\bibfnamefont {M.~L.}\ \bibnamefont {Katz}},\ }\bibfield  {title}
  {\bibinfo {title} {{Adiabatic waveforms for extreme mass-ratio inspirals via
  multivoice decomposition in time and frequency}},\ }\href
  {https://doi.org/10.1103/PhysRevD.103.104014} {\bibfield  {journal} {\bibinfo
   {journal} {Phys. Rev. D}\ }\textbf {\bibinfo {volume} {103}},\ \bibinfo
  {pages} {104014} (\bibinfo {year} {2021})},\ \bibinfo {note} {[Erratum:
  Phys.Rev.D 107, 089901 (2023)]},\ \Eprint {https://arxiv.org/abs/2102.02713}
  {arXiv:2102.02713 [gr-qc]} \BibitemShut {NoStop}%
\bibitem [{\citenamefont {{Shakura}}\ and\ \citenamefont
  {{Sunyaev}}(1973)}]{1973A&A....24..337S}%
  \BibitemOpen
  \bibfield  {author} {\bibinfo {author} {\bibfnamefont {N.~I.}\ \bibnamefont
  {{Shakura}}}\ and\ \bibinfo {author} {\bibfnamefont {R.~A.}\ \bibnamefont
  {{Sunyaev}}},\ }\bibfield  {title} {\bibinfo {title} {{Black holes in binary
  systems. Observational appearance.}},\ }\href@noop {} {\bibfield  {journal}
  {\bibinfo  {journal} {Astronomy and Astrophysics}\ }\textbf {\bibinfo
  {volume} {24}},\ \bibinfo {pages} {337} (\bibinfo {year} {1973})}\BibitemShut
  {NoStop}%
\bibitem [{\citenamefont {Novikov}\ and\ \citenamefont
  {Thorne}(1973)}]{Novikov:1973kta}%
  \BibitemOpen
  \bibfield  {author} {\bibinfo {author} {\bibfnamefont {I.~D.}\ \bibnamefont
  {Novikov}}\ and\ \bibinfo {author} {\bibfnamefont {K.~S.}\ \bibnamefont
  {Thorne}},\ }\bibfield  {title} {\bibinfo {title} {{Astrophysics and black
  holes}},\ }in\ \href@noop {} {\emph {\bibinfo {booktitle} {{Les Houches
  Summer School of Theoretical Physics}: {Black Holes}}}}\ (\bibinfo {year}
  {1973})\ pp.\ \bibinfo {pages} {343--550}\BibitemShut {NoStop}%
\bibitem [{\citenamefont {{Tanaka}}\ \emph
  {et~al.}(2002{\natexlab{a}})\citenamefont {{Tanaka}}, \citenamefont
  {{Takeuchi}},\ and\ \citenamefont {{Ward}}}]{2002ApJ...565.1257T}%
  \BibitemOpen
  \bibfield  {author} {\bibinfo {author} {\bibfnamefont {H.}~\bibnamefont
  {{Tanaka}}}, \bibinfo {author} {\bibfnamefont {T.}~\bibnamefont
  {{Takeuchi}}},\ and\ \bibinfo {author} {\bibfnamefont {W.~R.}\ \bibnamefont
  {{Ward}}},\ }\bibfield  {title} {\bibinfo {title} {{Three-Dimensional
  Interaction between a Planet and an Isothermal Gaseous Disk. I. Corotation
  and Lindblad Torques and Planet Migration}},\ }\href
  {https://doi.org/10.1086/324713} {\bibfield  {journal} {\bibinfo  {journal}
  {\apj}\ }\textbf {\bibinfo {volume} {565}},\ \bibinfo {pages} {1257}
  (\bibinfo {year} {2002}{\natexlab{a}})}\BibitemShut {NoStop}%
\bibitem [{\citenamefont {{Levin}}(2003)}]{YLevin:2003}%
  \BibitemOpen
  \bibfield  {author} {\bibinfo {author} {\bibfnamefont {Y.}~\bibnamefont
  {{Levin}}},\ }\bibfield  {title} {\bibinfo {title} {{Formation of massive
  stars and black holes in self-gravitating AGN discs, and gravitational waves
  in LISA band}},\ }\href {https://doi.org/10.48550/arXiv.astro-ph/0307084}
  {\bibfield  {journal} {\bibinfo  {journal} {arXiv e-prints}\ ,\ \bibinfo
  {eid} {astro-ph/0307084}} (\bibinfo {year} {2003})},\ \Eprint
  {https://arxiv.org/abs/astro-ph/0307084} {arXiv:astro-ph/0307084 [astro-ph]}
  \BibitemShut {NoStop}%
\bibitem [{\citenamefont {{Levin}}(2007)}]{YLevin:2007}%
  \BibitemOpen
  \bibfield  {author} {\bibinfo {author} {\bibfnamefont {Y.}~\bibnamefont
  {{Levin}}},\ }\bibfield  {title} {\bibinfo {title} {{Starbursts near
  supermassive black holes: young stars in the Galactic Centre, and
  gravitational waves in LISA band}},\ }\href
  {https://doi.org/10.1111/j.1365-2966.2006.11155.x} {\bibfield  {journal}
  {\bibinfo  {journal} {\mnras}\ }\textbf {\bibinfo {volume} {374}},\ \bibinfo
  {pages} {515} (\bibinfo {year} {2007})},\ \Eprint
  {https://arxiv.org/abs/astro-ph/0603583} {arXiv:astro-ph/0603583 [astro-ph]}
  \BibitemShut {NoStop}%
\bibitem [{\citenamefont {Barausse}\ and\ \citenamefont
  {Rezzolla}(2008)}]{Barausse:2007dy}%
  \BibitemOpen
  \bibfield  {author} {\bibinfo {author} {\bibfnamefont {E.}~\bibnamefont
  {Barausse}}\ and\ \bibinfo {author} {\bibfnamefont {L.}~\bibnamefont
  {Rezzolla}},\ }\bibfield  {title} {\bibinfo {title} {{The Influence of the
  hydrodynamic drag from an accretion torus on extreme mass-ratio inspirals}},\
  }\href {https://doi.org/10.1103/PhysRevD.77.104027} {\bibfield  {journal}
  {\bibinfo  {journal} {Phys. Rev. D}\ }\textbf {\bibinfo {volume} {77}},\
  \bibinfo {pages} {104027} (\bibinfo {year} {2008})},\ \Eprint
  {https://arxiv.org/abs/0711.4558} {arXiv:0711.4558 [gr-qc]} \BibitemShut
  {NoStop}%
\bibitem [{\citenamefont {Abramowicz}\ and\ \citenamefont
  {Fragile}(2013)}]{Abramowicz:2011xu}%
  \BibitemOpen
  \bibfield  {author} {\bibinfo {author} {\bibfnamefont {M.~A.}\ \bibnamefont
  {Abramowicz}}\ and\ \bibinfo {author} {\bibfnamefont {P.~C.}\ \bibnamefont
  {Fragile}},\ }\bibfield  {title} {\bibinfo {title} {{Foundations of Black
  Hole Accretion Disk Theory}},\ }\href {https://doi.org/10.12942/lrr-2013-1}
  {\bibfield  {journal} {\bibinfo  {journal} {Living Rev. Rel.}\ }\textbf
  {\bibinfo {volume} {16}},\ \bibinfo {pages} {1} (\bibinfo {year} {2013})},\
  \Eprint {https://arxiv.org/abs/1104.5499} {arXiv:1104.5499 [astro-ph.HE]}
  \BibitemShut {NoStop}%
\bibitem [{\citenamefont {{Yunes}}\ \emph {et~al.}(2011)\citenamefont
  {{Yunes}}, \citenamefont {{Kocsis}}, \citenamefont {{Loeb}},\ and\
  \citenamefont {{Haiman}}}]{Yunes:2011ws}%
  \BibitemOpen
  \bibfield  {author} {\bibinfo {author} {\bibfnamefont {N.}~\bibnamefont
  {{Yunes}}}, \bibinfo {author} {\bibfnamefont {B.}~\bibnamefont {{Kocsis}}},
  \bibinfo {author} {\bibfnamefont {A.}~\bibnamefont {{Loeb}}},\ and\ \bibinfo
  {author} {\bibfnamefont {Z.}~\bibnamefont {{Haiman}}},\ }\bibfield  {title}
  {\bibinfo {title} {{Imprint of Accretion Disk-Induced Migration on
  Gravitational Waves from Extreme Mass Ratio Inspirals}},\ }\href
  {https://doi.org/10.1103/PhysRevLett.107.171103} {\bibfield  {journal}
  {\bibinfo  {journal} {\prl}\ }\textbf {\bibinfo {volume} {107}},\ \bibinfo
  {eid} {171103} (\bibinfo {year} {2011})},\ \Eprint
  {https://arxiv.org/abs/1103.4609} {arXiv:1103.4609 [astro-ph.CO]}
  \BibitemShut {NoStop}%
\bibitem [{\citenamefont {{Kocsis}}\ \emph {et~al.}(2011)\citenamefont
  {{Kocsis}}, \citenamefont {{Yunes}},\ and\ \citenamefont
  {{Loeb}}}]{Kocsis:2011dr}%
  \BibitemOpen
  \bibfield  {author} {\bibinfo {author} {\bibfnamefont {B.}~\bibnamefont
  {{Kocsis}}}, \bibinfo {author} {\bibfnamefont {N.}~\bibnamefont {{Yunes}}},\
  and\ \bibinfo {author} {\bibfnamefont {A.}~\bibnamefont {{Loeb}}},\
  }\bibfield  {title} {\bibinfo {title} {{Observable signatures of extreme
  mass-ratio inspiral black hole binaries embedded in thin accretion disks}},\
  }\href {https://doi.org/10.1103/PhysRevD.84.024032} {\bibfield  {journal}
  {\bibinfo  {journal} {\prd}\ }\textbf {\bibinfo {volume} {84}},\ \bibinfo
  {eid} {024032} (\bibinfo {year} {2011})},\ \Eprint
  {https://arxiv.org/abs/1104.2322} {arXiv:1104.2322 [astro-ph.GA]}
  \BibitemShut {NoStop}%
\bibitem [{\citenamefont {Barausse}\ \emph {et~al.}(2014)\citenamefont
  {Barausse}, \citenamefont {Cardoso},\ and\ \citenamefont
  {Pani}}]{Barausse:2014tra}%
  \BibitemOpen
  \bibfield  {author} {\bibinfo {author} {\bibfnamefont {E.}~\bibnamefont
  {Barausse}}, \bibinfo {author} {\bibfnamefont {V.}~\bibnamefont {Cardoso}},\
  and\ \bibinfo {author} {\bibfnamefont {P.}~\bibnamefont {Pani}},\ }\bibfield
  {title} {\bibinfo {title} {{Can environmental effects spoil precision
  gravitational-wave astrophysics?}},\ }\href
  {https://doi.org/10.1103/PhysRevD.89.104059} {\bibfield  {journal} {\bibinfo
  {journal} {Phys. Rev. D}\ }\textbf {\bibinfo {volume} {89}},\ \bibinfo
  {pages} {104059} (\bibinfo {year} {2014})},\ \Eprint
  {https://arxiv.org/abs/1404.7149} {arXiv:1404.7149 [gr-qc]} \BibitemShut
  {NoStop}%
\bibitem [{\citenamefont {Derdzinski}\ \emph {et~al.}(2019)\citenamefont
  {Derdzinski}, \citenamefont {D'Orazio}, \citenamefont {Duffell},
  \citenamefont {Haiman},\ and\ \citenamefont
  {MacFadyen}}]{Derdzinski:2018qzv}%
  \BibitemOpen
  \bibfield  {author} {\bibinfo {author} {\bibfnamefont {A.~M.}\ \bibnamefont
  {Derdzinski}}, \bibinfo {author} {\bibfnamefont {D.}~\bibnamefont
  {D'Orazio}}, \bibinfo {author} {\bibfnamefont {P.}~\bibnamefont {Duffell}},
  \bibinfo {author} {\bibfnamefont {Z.}~\bibnamefont {Haiman}},\ and\ \bibinfo
  {author} {\bibfnamefont {A.}~\bibnamefont {MacFadyen}},\ }\bibfield  {title}
  {\bibinfo {title} {{Probing gas disc physics with LISA: simulations of an
  intermediate mass ratio inspiral in an accretion disc}},\ }\href
  {https://doi.org/10.1093/mnras/stz1026} {\bibfield  {journal} {\bibinfo
  {journal} {Mon. Not. Roy. Astron. Soc.}\ }\textbf {\bibinfo {volume} {486}},\
  \bibinfo {pages} {2754} (\bibinfo {year} {2019})},\ \bibinfo {note}
  {[Erratum: Mon.Not.Roy.Astron.Soc. 489, 4860--4861 (2019)]},\ \Eprint
  {https://arxiv.org/abs/1810.03623} {arXiv:1810.03623 [astro-ph.HE]}
  \BibitemShut {NoStop}%
\bibitem [{\citenamefont {Duffell}\ \emph {et~al.}(2020)\citenamefont
  {Duffell}, \citenamefont {D'Orazio}, \citenamefont {Derdzinski},
  \citenamefont {Haiman}, \citenamefont {MacFadyen}, \citenamefont {Rosen},\
  and\ \citenamefont {Zrake}}]{Duffell:2019uuk}%
  \BibitemOpen
  \bibfield  {author} {\bibinfo {author} {\bibfnamefont {P.~C.}\ \bibnamefont
  {Duffell}}, \bibinfo {author} {\bibfnamefont {D.}~\bibnamefont {D'Orazio}},
  \bibinfo {author} {\bibfnamefont {A.}~\bibnamefont {Derdzinski}}, \bibinfo
  {author} {\bibfnamefont {Z.}~\bibnamefont {Haiman}}, \bibinfo {author}
  {\bibfnamefont {A.}~\bibnamefont {MacFadyen}}, \bibinfo {author}
  {\bibfnamefont {A.~L.}\ \bibnamefont {Rosen}},\ and\ \bibinfo {author}
  {\bibfnamefont {J.}~\bibnamefont {Zrake}},\ }\bibfield  {title} {\bibinfo
  {title} {{Circumbinary Disks: Accretion and Torque as a Function of Mass
  Ratio and Disk Viscosity}},\ }\href
  {https://doi.org/10.3847/1538-4357/abab95} {\bibfield  {journal} {\bibinfo
  {journal} {Astrophys. J.}\ }\textbf {\bibinfo {volume} {901}},\ \bibinfo
  {pages} {25} (\bibinfo {year} {2020})},\ \Eprint
  {https://arxiv.org/abs/1911.05506} {arXiv:1911.05506 [astro-ph.SR]}
  \BibitemShut {NoStop}%
\bibitem [{\citenamefont {Derdzinski}\ \emph {et~al.}(2021)\citenamefont
  {Derdzinski}, \citenamefont {D'Orazio}, \citenamefont {Duffell},
  \citenamefont {Haiman},\ and\ \citenamefont
  {MacFadyen}}]{Derdzinski:2020wlw}%
  \BibitemOpen
  \bibfield  {author} {\bibinfo {author} {\bibfnamefont {A.}~\bibnamefont
  {Derdzinski}}, \bibinfo {author} {\bibfnamefont {D.}~\bibnamefont
  {D'Orazio}}, \bibinfo {author} {\bibfnamefont {P.}~\bibnamefont {Duffell}},
  \bibinfo {author} {\bibfnamefont {Z.}~\bibnamefont {Haiman}},\ and\ \bibinfo
  {author} {\bibfnamefont {A.}~\bibnamefont {MacFadyen}},\ }\bibfield  {title}
  {\bibinfo {title} {{Evolution of gas disc\textendash{}embedded intermediate
  mass ratio inspirals in the $LISA$ band}},\ }\href
  {https://doi.org/10.1093/mnras/staa3976} {\bibfield  {journal} {\bibinfo
  {journal} {Mon. Not. Roy. Astron. Soc.}\ }\textbf {\bibinfo {volume} {501}},\
  \bibinfo {pages} {3540} (\bibinfo {year} {2021})},\ \Eprint
  {https://arxiv.org/abs/2005.11333} {arXiv:2005.11333 [astro-ph.HE]}
  \BibitemShut {NoStop}%
\bibitem [{\citenamefont {Tagawa}\ \emph {et~al.}(2021)\citenamefont {Tagawa},
  \citenamefont {Kocsis}, \citenamefont {Haiman}, \citenamefont {Bartos},
  \citenamefont {Omukai},\ and\ \citenamefont {Samsing}}]{Tagawa:2020qll}%
  \BibitemOpen
  \bibfield  {author} {\bibinfo {author} {\bibfnamefont {H.}~\bibnamefont
  {Tagawa}}, \bibinfo {author} {\bibfnamefont {B.}~\bibnamefont {Kocsis}},
  \bibinfo {author} {\bibfnamefont {Z.}~\bibnamefont {Haiman}}, \bibinfo
  {author} {\bibfnamefont {I.}~\bibnamefont {Bartos}}, \bibinfo {author}
  {\bibfnamefont {K.}~\bibnamefont {Omukai}},\ and\ \bibinfo {author}
  {\bibfnamefont {J.}~\bibnamefont {Samsing}},\ }\bibfield  {title} {\bibinfo
  {title} {{Mass-gap Mergers in Active Galactic Nuclei}},\ }\href
  {https://doi.org/10.3847/1538-4357/abd555} {\bibfield  {journal} {\bibinfo
  {journal} {Astrophys. J.}\ }\textbf {\bibinfo {volume} {908}},\ \bibinfo
  {pages} {194} (\bibinfo {year} {2021})},\ \Eprint
  {https://arxiv.org/abs/2012.00011} {arXiv:2012.00011 [astro-ph.HE]}
  \BibitemShut {NoStop}%
\bibitem [{\citenamefont {Pan}\ and\ \citenamefont {Yang}(2021)}]{Pan:2021ksp}%
  \BibitemOpen
  \bibfield  {author} {\bibinfo {author} {\bibfnamefont {Z.}~\bibnamefont
  {Pan}}\ and\ \bibinfo {author} {\bibfnamefont {H.}~\bibnamefont {Yang}},\
  }\bibfield  {title} {\bibinfo {title} {{Formation Rate of Extreme Mass Ratio
  Inspirals in Active Galactic Nuclei}},\ }\href
  {https://doi.org/10.1103/PhysRevD.103.103018} {\bibfield  {journal} {\bibinfo
   {journal} {Phys. Rev. D}\ }\textbf {\bibinfo {volume} {103}},\ \bibinfo
  {pages} {103018} (\bibinfo {year} {2021})},\ \Eprint
  {https://arxiv.org/abs/2101.09146} {arXiv:2101.09146 [astro-ph.HE]}
  \BibitemShut {NoStop}%
\bibitem [{\citenamefont {Pan}\ \emph {et~al.}(2021)\citenamefont {Pan},
  \citenamefont {Lyu},\ and\ \citenamefont {Yang}}]{Pan:2021oob}%
  \BibitemOpen
  \bibfield  {author} {\bibinfo {author} {\bibfnamefont {Z.}~\bibnamefont
  {Pan}}, \bibinfo {author} {\bibfnamefont {Z.}~\bibnamefont {Lyu}},\ and\
  \bibinfo {author} {\bibfnamefont {H.}~\bibnamefont {Yang}},\ }\bibfield
  {title} {\bibinfo {title} {{Wet extreme mass ratio inspirals may be more
  common for spaceborne gravitational wave detection}},\ }\href
  {https://doi.org/10.1103/PhysRevD.104.063007} {\bibfield  {journal} {\bibinfo
   {journal} {Phys. Rev. D}\ }\textbf {\bibinfo {volume} {104}},\ \bibinfo
  {pages} {063007} (\bibinfo {year} {2021})},\ \Eprint
  {https://arxiv.org/abs/2104.01208} {arXiv:2104.01208 [astro-ph.HE]}
  \BibitemShut {NoStop}%
\bibitem [{\citenamefont {Zwick}\ \emph {et~al.}(2022)\citenamefont {Zwick},
  \citenamefont {Derdzinski}, \citenamefont {Garg}, \citenamefont {Capelo},\
  and\ \citenamefont {Mayer}}]{Zwick:2021dlg}%
  \BibitemOpen
  \bibfield  {author} {\bibinfo {author} {\bibfnamefont {L.}~\bibnamefont
  {Zwick}}, \bibinfo {author} {\bibfnamefont {A.}~\bibnamefont {Derdzinski}},
  \bibinfo {author} {\bibfnamefont {M.}~\bibnamefont {Garg}}, \bibinfo {author}
  {\bibfnamefont {P.~R.}\ \bibnamefont {Capelo}},\ and\ \bibinfo {author}
  {\bibfnamefont {L.}~\bibnamefont {Mayer}},\ }\bibfield  {title} {\bibinfo
  {title} {{Dirty waveforms: multiband harmonic content of gas-embedded
  gravitational wave sources}},\ }\href {https://doi.org/10.1093/mnras/stac299}
  {\bibfield  {journal} {\bibinfo  {journal} {Mon. Not. Roy. Astron. Soc.}\
  }\textbf {\bibinfo {volume} {511}},\ \bibinfo {pages} {6143} (\bibinfo {year}
  {2022})},\ \Eprint {https://arxiv.org/abs/2110.09097} {arXiv:2110.09097
  [astro-ph.HE]} \BibitemShut {NoStop}%
\bibitem [{\citenamefont {Derdzinski}\ and\ \citenamefont
  {Mayer}(2023)}]{Derdzinski:2022ltb}%
  \BibitemOpen
  \bibfield  {author} {\bibinfo {author} {\bibfnamefont {A.}~\bibnamefont
  {Derdzinski}}\ and\ \bibinfo {author} {\bibfnamefont {L.}~\bibnamefont
  {Mayer}},\ }\bibfield  {title} {\bibinfo {title} {{In situ extreme mass ratio
  inspirals via subparsec formation and migration of stars in thin,
  gravitationally unstable AGN discs}},\ }\href
  {https://doi.org/10.1093/mnras/stad749} {\bibfield  {journal} {\bibinfo
  {journal} {Mon. Not. Roy. Astron. Soc.}\ }\textbf {\bibinfo {volume} {521}},\
  \bibinfo {pages} {4522} (\bibinfo {year} {2023})},\ \Eprint
  {https://arxiv.org/abs/2205.10382} {arXiv:2205.10382 [astro-ph.GA]}
  \BibitemShut {NoStop}%
\bibitem [{\citenamefont {Cole}\ \emph {et~al.}(2023)\citenamefont {Cole},
  \citenamefont {Bertone}, \citenamefont {Coogan}, \citenamefont {Gaggero},
  \citenamefont {Karydas}, \citenamefont {Kavanagh}, \citenamefont {Spieksma},\
  and\ \citenamefont {Tomaselli}}]{Cole:2022yzw}%
  \BibitemOpen
  \bibfield  {author} {\bibinfo {author} {\bibfnamefont {P.~S.}\ \bibnamefont
  {Cole}}, \bibinfo {author} {\bibfnamefont {G.}~\bibnamefont {Bertone}},
  \bibinfo {author} {\bibfnamefont {A.}~\bibnamefont {Coogan}}, \bibinfo
  {author} {\bibfnamefont {D.}~\bibnamefont {Gaggero}}, \bibinfo {author}
  {\bibfnamefont {T.}~\bibnamefont {Karydas}}, \bibinfo {author} {\bibfnamefont
  {B.~J.}\ \bibnamefont {Kavanagh}}, \bibinfo {author} {\bibfnamefont
  {T.~F.~M.}\ \bibnamefont {Spieksma}},\ and\ \bibinfo {author} {\bibfnamefont
  {G.~M.}\ \bibnamefont {Tomaselli}},\ }\bibfield  {title} {\bibinfo {title}
  {{Distinguishing environmental effects on binary black hole gravitational
  waveforms}},\ }\href {https://doi.org/10.1038/s41550-023-01990-2} {\bibfield
  {journal} {\bibinfo  {journal} {Nature Astron.}\ }\textbf {\bibinfo {volume}
  {7}},\ \bibinfo {pages} {943} (\bibinfo {year} {2023})},\ \Eprint
  {https://arxiv.org/abs/2211.01362} {arXiv:2211.01362 [gr-qc]} \BibitemShut
  {NoStop}%
\bibitem [{\citenamefont {Speri}\ \emph {et~al.}(2023)\citenamefont {Speri},
  \citenamefont {Antonelli}, \citenamefont {Sberna}, \citenamefont {Babak},
  \citenamefont {Barausse}, \citenamefont {Gair},\ and\ \citenamefont
  {Katz}}]{Speri:2022upm}%
  \BibitemOpen
  \bibfield  {author} {\bibinfo {author} {\bibfnamefont {L.}~\bibnamefont
  {Speri}}, \bibinfo {author} {\bibfnamefont {A.}~\bibnamefont {Antonelli}},
  \bibinfo {author} {\bibfnamefont {L.}~\bibnamefont {Sberna}}, \bibinfo
  {author} {\bibfnamefont {S.}~\bibnamefont {Babak}}, \bibinfo {author}
  {\bibfnamefont {E.}~\bibnamefont {Barausse}}, \bibinfo {author}
  {\bibfnamefont {J.~R.}\ \bibnamefont {Gair}},\ and\ \bibinfo {author}
  {\bibfnamefont {M.~L.}\ \bibnamefont {Katz}},\ }\bibfield  {title} {\bibinfo
  {title} {{Probing Accretion Physics with Gravitational Waves}},\ }\href
  {https://doi.org/10.1103/PhysRevX.13.021035} {\bibfield  {journal} {\bibinfo
  {journal} {Phys. Rev. X}\ }\textbf {\bibinfo {volume} {13}},\ \bibinfo
  {pages} {021035} (\bibinfo {year} {2023})},\ \Eprint
  {https://arxiv.org/abs/2207.10086} {arXiv:2207.10086 [gr-qc]} \BibitemShut
  {NoStop}%
\bibitem [{\citenamefont {Garg}\ \emph {et~al.}(2022)\citenamefont {Garg},
  \citenamefont {Derdzinski}, \citenamefont {Zwick}, \citenamefont {Capelo},\
  and\ \citenamefont {Mayer}}]{Garg:2022nko}%
  \BibitemOpen
  \bibfield  {author} {\bibinfo {author} {\bibfnamefont {M.}~\bibnamefont
  {Garg}}, \bibinfo {author} {\bibfnamefont {A.}~\bibnamefont {Derdzinski}},
  \bibinfo {author} {\bibfnamefont {L.}~\bibnamefont {Zwick}}, \bibinfo
  {author} {\bibfnamefont {P.~R.}\ \bibnamefont {Capelo}},\ and\ \bibinfo
  {author} {\bibfnamefont {L.}~\bibnamefont {Mayer}},\ }\bibfield  {title}
  {\bibinfo {title} {{The imprint of gas on gravitational waves from LISA
  intermediate-mass black hole binaries}},\ }\href
  {https://doi.org/10.1093/mnras/stac2711} {\bibfield  {journal} {\bibinfo
  {journal} {Mon. Not. Roy. Astron. Soc.}\ }\textbf {\bibinfo {volume} {517}},\
  \bibinfo {pages} {1339} (\bibinfo {year} {2022})},\ \Eprint
  {https://arxiv.org/abs/2206.05292} {arXiv:2206.05292 [astro-ph.GA]}
  \BibitemShut {NoStop}%
\bibitem [{\citenamefont {Morton}\ \emph {et~al.}(2023)\citenamefont {Morton},
  \citenamefont {Rinaldi}, \citenamefont {Torres-Orjuela}, \citenamefont
  {Derdzinski}, \citenamefont {Vaccaro},\ and\ \citenamefont
  {Del~Pozzo}}]{Morton:2023wxg}%
  \BibitemOpen
  \bibfield  {author} {\bibinfo {author} {\bibfnamefont {S.~L.}\ \bibnamefont
  {Morton}}, \bibinfo {author} {\bibfnamefont {S.}~\bibnamefont {Rinaldi}},
  \bibinfo {author} {\bibfnamefont {A.}~\bibnamefont {Torres-Orjuela}},
  \bibinfo {author} {\bibfnamefont {A.}~\bibnamefont {Derdzinski}}, \bibinfo
  {author} {\bibfnamefont {M.~P.}\ \bibnamefont {Vaccaro}},\ and\ \bibinfo
  {author} {\bibfnamefont {W.}~\bibnamefont {Del~Pozzo}},\ }\bibfield  {title}
  {\bibinfo {title} {{GW190521: A binary black hole merger inside an active
  galactic nucleus?}},\ }\href {https://doi.org/10.1103/PhysRevD.108.123039}
  {\bibfield  {journal} {\bibinfo  {journal} {Phys. Rev. D}\ }\textbf {\bibinfo
  {volume} {108}},\ \bibinfo {pages} {123039} (\bibinfo {year} {2023})},\
  \Eprint {https://arxiv.org/abs/2310.16025} {arXiv:2310.16025 [gr-qc]}
  \BibitemShut {NoStop}%
\bibitem [{\citenamefont {Tiede}\ \emph {et~al.}(2024)\citenamefont {Tiede},
  \citenamefont {D'Orazio}, \citenamefont {Zwick},\ and\ \citenamefont
  {Duffell}}]{Tiede:2023cje}%
  \BibitemOpen
  \bibfield  {author} {\bibinfo {author} {\bibfnamefont {C.}~\bibnamefont
  {Tiede}}, \bibinfo {author} {\bibfnamefont {D.~J.}\ \bibnamefont {D'Orazio}},
  \bibinfo {author} {\bibfnamefont {L.}~\bibnamefont {Zwick}},\ and\ \bibinfo
  {author} {\bibfnamefont {P.~C.}\ \bibnamefont {Duffell}},\ }\bibfield
  {title} {\bibinfo {title} {{Disk-induced Binary Precession: Implications for
  Dynamics and Multimessenger Observations of Black Hole Binaries}},\ }\href
  {https://doi.org/10.3847/1538-4357/ad2613} {\bibfield  {journal} {\bibinfo
  {journal} {Astrophys. J.}\ }\textbf {\bibinfo {volume} {964}},\ \bibinfo
  {pages} {46} (\bibinfo {year} {2024})},\ \Eprint
  {https://arxiv.org/abs/2312.01805} {arXiv:2312.01805 [astro-ph.HE]}
  \BibitemShut {NoStop}%
\bibitem [{\citenamefont {Zwick}\ \emph {et~al.}(2024)\citenamefont {Zwick},
  \citenamefont {Tiede}, \citenamefont {Trani}, \citenamefont {Derdzinski},
  \citenamefont {Haiman}, \citenamefont {D'Orazio},\ and\ \citenamefont
  {Samsing}}]{Zwick:2024yzh}%
  \BibitemOpen
  \bibfield  {author} {\bibinfo {author} {\bibfnamefont {L.}~\bibnamefont
  {Zwick}}, \bibinfo {author} {\bibfnamefont {C.}~\bibnamefont {Tiede}},
  \bibinfo {author} {\bibfnamefont {A.~A.}\ \bibnamefont {Trani}}, \bibinfo
  {author} {\bibfnamefont {A.}~\bibnamefont {Derdzinski}}, \bibinfo {author}
  {\bibfnamefont {Z.}~\bibnamefont {Haiman}}, \bibinfo {author} {\bibfnamefont
  {D.~J.}\ \bibnamefont {D'Orazio}},\ and\ \bibinfo {author} {\bibfnamefont
  {J.}~\bibnamefont {Samsing}},\ }\bibfield  {title} {\bibinfo {title} {{Novel
  category of environmental effects on gravitational waves from binaries
  perturbed by periodic forces}},\ }\href
  {https://doi.org/10.1103/PhysRevD.110.103005} {\bibfield  {journal} {\bibinfo
   {journal} {Phys. Rev. D}\ }\textbf {\bibinfo {volume} {110}},\ \bibinfo
  {pages} {103005} (\bibinfo {year} {2024})},\ \Eprint
  {https://arxiv.org/abs/2405.05698} {arXiv:2405.05698 [gr-qc]} \BibitemShut
  {NoStop}%
\bibitem [{\citenamefont {Garg}\ \emph
  {et~al.}(2024{\natexlab{a}})\citenamefont {Garg}, \citenamefont {Derdzinski},
  \citenamefont {Tiwari}, \citenamefont {Gair},\ and\ \citenamefont
  {Mayer}}]{Garg:2024oeu}%
  \BibitemOpen
  \bibfield  {author} {\bibinfo {author} {\bibfnamefont {M.}~\bibnamefont
  {Garg}}, \bibinfo {author} {\bibfnamefont {A.}~\bibnamefont {Derdzinski}},
  \bibinfo {author} {\bibfnamefont {S.}~\bibnamefont {Tiwari}}, \bibinfo
  {author} {\bibfnamefont {J.}~\bibnamefont {Gair}},\ and\ \bibinfo {author}
  {\bibfnamefont {L.}~\bibnamefont {Mayer}},\ }\bibfield  {title} {\bibinfo
  {title} {{Measuring eccentricity and gas-induced perturbation from
  gravitational waves of LISA massive black hole binaries}},\ }\href
  {https://doi.org/10.1093/mnras/stae1764} {\bibfield  {journal} {\bibinfo
  {journal} {Mon. Not. Roy. Astron. Soc.}\ }\textbf {\bibinfo {volume} {532}},\
  \bibinfo {pages} {4060} (\bibinfo {year} {2024}{\natexlab{a}})},\ \Eprint
  {https://arxiv.org/abs/2402.14058} {arXiv:2402.14058 [astro-ph.GA]}
  \BibitemShut {NoStop}%
\bibitem [{\citenamefont {Garg}\ \emph
  {et~al.}(2024{\natexlab{b}})\citenamefont {Garg}, \citenamefont {Tiede},\
  and\ \citenamefont {D'Orazio}}]{Garg:2024yrs}%
  \BibitemOpen
  \bibfield  {author} {\bibinfo {author} {\bibfnamefont {M.}~\bibnamefont
  {Garg}}, \bibinfo {author} {\bibfnamefont {C.}~\bibnamefont {Tiede}},\ and\
  \bibinfo {author} {\bibfnamefont {D.~J.}\ \bibnamefont {D'Orazio}},\
  }\bibfield  {title} {\bibinfo {title} {{Accretion-mediated
  spin\textendash{}eccentricity correlations in LISA massive black hole
  binaries}},\ }\href {https://doi.org/10.1093/mnras/stae2357} {\bibfield
  {journal} {\bibinfo  {journal} {Mon. Not. Roy. Astron. Soc.}\ }\textbf
  {\bibinfo {volume} {534}},\ \bibinfo {pages} {3705} (\bibinfo {year}
  {2024}{\natexlab{b}})},\ \Eprint {https://arxiv.org/abs/2405.04411}
  {arXiv:2405.04411 [astro-ph.HE]} \BibitemShut {NoStop}%
\bibitem [{\citenamefont {{Garg}}\ \emph {et~al.}(2025)\citenamefont {{Garg}},
  \citenamefont {{Franchini}}, \citenamefont {{Lupi}}, \citenamefont
  {{Bonetti}},\ and\ \citenamefont {{Mayer}}}]{Garg:2024zku}%
  \BibitemOpen
  \bibfield  {author} {\bibinfo {author} {\bibfnamefont {M.}~\bibnamefont
  {{Garg}}}, \bibinfo {author} {\bibfnamefont {A.}~\bibnamefont {{Franchini}}},
  \bibinfo {author} {\bibfnamefont {A.}~\bibnamefont {{Lupi}}}, \bibinfo
  {author} {\bibfnamefont {M.}~\bibnamefont {{Bonetti}}},\ and\ \bibinfo
  {author} {\bibfnamefont {L.}~\bibnamefont {{Mayer}}},\ }\bibfield  {title}
  {\bibinfo {title} {{Gas-induced Perturbations on the Gravitational Wave
  Inspiral of Live Post-Newtonian LISA Massive Black Hole Binaries}},\ }\href
  {https://doi.org/10.3847/1538-4357/ae10b4} {\bibfield  {journal} {\bibinfo
  {journal} {\apj}\ }\textbf {\bibinfo {volume} {993}},\ \bibinfo {eid} {145}
  (\bibinfo {year} {2025})},\ \Eprint {https://arxiv.org/abs/2410.17305}
  {arXiv:2410.17305 [astro-ph.HE]} \BibitemShut {NoStop}%
\bibitem [{\citenamefont {{Khalvati}}\ \emph {et~al.}(2025)\citenamefont
  {{Khalvati}}, \citenamefont {{Santini}}, \citenamefont {{Duque}},
  \citenamefont {{Speri}}, \citenamefont {{Gair}}, \citenamefont {{Yang}},\
  and\ \citenamefont {{Brito}}}]{Khalvati:2024tzz}%
  \BibitemOpen
  \bibfield  {author} {\bibinfo {author} {\bibfnamefont {H.}~\bibnamefont
  {{Khalvati}}}, \bibinfo {author} {\bibfnamefont {A.}~\bibnamefont
  {{Santini}}}, \bibinfo {author} {\bibfnamefont {F.}~\bibnamefont {{Duque}}},
  \bibinfo {author} {\bibfnamefont {L.}~\bibnamefont {{Speri}}}, \bibinfo
  {author} {\bibfnamefont {J.}~\bibnamefont {{Gair}}}, \bibinfo {author}
  {\bibfnamefont {H.}~\bibnamefont {{Yang}}},\ and\ \bibinfo {author}
  {\bibfnamefont {R.}~\bibnamefont {{Brito}}},\ }\bibfield  {title} {\bibinfo
  {title} {{Impact of relativistic waveforms in LISA's science objectives with
  extreme-mass-ratio inspirals}},\ }\href
  {https://doi.org/10.1103/PhysRevD.111.082010} {\bibfield  {journal} {\bibinfo
   {journal} {\prd}\ }\textbf {\bibinfo {volume} {111}},\ \bibinfo {eid}
  {082010} (\bibinfo {year} {2025})},\ \Eprint
  {https://arxiv.org/abs/2410.17310} {arXiv:2410.17310 [gr-qc]} \BibitemShut
  {NoStop}%
\bibitem [{\citenamefont {{Duque}}\ \emph
  {et~al.}(2025{\natexlab{a}})\citenamefont {{Duque}}, \citenamefont
  {{Kejriwal}}, \citenamefont {{Sberna}}, \citenamefont {{Speri}},\ and\
  \citenamefont {{Gair}}}]{Duque:2024mfw}%
  \BibitemOpen
  \bibfield  {author} {\bibinfo {author} {\bibfnamefont {F.}~\bibnamefont
  {{Duque}}}, \bibinfo {author} {\bibfnamefont {S.}~\bibnamefont {{Kejriwal}}},
  \bibinfo {author} {\bibfnamefont {L.}~\bibnamefont {{Sberna}}}, \bibinfo
  {author} {\bibfnamefont {L.}~\bibnamefont {{Speri}}},\ and\ \bibinfo {author}
  {\bibfnamefont {J.}~\bibnamefont {{Gair}}},\ }\bibfield  {title} {\bibinfo
  {title} {{Constraining accretion physics with gravitational waves from
  eccentric extreme-mass-ratio inspirals}},\ }\href
  {https://doi.org/10.1103/PhysRevD.111.084006} {\bibfield  {journal} {\bibinfo
   {journal} {\prd}\ }\textbf {\bibinfo {volume} {111}},\ \bibinfo {eid}
  {084006} (\bibinfo {year} {2025}{\natexlab{a}})},\ \Eprint
  {https://arxiv.org/abs/2411.03436} {arXiv:2411.03436 [gr-qc]} \BibitemShut
  {NoStop}%
\bibitem [{\citenamefont {{Spieksma}}\ and\ \citenamefont
  {{Cannizzaro}}(2025)}]{Spieksma:2025wex}%
  \BibitemOpen
  \bibfield  {author} {\bibinfo {author} {\bibfnamefont {T.~F.~M.}\
  \bibnamefont {{Spieksma}}}\ and\ \bibinfo {author} {\bibfnamefont
  {E.}~\bibnamefont {{Cannizzaro}}},\ }\bibfield  {title} {\bibinfo {title}
  {{In the grip of the disk: dragging the companion through an AGN}},\ }\href
  {https://doi.org/10.48550/arXiv.2504.08033} {\bibfield  {journal} {\bibinfo
  {journal} {arXiv e-prints}\ ,\ \bibinfo {eid} {arXiv:2504.08033}} (\bibinfo
  {year} {2025})},\ \Eprint {https://arxiv.org/abs/2504.08033}
  {arXiv:2504.08033 [astro-ph.GA]} \BibitemShut {NoStop}%
\bibitem [{\citenamefont {{Zwick}}\ \emph {et~al.}(2025)\citenamefont
  {{Zwick}}, \citenamefont {{Hendriks}}, \citenamefont {{O'Neill}},
  \citenamefont {{Tak{\'a}tsy}}, \citenamefont {{Kirkeberg}}, \citenamefont
  {{Tiede}}, \citenamefont {{Stegmann}}, \citenamefont {{Samsing}},\ and\
  \citenamefont {{D'Orazio}}}]{Zwick_eccenv+2025}%
  \BibitemOpen
  \bibfield  {author} {\bibinfo {author} {\bibfnamefont {L.}~\bibnamefont
  {{Zwick}}}, \bibinfo {author} {\bibfnamefont {K.}~\bibnamefont {{Hendriks}}},
  \bibinfo {author} {\bibfnamefont {D.}~\bibnamefont {{O'Neill}}}, \bibinfo
  {author} {\bibfnamefont {J.}~\bibnamefont {{Tak{\'a}tsy}}}, \bibinfo {author}
  {\bibfnamefont {P.}~\bibnamefont {{Kirkeberg}}}, \bibinfo {author}
  {\bibfnamefont {C.}~\bibnamefont {{Tiede}}}, \bibinfo {author} {\bibfnamefont
  {J.}~\bibnamefont {{Stegmann}}}, \bibinfo {author} {\bibfnamefont
  {J.}~\bibnamefont {{Samsing}}},\ and\ \bibinfo {author} {\bibfnamefont
  {D.~J.}\ \bibnamefont {{D'Orazio}}},\ }\bibfield  {title} {\bibinfo {title}
  {{Dissecting environmental effects with eccentric gravitational wave
  sources}},\ }\href {https://doi.org/10.1103/lz7k-bvjf} {\bibfield  {journal}
  {\bibinfo  {journal} {\prd}\ }\textbf {\bibinfo {volume} {112}},\ \bibinfo
  {eid} {063005} (\bibinfo {year} {2025})},\ \Eprint
  {https://arxiv.org/abs/2506.09140} {arXiv:2506.09140 [astro-ph.HE]}
  \BibitemShut {NoStop}%
\bibitem [{\citenamefont {{Zwick}}(2026)}]{Zwick:2026}%
  \BibitemOpen
  \bibfield  {author} {\bibinfo {author} {\bibfnamefont {L.}~\bibnamefont
  {{Zwick}}},\ }\bibfield  {title} {\bibinfo {title} {{Limits of
  vacuum-template subtraction for LISA massive black hole binary sources in
  realistic environments}},\ }\href {https://doi.org/10.48550/arXiv.2601.06684}
  {\bibfield  {journal} {\bibinfo  {journal} {arXiv e-prints}\ ,\ \bibinfo
  {eid} {arXiv:2601.06684}} (\bibinfo {year} {2026})},\ \Eprint
  {https://arxiv.org/abs/2601.06684} {arXiv:2601.06684 [gr-qc]} \BibitemShut
  {NoStop}%
\bibitem [{\citenamefont {Zwick}\ \emph {et~al.}(2023)\citenamefont {Zwick},
  \citenamefont {Capelo},\ and\ \citenamefont {Mayer}}]{Zwick:2022dih}%
  \BibitemOpen
  \bibfield  {author} {\bibinfo {author} {\bibfnamefont {L.}~\bibnamefont
  {Zwick}}, \bibinfo {author} {\bibfnamefont {P.~R.}\ \bibnamefont {Capelo}},\
  and\ \bibinfo {author} {\bibfnamefont {L.}~\bibnamefont {Mayer}},\ }\bibfield
   {title} {\bibinfo {title} {{Priorities in gravitational waveforms for future
  space-borne detectors: vacuum accuracy or environment?}},\ }\href
  {https://doi.org/10.1093/mnras/stad707} {\bibfield  {journal} {\bibinfo
  {journal} {Mon. Not. Roy. Astron. Soc.}\ }\textbf {\bibinfo {volume} {521}},\
  \bibinfo {pages} {4645} (\bibinfo {year} {2023})},\ \Eprint
  {https://arxiv.org/abs/2209.04060} {arXiv:2209.04060 [gr-qc]} \BibitemShut
  {NoStop}%
\bibitem [{\citenamefont {Garg}\ \emph
  {et~al.}(2024{\natexlab{c}})\citenamefont {Garg}, \citenamefont {Sberna},
  \citenamefont {Speri}, \citenamefont {Duque},\ and\ \citenamefont
  {Gair}}]{Garg:2024qxq}%
  \BibitemOpen
  \bibfield  {author} {\bibinfo {author} {\bibfnamefont {M.}~\bibnamefont
  {Garg}}, \bibinfo {author} {\bibfnamefont {L.}~\bibnamefont {Sberna}},
  \bibinfo {author} {\bibfnamefont {L.}~\bibnamefont {Speri}}, \bibinfo
  {author} {\bibfnamefont {F.}~\bibnamefont {Duque}},\ and\ \bibinfo {author}
  {\bibfnamefont {J.}~\bibnamefont {Gair}},\ }\bibfield  {title} {\bibinfo
  {title} {{Systematics in tests of general relativity using LISA massive black
  hole binaries}},\ }\href@noop {} {\bibfield  {journal} {\bibinfo  {journal}
  {Monthly Notices of the Royal Astronomical Society}\ }\textbf {\bibinfo
  {volume} {535}},\ \bibinfo {pages} {3283} (\bibinfo {year}
  {2024}{\natexlab{c}})},\ \Eprint {https://arxiv.org/abs/2410.02910}
  {arXiv:2410.02910 [astro-ph.GA]} \BibitemShut {NoStop}%
\bibitem [{\citenamefont {Pound}(2012)}]{Pound:2012dk}%
  \BibitemOpen
  \bibfield  {author} {\bibinfo {author} {\bibfnamefont {A.}~\bibnamefont
  {Pound}},\ }\bibfield  {title} {\bibinfo {title} {{Nonlinear gravitational
  self-force. I. Field outside a small body}},\ }\href
  {https://doi.org/10.1103/PhysRevD.86.084019} {\bibfield  {journal} {\bibinfo
  {journal} {Phys. Rev. D}\ }\textbf {\bibinfo {volume} {86}},\ \bibinfo
  {pages} {084019} (\bibinfo {year} {2012})},\ \Eprint
  {https://arxiv.org/abs/1206.6538} {arXiv:1206.6538 [gr-qc]} \BibitemShut
  {NoStop}%
\bibitem [{\citenamefont {Chapman-Bird}\ \emph {et~al.}(2025)\citenamefont
  {Chapman-Bird} \emph {et~al.}}]{Chapman-Bird:2025xtd}%
  \BibitemOpen
  \bibfield  {author} {\bibinfo {author} {\bibfnamefont {C.~E.~A.}\
  \bibnamefont {Chapman-Bird}} \emph {et~al.},\ }\bibfield  {title} {\bibinfo
  {title} {{Efficient waveforms for asymmetric-mass eccentric equatorial
  inspirals into rapidly spinning black holes}},\ }\href
  {https://doi.org/10.1103/scbp-75pf} {\bibfield  {journal} {\bibinfo
  {journal} {Phys. Rev. D}\ }\textbf {\bibinfo {volume} {112}},\ \bibinfo
  {pages} {104023} (\bibinfo {year} {2025})},\ \Eprint
  {https://arxiv.org/abs/2506.09470} {arXiv:2506.09470 [gr-qc]} \BibitemShut
  {NoStop}%
\bibitem [{\citenamefont {{Speri}}\ \emph {et~al.}(2023)\citenamefont
  {{Speri}}, \citenamefont {{Katz}}, \citenamefont {{Chua}}, \citenamefont
  {{Hughes}}, \citenamefont {{Warburton}}, \citenamefont {{Thompson}},
  \citenamefont {{Chapman-Bird}},\ and\ \citenamefont
  {{Gair}}}]{Speri:2023jte}%
  \BibitemOpen
  \bibfield  {author} {\bibinfo {author} {\bibfnamefont {L.}~\bibnamefont
  {{Speri}}}, \bibinfo {author} {\bibfnamefont {M.~L.}\ \bibnamefont {{Katz}}},
  \bibinfo {author} {\bibfnamefont {A.~J.~K.}\ \bibnamefont {{Chua}}}, \bibinfo
  {author} {\bibfnamefont {S.~A.}\ \bibnamefont {{Hughes}}}, \bibinfo {author}
  {\bibfnamefont {N.}~\bibnamefont {{Warburton}}}, \bibinfo {author}
  {\bibfnamefont {J.~E.}\ \bibnamefont {{Thompson}}}, \bibinfo {author}
  {\bibfnamefont {C.~E.~A.}\ \bibnamefont {{Chapman-Bird}}},\ and\ \bibinfo
  {author} {\bibfnamefont {J.~R.}\ \bibnamefont {{Gair}}},\ }\bibfield  {title}
  {\bibinfo {title} {{Fast and Fourier: Extreme Mass Ratio Inspiral Waveforms
  in the Frequency Domain}},\ }\href
  {https://doi.org/10.48550/arXiv.2307.12585} {\bibfield  {journal} {\bibinfo
  {journal} {arXiv e-prints}\ ,\ \bibinfo {eid} {arXiv:2307.12585}} (\bibinfo
  {year} {2023})},\ \Eprint {https://arxiv.org/abs/2307.12585}
  {arXiv:2307.12585 [gr-qc]} \BibitemShut {NoStop}%
\bibitem [{\citenamefont {Pound}(2017)}]{Pound:2017psq}%
  \BibitemOpen
  \bibfield  {author} {\bibinfo {author} {\bibfnamefont {A.}~\bibnamefont
  {Pound}},\ }\bibfield  {title} {\bibinfo {title} {{Nonlinear gravitational
  self-force: second-order equation of motion}},\ }\href
  {https://doi.org/10.1103/PhysRevD.95.104056} {\bibfield  {journal} {\bibinfo
  {journal} {Phys. Rev. D}\ }\textbf {\bibinfo {volume} {95}},\ \bibinfo
  {pages} {104056} (\bibinfo {year} {2017})},\ \Eprint
  {https://arxiv.org/abs/1703.02836} {arXiv:1703.02836 [gr-qc]} \BibitemShut
  {NoStop}%
\bibitem [{\citenamefont {van~de Meent}(2018)}]{vandeMeent:2017bcc}%
  \BibitemOpen
  \bibfield  {author} {\bibinfo {author} {\bibfnamefont {M.}~\bibnamefont
  {van~de Meent}},\ }\bibfield  {title} {\bibinfo {title} {{Gravitational
  self-force on generic bound geodesics in Kerr spacetime}},\ }\href
  {https://doi.org/10.1103/PhysRevD.97.104033} {\bibfield  {journal} {\bibinfo
  {journal} {Phys. Rev. D}\ }\textbf {\bibinfo {volume} {97}},\ \bibinfo
  {pages} {104033} (\bibinfo {year} {2018})},\ \Eprint
  {https://arxiv.org/abs/1711.09607} {arXiv:1711.09607 [gr-qc]} \BibitemShut
  {NoStop}%
\bibitem [{\citenamefont {Lynch}\ \emph {et~al.}(2024)\citenamefont {Lynch},
  \citenamefont {Witzany}, \citenamefont {van~de Meent},\ and\ \citenamefont
  {Warburton}}]{Lynch:2024ohd}%
  \BibitemOpen
  \bibfield  {author} {\bibinfo {author} {\bibfnamefont {P.}~\bibnamefont
  {Lynch}}, \bibinfo {author} {\bibfnamefont {V.}~\bibnamefont {Witzany}},
  \bibinfo {author} {\bibfnamefont {M.}~\bibnamefont {van~de Meent}},\ and\
  \bibinfo {author} {\bibfnamefont {N.}~\bibnamefont {Warburton}},\ }\bibfield
  {title} {\bibinfo {title} {{Fast inspirals and the treatment of orbital
  resonances}},\ }\href {https://doi.org/10.1088/1361-6382/ad7dc9} {\bibfield
  {journal} {\bibinfo  {journal} {Class. Quant. Grav.}\ }\textbf {\bibinfo
  {volume} {41}},\ \bibinfo {pages} {225002} (\bibinfo {year} {2024})},\
  \Eprint {https://arxiv.org/abs/2405.21072} {arXiv:2405.21072 [gr-qc]}
  \BibitemShut {NoStop}%
\bibitem [{\citenamefont {Upton}(2024)}]{Upton:2023tcv}%
  \BibitemOpen
  \bibfield  {author} {\bibinfo {author} {\bibfnamefont {S.~D.}\ \bibnamefont
  {Upton}},\ }\bibfield  {title} {\bibinfo {title} {{Second-order gravitational
  self-force in a highly regular gauge: Covariant and coordinate punctures}},\
  }\href {https://doi.org/10.1103/PhysRevD.109.044021} {\bibfield  {journal}
  {\bibinfo  {journal} {Phys. Rev. D}\ }\textbf {\bibinfo {volume} {109}},\
  \bibinfo {pages} {044021} (\bibinfo {year} {2024})},\ \Eprint
  {https://arxiv.org/abs/2309.03778} {arXiv:2309.03778 [gr-qc]} \BibitemShut
  {NoStop}%
\bibitem [{\citenamefont {Lynch}\ \emph {et~al.}(2022)\citenamefont {Lynch},
  \citenamefont {van~de Meent},\ and\ \citenamefont
  {Warburton}}]{Lynch:2021ogr}%
  \BibitemOpen
  \bibfield  {author} {\bibinfo {author} {\bibfnamefont {P.}~\bibnamefont
  {Lynch}}, \bibinfo {author} {\bibfnamefont {M.}~\bibnamefont {van~de
  Meent}},\ and\ \bibinfo {author} {\bibfnamefont {N.}~\bibnamefont
  {Warburton}},\ }\bibfield  {title} {\bibinfo {title} {{Eccentric self-forced
  inspirals into a rotating black hole}},\ }\href
  {https://doi.org/10.1088/1361-6382/ac7507} {\bibfield  {journal} {\bibinfo
  {journal} {Class. Quant. Grav.}\ }\textbf {\bibinfo {volume} {39}},\ \bibinfo
  {pages} {145004} (\bibinfo {year} {2022})},\ \Eprint
  {https://arxiv.org/abs/2112.05651} {arXiv:2112.05651 [gr-qc]} \BibitemShut
  {NoStop}%
\bibitem [{\citenamefont {Leather}(2025)}]{Leather:2024mls}%
  \BibitemOpen
  \bibfield  {author} {\bibinfo {author} {\bibfnamefont {B.}~\bibnamefont
  {Leather}},\ }\bibfield  {title} {\bibinfo {title} {{Gravitational self-force
  with hyperboloidal slicing and spectral methods}},\ }\href
  {https://doi.org/10.1007/s10714-025-03443-9} {\bibfield  {journal} {\bibinfo
  {journal} {Gen. Rel. Grav.}\ }\textbf {\bibinfo {volume} {57}},\ \bibinfo
  {pages} {112} (\bibinfo {year} {2025})},\ \Eprint
  {https://arxiv.org/abs/2411.14976} {arXiv:2411.14976 [gr-qc]} \BibitemShut
  {NoStop}%
\bibitem [{\citenamefont {Bourg}\ \emph
  {et~al.}(2024{\natexlab{a}})\citenamefont {Bourg}, \citenamefont {Leather},
  \citenamefont {Casals}, \citenamefont {Pound},\ and\ \citenamefont
  {Wardell}}]{Bourg:2024vre}%
  \BibitemOpen
  \bibfield  {author} {\bibinfo {author} {\bibfnamefont {P.}~\bibnamefont
  {Bourg}}, \bibinfo {author} {\bibfnamefont {B.}~\bibnamefont {Leather}},
  \bibinfo {author} {\bibfnamefont {M.}~\bibnamefont {Casals}}, \bibinfo
  {author} {\bibfnamefont {A.}~\bibnamefont {Pound}},\ and\ \bibinfo {author}
  {\bibfnamefont {B.}~\bibnamefont {Wardell}},\ }\bibfield  {title} {\bibinfo
  {title} {{Implementation of a Green-Hollands-Zimmerman-Teukolsky puncture
  scheme for gravitational self-force calculations}},\ }\href
  {https://doi.org/10.1103/PhysRevD.110.044007} {\bibfield  {journal} {\bibinfo
   {journal} {Phys. Rev. D}\ }\textbf {\bibinfo {volume} {110}},\ \bibinfo
  {pages} {044007} (\bibinfo {year} {2024}{\natexlab{a}})},\ \Eprint
  {https://arxiv.org/abs/2403.12634} {arXiv:2403.12634 [gr-qc]} \BibitemShut
  {NoStop}%
\bibitem [{\citenamefont {Miller}\ and\ \citenamefont
  {Pound}(2021)}]{Miller:2020bft}%
  \BibitemOpen
  \bibfield  {author} {\bibinfo {author} {\bibfnamefont {J.}~\bibnamefont
  {Miller}}\ and\ \bibinfo {author} {\bibfnamefont {A.}~\bibnamefont {Pound}},\
  }\bibfield  {title} {\bibinfo {title} {{Two-timescale evolution of
  extreme-mass-ratio inspirals: waveform generation scheme for quasicircular
  orbits in Schwarzschild spacetime}},\ }\href
  {https://doi.org/10.1103/PhysRevD.103.064048} {\bibfield  {journal} {\bibinfo
   {journal} {Phys. Rev. D}\ }\textbf {\bibinfo {volume} {103}},\ \bibinfo
  {pages} {064048} (\bibinfo {year} {2021})},\ \Eprint
  {https://arxiv.org/abs/2006.11263} {arXiv:2006.11263 [gr-qc]} \BibitemShut
  {NoStop}%
\bibitem [{\citenamefont {Nasipak}(2024)}]{Nasipak:2023kuf}%
  \BibitemOpen
  \bibfield  {author} {\bibinfo {author} {\bibfnamefont {Z.}~\bibnamefont
  {Nasipak}},\ }\bibfield  {title} {\bibinfo {title} {{Adiabatic gravitational
  waveform model for compact objects undergoing quasicircular inspirals into
  rotating massive black holes}},\ }\href
  {https://doi.org/10.1103/PhysRevD.109.044020} {\bibfield  {journal} {\bibinfo
   {journal} {Phys. Rev. D}\ }\textbf {\bibinfo {volume} {109}},\ \bibinfo
  {pages} {044020} (\bibinfo {year} {2024})},\ \Eprint
  {https://arxiv.org/abs/2310.19706} {arXiv:2310.19706 [gr-qc]} \BibitemShut
  {NoStop}%
\bibitem [{\citenamefont {Bourg}\ \emph
  {et~al.}(2024{\natexlab{b}})\citenamefont {Bourg}, \citenamefont {Pound},
  \citenamefont {Upton},\ and\ \citenamefont {Panosso~Macedo}}]{Bourg:2024cgh}%
  \BibitemOpen
  \bibfield  {author} {\bibinfo {author} {\bibfnamefont {P.}~\bibnamefont
  {Bourg}}, \bibinfo {author} {\bibfnamefont {A.}~\bibnamefont {Pound}},
  \bibinfo {author} {\bibfnamefont {S.~D.}\ \bibnamefont {Upton}},\ and\
  \bibinfo {author} {\bibfnamefont {R.}~\bibnamefont {Panosso~Macedo}},\
  }\bibfield  {title} {\bibinfo {title} {{Simple, efficient method of
  calculating the Detweiler-Whiting singular field to very high order}},\
  }\href {https://doi.org/10.1103/PhysRevD.110.084007} {\bibfield  {journal}
  {\bibinfo  {journal} {Phys. Rev. D}\ }\textbf {\bibinfo {volume} {110}},\
  \bibinfo {pages} {084007} (\bibinfo {year} {2024}{\natexlab{b}})},\ \Eprint
  {https://arxiv.org/abs/2404.10082} {arXiv:2404.10082 [gr-qc]} \BibitemShut
  {NoStop}%
\bibitem [{\citenamefont {{Upton}}\ \emph {et~al.}(2025)\citenamefont
  {{Upton}}, \citenamefont {{Wardell}}, \citenamefont {{Pound}}, \citenamefont
  {{Warburton}},\ and\ \citenamefont {{Barack}}}]{Upton:2025bja}%
  \BibitemOpen
  \bibfield  {author} {\bibinfo {author} {\bibfnamefont {S.~D.}\ \bibnamefont
  {{Upton}}}, \bibinfo {author} {\bibfnamefont {B.}~\bibnamefont {{Wardell}}},
  \bibinfo {author} {\bibfnamefont {A.}~\bibnamefont {{Pound}}}, \bibinfo
  {author} {\bibfnamefont {N.}~\bibnamefont {{Warburton}}},\ and\ \bibinfo
  {author} {\bibfnamefont {L.}~\bibnamefont {{Barack}}},\ }\bibfield  {title}
  {\bibinfo {title} {{Effective source for second-order self-force
  calculations: quasicircular orbits in Schwarzschild spacetime}},\ }\href
  {https://doi.org/10.48550/arXiv.2508.00087} {\bibfield  {journal} {\bibinfo
  {journal} {arXiv e-prints}\ ,\ \bibinfo {eid} {arXiv:2508.00087}} (\bibinfo
  {year} {2025})},\ \Eprint {https://arxiv.org/abs/2508.00087}
  {arXiv:2508.00087 [gr-qc]} \BibitemShut {NoStop}%
\bibitem [{\citenamefont {Vicente}\ \emph {et~al.}(2025)\citenamefont
  {Vicente}, \citenamefont {Karydas},\ and\ \citenamefont
  {Bertone}}]{Vicente:2025gsg}%
  \BibitemOpen
  \bibfield  {author} {\bibinfo {author} {\bibfnamefont {R.}~\bibnamefont
  {Vicente}}, \bibinfo {author} {\bibfnamefont {T.~K.}\ \bibnamefont
  {Karydas}},\ and\ \bibinfo {author} {\bibfnamefont {G.}~\bibnamefont
  {Bertone}},\ }\bibfield  {title} {\bibinfo {title} {{Fully Relativistic
  Treatment of Extreme Mass-Ratio Inspirals in Collisionless Environments}},\
  }\href {https://doi.org/10.1103/s4wh-x6c4} {\bibfield  {journal} {\bibinfo
  {journal} {Phys. Rev. Lett.}\ }\textbf {\bibinfo {volume} {135}},\ \bibinfo
  {pages} {211401} (\bibinfo {year} {2025})},\ \Eprint
  {https://arxiv.org/abs/2505.09715} {arXiv:2505.09715 [gr-qc]} \BibitemShut
  {NoStop}%
\bibitem [{\citenamefont {Brito}\ \emph {et~al.}(2015)\citenamefont {Brito},
  \citenamefont {Cardoso},\ and\ \citenamefont {Pani}}]{Brito:2014wla}%
  \BibitemOpen
  \bibfield  {author} {\bibinfo {author} {\bibfnamefont {R.}~\bibnamefont
  {Brito}}, \bibinfo {author} {\bibfnamefont {V.}~\bibnamefont {Cardoso}},\
  and\ \bibinfo {author} {\bibfnamefont {P.}~\bibnamefont {Pani}},\ }\bibfield
  {title} {\bibinfo {title} {{Black holes as particle detectors: evolution of
  superradiant instabilities}},\ }\href
  {https://doi.org/10.1088/0264-9381/32/13/134001} {\bibfield  {journal}
  {\bibinfo  {journal} {Class. Quant. Grav.}\ }\textbf {\bibinfo {volume}
  {32}},\ \bibinfo {pages} {134001} (\bibinfo {year} {2015})},\ \Eprint
  {https://arxiv.org/abs/1411.0686} {arXiv:1411.0686 [gr-qc]} \BibitemShut
  {NoStop}%
\bibitem [{\citenamefont {{Rahman}}\ and\ \citenamefont
  {{Takahashi}}(2025)}]{Rahman:2025mip}%
  \BibitemOpen
  \bibfield  {author} {\bibinfo {author} {\bibfnamefont {M.}~\bibnamefont
  {{Rahman}}}\ and\ \bibinfo {author} {\bibfnamefont {T.}~\bibnamefont
  {{Takahashi}}},\ }\bibfield  {title} {\bibinfo {title} {{Post-adiabatic
  waveforms from extreme mass ratio inspirals in the presence of dark
  matter}},\ }\href {https://doi.org/10.48550/arXiv.2507.06923} {\bibfield
  {journal} {\bibinfo  {journal} {arXiv e-prints}\ ,\ \bibinfo {eid}
  {arXiv:2507.06923}} (\bibinfo {year} {2025})},\ \Eprint
  {https://arxiv.org/abs/2507.06923} {arXiv:2507.06923 [gr-qc]} \BibitemShut
  {NoStop}%
\bibitem [{\citenamefont {Li}\ \emph {et~al.}(2025{\natexlab{b}})\citenamefont
  {Li}, \citenamefont {Weller}, \citenamefont {Bourg}, \citenamefont {LaHaye},
  \citenamefont {Yunes},\ and\ \citenamefont {Yang}}]{Li:2025ffh}%
  \BibitemOpen
  \bibfield  {author} {\bibinfo {author} {\bibfnamefont {D.}~\bibnamefont
  {Li}}, \bibinfo {author} {\bibfnamefont {C.}~\bibnamefont {Weller}}, \bibinfo
  {author} {\bibfnamefont {P.}~\bibnamefont {Bourg}}, \bibinfo {author}
  {\bibfnamefont {M.}~\bibnamefont {LaHaye}}, \bibinfo {author} {\bibfnamefont
  {N.}~\bibnamefont {Yunes}},\ and\ \bibinfo {author} {\bibfnamefont
  {H.}~\bibnamefont {Yang}},\ }\bibfield  {title} {\bibinfo {title} {{Extreme
  mass-ratio inspiral within an ultralight scalar cloud: Scalar radiation}},\
  }\href {https://doi.org/10.1103/7l9s-g21j} {\bibfield  {journal} {\bibinfo
  {journal} {Phys. Rev. D}\ }\textbf {\bibinfo {volume} {112}},\ \bibinfo
  {pages} {084057} (\bibinfo {year} {2025}{\natexlab{b}})},\ \Eprint
  {https://arxiv.org/abs/2507.02045} {arXiv:2507.02045 [gr-qc]} \BibitemShut
  {NoStop}%
\bibitem [{\citenamefont {{Dyson}}\ \emph {et~al.}(2025)\citenamefont
  {{Dyson}}, \citenamefont {{Spieksma}}, \citenamefont {{Brito}}, \citenamefont
  {{van de Meent}},\ and\ \citenamefont {{Dolan}}}]{Dyson:2025dlj}%
  \BibitemOpen
  \bibfield  {author} {\bibinfo {author} {\bibfnamefont {C.}~\bibnamefont
  {{Dyson}}}, \bibinfo {author} {\bibfnamefont {T.~F.~M.}\ \bibnamefont
  {{Spieksma}}}, \bibinfo {author} {\bibfnamefont {R.}~\bibnamefont {{Brito}}},
  \bibinfo {author} {\bibfnamefont {M.}~\bibnamefont {{van de Meent}}},\ and\
  \bibinfo {author} {\bibfnamefont {S.}~\bibnamefont {{Dolan}}},\ }\bibfield
  {title} {\bibinfo {title} {{Environmental Effects in Extreme-Mass-Ratio
  Inspirals: Perturbations to the Environment in Kerr Spacetimes}},\ }\href
  {https://doi.org/10.1103/PhysRevLett.134.211403} {\bibfield  {journal}
  {\bibinfo  {journal} {\prl}\ }\textbf {\bibinfo {volume} {134}},\ \bibinfo
  {eid} {211403} (\bibinfo {year} {2025})},\ \Eprint
  {https://arxiv.org/abs/2501.09806} {arXiv:2501.09806 [gr-qc]} \BibitemShut
  {NoStop}%
\bibitem [{\citenamefont {Polcar}\ and\ \citenamefont
  {Witzany}(2025)}]{Polcar:2025yto}%
  \BibitemOpen
  \bibfield  {author} {\bibinfo {author} {\bibfnamefont {L.}~\bibnamefont
  {Polcar}}\ and\ \bibinfo {author} {\bibfnamefont {V.}~\bibnamefont
  {Witzany}},\ }\bibfield  {title} {\bibinfo {title} {{Toward relativistic
  inspirals into black holes surrounded by matter}},\ }\href
  {https://doi.org/10.1103/79w6-s8sk} {\bibfield  {journal} {\bibinfo
  {journal} {Phys. Rev. D}\ }\textbf {\bibinfo {volume} {112}},\ \bibinfo
  {pages} {104003} (\bibinfo {year} {2025})},\ \Eprint
  {https://arxiv.org/abs/2507.15720} {arXiv:2507.15720 [gr-qc]} \BibitemShut
  {NoStop}%
\bibitem [{\citenamefont {{Duque}}\ \emph
  {et~al.}(2025{\natexlab{b}})\citenamefont {{Duque}}, \citenamefont
  {{Sberna}}, \citenamefont {{Spiers}},\ and\ \citenamefont
  {{Vicente}}}]{Duque:2025yfm}%
  \BibitemOpen
  \bibfield  {author} {\bibinfo {author} {\bibfnamefont {F.}~\bibnamefont
  {{Duque}}}, \bibinfo {author} {\bibfnamefont {L.}~\bibnamefont {{Sberna}}},
  \bibinfo {author} {\bibfnamefont {A.}~\bibnamefont {{Spiers}}},\ and\
  \bibinfo {author} {\bibfnamefont {R.}~\bibnamefont {{Vicente}}},\ }\bibfield
  {title} {\bibinfo {title} {{Extreme-mass-ratio inspirals in relativistic
  accretion discs}},\ }\href {https://doi.org/10.48550/arXiv.2510.02433}
  {\bibfield  {journal} {\bibinfo  {journal} {arXiv e-prints}\ ,\ \bibinfo
  {eid} {arXiv:2510.02433}} (\bibinfo {year} {2025}{\natexlab{b}})},\ \Eprint
  {https://arxiv.org/abs/2510.02433} {arXiv:2510.02433 [gr-qc]} \BibitemShut
  {NoStop}%
\bibitem [{\citenamefont {Duque}\ \emph {et~al.}(2024)\citenamefont {Duque},
  \citenamefont {Macedo}, \citenamefont {Vicente},\ and\ \citenamefont
  {Cardoso}}]{Duque:2023seg}%
  \BibitemOpen
  \bibfield  {author} {\bibinfo {author} {\bibfnamefont {F.}~\bibnamefont
  {Duque}}, \bibinfo {author} {\bibfnamefont {C.~F.~B.}\ \bibnamefont
  {Macedo}}, \bibinfo {author} {\bibfnamefont {R.}~\bibnamefont {Vicente}},\
  and\ \bibinfo {author} {\bibfnamefont {V.}~\bibnamefont {Cardoso}},\
  }\bibfield  {title} {\bibinfo {title} {{Extreme-Mass-Ratio Inspirals in
  Ultralight Dark Matter}},\ }\href
  {https://doi.org/10.1103/PhysRevLett.133.121404} {\bibfield  {journal}
  {\bibinfo  {journal} {Phys. Rev. Lett.}\ }\textbf {\bibinfo {volume} {133}},\
  \bibinfo {pages} {121404} (\bibinfo {year} {2024})},\ \Eprint
  {https://arxiv.org/abs/2312.06767} {arXiv:2312.06767 [gr-qc]} \BibitemShut
  {NoStop}%
\bibitem [{\citenamefont {Hegade K.~R.}\ \emph
  {et~al.}(2025{\natexlab{a}})\citenamefont {Hegade K.~R.}, \citenamefont
  {Gammie},\ and\ \citenamefont {Yunes}}]{HegadeKR:2025dur}%
  \BibitemOpen
  \bibfield  {author} {\bibinfo {author} {\bibfnamefont {A.}~\bibnamefont
  {Hegade K.~R.}}, \bibinfo {author} {\bibfnamefont {C.~F.}\ \bibnamefont
  {Gammie}},\ and\ \bibinfo {author} {\bibfnamefont {N.}~\bibnamefont
  {Yunes}},\ }\bibfield  {title} {\bibinfo {title} {{Relativistic treatment of
  accretion disk torques on extreme mass-ratio inspirals around nonspinning
  black holes}},\ }\href {https://doi.org/10.1103/9src-p7sp} {\bibfield
  {journal} {\bibinfo  {journal} {Phys. Rev. D}\ }\textbf {\bibinfo {volume}
  {112}},\ \bibinfo {pages} {124012} (\bibinfo {year} {2025}{\natexlab{a}})},\
  \Eprint {https://arxiv.org/abs/2509.20457} {arXiv:2509.20457 [gr-qc]}
  \BibitemShut {NoStop}%
\bibitem [{\citenamefont {Hegade K.~R.}\ \emph
  {et~al.}(2025{\natexlab{b}})\citenamefont {Hegade K.~R.}, \citenamefont
  {Gammie},\ and\ \citenamefont {Yunes}}]{HegadeKR:2025rpr}%
  \BibitemOpen
  \bibfield  {author} {\bibinfo {author} {\bibfnamefont {A.}~\bibnamefont
  {Hegade K.~R.}}, \bibinfo {author} {\bibfnamefont {C.~F.}\ \bibnamefont
  {Gammie}},\ and\ \bibinfo {author} {\bibfnamefont {N.}~\bibnamefont
  {Yunes}},\ }\bibfield  {title} {\bibinfo {title} {{Relativistic treatment of
  accretion disk torques on extreme mass ratio inspirals around spinning black
  holes}},\ }\href {https://doi.org/10.1103/g83s-jdld} {\bibfield  {journal}
  {\bibinfo  {journal} {Phys. Rev. D}\ }\textbf {\bibinfo {volume} {112}},\
  \bibinfo {pages} {124068} (\bibinfo {year} {2025}{\natexlab{b}})},\ \Eprint
  {https://arxiv.org/abs/2510.03564} {arXiv:2510.03564 [gr-qc]} \BibitemShut
  {NoStop}%
\bibitem [{\citenamefont {Chandrasekhar}(1943)}]{chandrasekhar1943dynamical}%
  \BibitemOpen
  \bibfield  {author} {\bibinfo {author} {\bibfnamefont {S.}~\bibnamefont
  {Chandrasekhar}},\ }\bibfield  {title} {\bibinfo {title} {Dynamical friction.
  ii. the rate of escape of stars from clusters and the evidence for the
  operation of dynamical friction},\ }\href@noop {} {\bibfield  {journal}
  {\bibinfo  {journal} {Astrophysical journal}\ }\textbf {\bibinfo {volume}
  {97}},\ \bibinfo {pages} {263} (\bibinfo {year} {1943})}\BibitemShut
  {NoStop}%
\bibitem [{\citenamefont {Ostriker}(1999)}]{Ostriker_1999}%
  \BibitemOpen
  \bibfield  {author} {\bibinfo {author} {\bibfnamefont {E.~C.}\ \bibnamefont
  {Ostriker}},\ }\bibfield  {title} {\bibinfo {title} {Dynamical friction in a
  gaseous medium},\ }\href {https://doi.org/10.1086/306858} {\bibfield
  {journal} {\bibinfo  {journal} {The Astrophysical Journal}\ }\textbf
  {\bibinfo {volume} {513}},\ \bibinfo {pages} {252} (\bibinfo {year}
  {1999})}\BibitemShut {NoStop}%
\bibitem [{\citenamefont {O’Neill}\ \emph {et~al.}(2025)\citenamefont
  {O’Neill}, \citenamefont {Tiede}, \citenamefont {D’Orazio}, \citenamefont
  {Haiman},\ and\ \citenamefont {MacFadyen}}]{O’Neill_2025}%
  \BibitemOpen
  \bibfield  {author} {\bibinfo {author} {\bibfnamefont {D.}~\bibnamefont
  {O’Neill}}, \bibinfo {author} {\bibfnamefont {C.}~\bibnamefont {Tiede}},
  \bibinfo {author} {\bibfnamefont {D.~J.}\ \bibnamefont {D’Orazio}},
  \bibinfo {author} {\bibfnamefont {Z.}~\bibnamefont {Haiman}},\ and\ \bibinfo
  {author} {\bibfnamefont {A.}~\bibnamefont {MacFadyen}},\ }\bibfield  {title}
  {\bibinfo {title} {Gravitational wave decoupling in retrograde circumbinary
  disks},\ }\href {https://doi.org/10.3847/1538-4357/ae0ca8} {\bibfield
  {journal} {\bibinfo  {journal} {The Astrophysical Journal}\ }\textbf
  {\bibinfo {volume} {993}},\ \bibinfo {pages} {206} (\bibinfo {year}
  {2025})}\BibitemShut {NoStop}%
\bibitem [{\citenamefont {{Kim}}(2010)}]{Kim_NLGDF:2010}%
  \BibitemOpen
  \bibfield  {author} {\bibinfo {author} {\bibfnamefont {W.-T.}\ \bibnamefont
  {{Kim}}},\ }\bibfield  {title} {\bibinfo {title} {{Nonlinear Dynamical
  Friction of a Circular-orbit Perturber in a Gaseous Medium}},\ }\href
  {https://doi.org/10.1088/0004-637X/725/1/1069} {\bibfield  {journal}
  {\bibinfo  {journal} {\apj}\ }\textbf {\bibinfo {volume} {725}},\ \bibinfo
  {pages} {1069} (\bibinfo {year} {2010})},\ \Eprint
  {https://arxiv.org/abs/1010.1995} {arXiv:1010.1995 [astro-ph.GA]}
  \BibitemShut {NoStop}%
\bibitem [{\citenamefont {Barausse}(2007)}]{Barausse:2007ph}%
  \BibitemOpen
  \bibfield  {author} {\bibinfo {author} {\bibfnamefont {E.}~\bibnamefont
  {Barausse}},\ }\bibfield  {title} {\bibinfo {title} {{Relativistic dynamical
  friction in a collisional fluid}},\ }\href
  {https://doi.org/10.1111/j.1365-2966.2007.12408.x} {\bibfield  {journal}
  {\bibinfo  {journal} {Mon. Not. Roy. Astron. Soc.}\ }\textbf {\bibinfo
  {volume} {382}},\ \bibinfo {pages} {826} (\bibinfo {year} {2007})},\ \Eprint
  {https://arxiv.org/abs/0709.0211} {arXiv:0709.0211 [astro-ph]} \BibitemShut
  {NoStop}%
\bibitem [{\citenamefont {Vicente}\ and\ \citenamefont
  {Cardoso}(2022)}]{Vicente:2022ivh}%
  \BibitemOpen
  \bibfield  {author} {\bibinfo {author} {\bibfnamefont {R.}~\bibnamefont
  {Vicente}}\ and\ \bibinfo {author} {\bibfnamefont {V.}~\bibnamefont
  {Cardoso}},\ }\bibfield  {title} {\bibinfo {title} {{Dynamical friction of
  black holes in ultralight dark matter}},\ }\href
  {https://doi.org/10.1103/PhysRevD.105.083008} {\bibfield  {journal} {\bibinfo
   {journal} {Phys. Rev. D}\ }\textbf {\bibinfo {volume} {105}},\ \bibinfo
  {pages} {083008} (\bibinfo {year} {2022})},\ \Eprint
  {https://arxiv.org/abs/2201.08854} {arXiv:2201.08854 [gr-qc]} \BibitemShut
  {NoStop}%
\bibitem [{\citenamefont {Traykova}\ \emph {et~al.}(2021)\citenamefont
  {Traykova}, \citenamefont {Clough}, \citenamefont {Helfer}, \citenamefont
  {Berti}, \citenamefont {Ferreira},\ and\ \citenamefont
  {Hui}}]{Traykova:2021dua}%
  \BibitemOpen
  \bibfield  {author} {\bibinfo {author} {\bibfnamefont {D.}~\bibnamefont
  {Traykova}}, \bibinfo {author} {\bibfnamefont {K.}~\bibnamefont {Clough}},
  \bibinfo {author} {\bibfnamefont {T.}~\bibnamefont {Helfer}}, \bibinfo
  {author} {\bibfnamefont {E.}~\bibnamefont {Berti}}, \bibinfo {author}
  {\bibfnamefont {P.~G.}\ \bibnamefont {Ferreira}},\ and\ \bibinfo {author}
  {\bibfnamefont {L.}~\bibnamefont {Hui}},\ }\bibfield  {title} {\bibinfo
  {title} {{Dynamical friction from scalar dark matter in the relativistic
  regime}},\ }\href {https://doi.org/10.1103/PhysRevD.104.103014} {\bibfield
  {journal} {\bibinfo  {journal} {Phys. Rev. D}\ }\textbf {\bibinfo {volume}
  {104}},\ \bibinfo {pages} {103014} (\bibinfo {year} {2021})},\ \Eprint
  {https://arxiv.org/abs/2106.08280} {arXiv:2106.08280 [gr-qc]} \BibitemShut
  {NoStop}%
\bibitem [{\citenamefont {Traykova}\ \emph {et~al.}(2023)\citenamefont
  {Traykova}, \citenamefont {Vicente}, \citenamefont {Clough}, \citenamefont
  {Helfer}, \citenamefont {Berti}, \citenamefont {Ferreira},\ and\
  \citenamefont {Hui}}]{Traykova:2023qyv}%
  \BibitemOpen
  \bibfield  {author} {\bibinfo {author} {\bibfnamefont {D.}~\bibnamefont
  {Traykova}}, \bibinfo {author} {\bibfnamefont {R.}~\bibnamefont {Vicente}},
  \bibinfo {author} {\bibfnamefont {K.}~\bibnamefont {Clough}}, \bibinfo
  {author} {\bibfnamefont {T.}~\bibnamefont {Helfer}}, \bibinfo {author}
  {\bibfnamefont {E.}~\bibnamefont {Berti}}, \bibinfo {author} {\bibfnamefont
  {P.~G.}\ \bibnamefont {Ferreira}},\ and\ \bibinfo {author} {\bibfnamefont
  {L.}~\bibnamefont {Hui}},\ }\bibfield  {title} {\bibinfo {title}
  {{Relativistic drag forces on black holes from scalar dark matter clouds of
  all sizes}},\ }\href {https://doi.org/10.1103/PhysRevD.108.L121502}
  {\bibfield  {journal} {\bibinfo  {journal} {Phys. Rev. D}\ }\textbf {\bibinfo
  {volume} {108}},\ \bibinfo {pages} {L121502} (\bibinfo {year} {2023})},\
  \Eprint {https://arxiv.org/abs/2305.10492} {arXiv:2305.10492 [gr-qc]}
  \BibitemShut {NoStop}%
\bibitem [{\citenamefont {Dyson}\ and\ \citenamefont {van~de
  Meent}(2023)}]{Dyson:2023fws}%
  \BibitemOpen
  \bibfield  {author} {\bibinfo {author} {\bibfnamefont {C.}~\bibnamefont
  {Dyson}}\ and\ \bibinfo {author} {\bibfnamefont {M.}~\bibnamefont {van~de
  Meent}},\ }\bibfield  {title} {\bibinfo {title} {{Kerr-fully diving into the
  abyss: analytic solutions to plunging geodesics in Kerr}},\ }\href
  {https://doi.org/10.1088/1361-6382/acf552} {\bibfield  {journal} {\bibinfo
  {journal} {Class. Quant. Grav.}\ }\textbf {\bibinfo {volume} {40}},\ \bibinfo
  {pages} {195026} (\bibinfo {year} {2023})},\ \Eprint
  {https://arxiv.org/abs/2302.03704} {arXiv:2302.03704 [gr-qc]} \BibitemShut
  {NoStop}%
\bibitem [{\citenamefont {Wang}\ \emph {et~al.}(2024)\citenamefont {Wang},
  \citenamefont {Helfer}, \citenamefont {Traykova}, \citenamefont {Clough},\
  and\ \citenamefont {Berti}}]{Wang23}%
  \BibitemOpen
  \bibfield  {author} {\bibinfo {author} {\bibfnamefont {Z.}~\bibnamefont
  {Wang}}, \bibinfo {author} {\bibfnamefont {T.}~\bibnamefont {Helfer}},
  \bibinfo {author} {\bibfnamefont {D.}~\bibnamefont {Traykova}}, \bibinfo
  {author} {\bibfnamefont {K.}~\bibnamefont {Clough}},\ and\ \bibinfo {author}
  {\bibfnamefont {E.}~\bibnamefont {Berti}},\ }\bibfield  {title} {\bibinfo
  {title} {Gravitational magnus effect from scalar dark matter},\ }\href
  {https://doi.org/10.1103/PhysRevD.110.024009} {\bibfield  {journal} {\bibinfo
   {journal} {Phys. Rev. D}\ }\textbf {\bibinfo {volume} {110}},\ \bibinfo
  {pages} {024009} (\bibinfo {year} {2024})}\BibitemShut {NoStop}%
\bibitem [{\citenamefont {Cardoso}\ \emph
  {et~al.}(2022{\natexlab{a}})\citenamefont {Cardoso}, \citenamefont
  {Destounis}, \citenamefont {Duque}, \citenamefont {Panosso~Macedo},\ and\
  \citenamefont {Maselli}}]{Cardoso:2022whc}%
  \BibitemOpen
  \bibfield  {author} {\bibinfo {author} {\bibfnamefont {V.}~\bibnamefont
  {Cardoso}}, \bibinfo {author} {\bibfnamefont {K.}~\bibnamefont {Destounis}},
  \bibinfo {author} {\bibfnamefont {F.}~\bibnamefont {Duque}}, \bibinfo
  {author} {\bibfnamefont {R.}~\bibnamefont {Panosso~Macedo}},\ and\ \bibinfo
  {author} {\bibfnamefont {A.}~\bibnamefont {Maselli}},\ }\bibfield  {title}
  {\bibinfo {title} {{Gravitational Waves from Extreme-Mass-Ratio Systems in
  Astrophysical Environments}},\ }\href
  {https://doi.org/10.1103/PhysRevLett.129.241103} {\bibfield  {journal}
  {\bibinfo  {journal} {Phys. Rev. Lett.}\ }\textbf {\bibinfo {volume} {129}},\
  \bibinfo {pages} {241103} (\bibinfo {year} {2022}{\natexlab{a}})},\ \Eprint
  {https://arxiv.org/abs/2210.01133} {arXiv:2210.01133 [gr-qc]} \BibitemShut
  {NoStop}%
\bibitem [{\citenamefont {{Lin}}\ and\ \citenamefont
  {{Papaloizou}}(1979)}]{LinPapa:1979}%
  \BibitemOpen
  \bibfield  {author} {\bibinfo {author} {\bibfnamefont {D.~N.~C.}\
  \bibnamefont {{Lin}}}\ and\ \bibinfo {author} {\bibfnamefont
  {J.}~\bibnamefont {{Papaloizou}}},\ }\bibfield  {title} {\bibinfo {title}
  {{Tidal torques on accretion discs in binary systems with extreme mass
  ratios.}},\ }\href {https://doi.org/10.1093/mnras/186.4.799} {\bibfield
  {journal} {\bibinfo  {journal} {\mnras}\ }\textbf {\bibinfo {volume} {186}},\
  \bibinfo {pages} {799} (\bibinfo {year} {1979})}\BibitemShut {NoStop}%
\bibitem [{\citenamefont {{Goldreich}}\ and\ \citenamefont
  {{Tremaine}}(1979)}]{Goldreich79}%
  \BibitemOpen
  \bibfield  {author} {\bibinfo {author} {\bibfnamefont {P.}~\bibnamefont
  {{Goldreich}}}\ and\ \bibinfo {author} {\bibfnamefont {S.}~\bibnamefont
  {{Tremaine}}},\ }\bibfield  {title} {\bibinfo {title} {{The excitation of
  density waves at the Lindblad and corotation resonances by an external
  potential.}},\ }\href {https://doi.org/10.1086/157448} {\bibfield  {journal}
  {\bibinfo  {journal} {\apj}\ }\textbf {\bibinfo {volume} {233}},\ \bibinfo
  {pages} {857} (\bibinfo {year} {1979})}\BibitemShut {NoStop}%
\bibitem [{\citenamefont {{Goldreich}}\ and\ \citenamefont
  {{Tremaine}}(1980)}]{Goldreich80}%
  \BibitemOpen
  \bibfield  {author} {\bibinfo {author} {\bibfnamefont {P.}~\bibnamefont
  {{Goldreich}}}\ and\ \bibinfo {author} {\bibfnamefont {S.}~\bibnamefont
  {{Tremaine}}},\ }\bibfield  {title} {\bibinfo {title} {{Disk-satellite
  interactions.}},\ }\href {https://doi.org/10.1086/158356} {\bibfield
  {journal} {\bibinfo  {journal} {\apj}\ }\textbf {\bibinfo {volume} {241}},\
  \bibinfo {pages} {425} (\bibinfo {year} {1980})}\BibitemShut {NoStop}%
\bibitem [{\citenamefont {{Ward}}(1986)}]{Ward:1986}%
  \BibitemOpen
  \bibfield  {author} {\bibinfo {author} {\bibfnamefont {W.~R.}\ \bibnamefont
  {{Ward}}},\ }\bibfield  {title} {\bibinfo {title} {{Density waves in the
  solar nebula: Diffential Lindblad torque}},\ }\href
  {https://doi.org/10.1016/0019-1035(86)90182-X} {\bibfield  {journal}
  {\bibinfo  {journal} {\icarus}\ }\textbf {\bibinfo {volume} {67}},\ \bibinfo
  {pages} {164} (\bibinfo {year} {1986})}\BibitemShut {NoStop}%
\bibitem [{\citenamefont {{Meyer-Vernet}}\ and\ \citenamefont
  {{Sicardy}}(1987)}]{MeyerSicardy:1987}%
  \BibitemOpen
  \bibfield  {author} {\bibinfo {author} {\bibfnamefont {N.}~\bibnamefont
  {{Meyer-Vernet}}}\ and\ \bibinfo {author} {\bibfnamefont {B.}~\bibnamefont
  {{Sicardy}}},\ }\bibfield  {title} {\bibinfo {title} {{On the physics of
  resonant disk-satellite interaction}},\ }\href
  {https://doi.org/10.1016/0019-1035(87)90011-X} {\bibfield  {journal}
  {\bibinfo  {journal} {\icarus}\ }\textbf {\bibinfo {volume} {69}},\ \bibinfo
  {pages} {157} (\bibinfo {year} {1987})}\BibitemShut {NoStop}%
\bibitem [{\citenamefont {{Lubow}}(1991)}]{Lubow:1991}%
  \BibitemOpen
  \bibfield  {author} {\bibinfo {author} {\bibfnamefont {S.~H.}\ \bibnamefont
  {{Lubow}}},\ }\bibfield  {title} {\bibinfo {title} {{A Model for Tidally
  Driven Eccentric Instabilities in Fluid Disks}},\ }\href
  {https://doi.org/10.1086/170647} {\bibfield  {journal} {\bibinfo  {journal}
  {\apj}\ }\textbf {\bibinfo {volume} {381}},\ \bibinfo {pages} {259} (\bibinfo
  {year} {1991})}\BibitemShut {NoStop}%
\bibitem [{\citenamefont {{Artymowicz}}\ and\ \citenamefont
  {{Lubow}}(1994)}]{AL:1994}%
  \BibitemOpen
  \bibfield  {author} {\bibinfo {author} {\bibfnamefont {P.}~\bibnamefont
  {{Artymowicz}}}\ and\ \bibinfo {author} {\bibfnamefont {S.~H.}\ \bibnamefont
  {{Lubow}}},\ }\bibfield  {title} {\bibinfo {title} {{Dynamics of Binary-Disk
  Interaction. I. Resonances and Disk Gap Sizes}},\ }\href
  {https://doi.org/10.1086/173679} {\bibfield  {journal} {\bibinfo  {journal}
  {\apj}\ }\textbf {\bibinfo {volume} {421}},\ \bibinfo {pages} {651} (\bibinfo
  {year} {1994})}\BibitemShut {NoStop}%
\bibitem [{\citenamefont {{Ward}}(1997)}]{Ward:1997}%
  \BibitemOpen
  \bibfield  {author} {\bibinfo {author} {\bibfnamefont {W.~R.}\ \bibnamefont
  {{Ward}}},\ }\bibfield  {title} {\bibinfo {title} {{Protoplanet Migration by
  Nebula Tides}},\ }\href {https://doi.org/10.1006/icar.1996.5647} {\bibfield
  {journal} {\bibinfo  {journal} {\icarus}\ }\textbf {\bibinfo {volume}
  {126}},\ \bibinfo {pages} {261} (\bibinfo {year} {1997})}\BibitemShut
  {NoStop}%
\bibitem [{\citenamefont {{Tanaka}}\ \emph
  {et~al.}(2002{\natexlab{b}})\citenamefont {{Tanaka}}, \citenamefont
  {{Takeuchi}},\ and\ \citenamefont {{Ward}}}]{Tanaka2002}%
  \BibitemOpen
  \bibfield  {author} {\bibinfo {author} {\bibfnamefont {H.}~\bibnamefont
  {{Tanaka}}}, \bibinfo {author} {\bibfnamefont {T.}~\bibnamefont
  {{Takeuchi}}},\ and\ \bibinfo {author} {\bibfnamefont {W.~R.}\ \bibnamefont
  {{Ward}}},\ }\bibfield  {title} {\bibinfo {title} {{Three-Dimensional
  Interaction between a Planet and an Isothermal Gaseous Disk. I. Corotation
  and Lindblad Torques and Planet Migration}},\ }\href
  {https://doi.org/10.1086/324713} {\bibfield  {journal} {\bibinfo  {journal}
  {\apj}\ }\textbf {\bibinfo {volume} {565}},\ \bibinfo {pages} {1257}
  (\bibinfo {year} {2002}{\natexlab{b}})}\BibitemShut {NoStop}%
\bibitem [{\citenamefont {{Tanaka}}\ and\ \citenamefont
  {{Ward}}(2004)}]{Tanaka2004}%
  \BibitemOpen
  \bibfield  {author} {\bibinfo {author} {\bibfnamefont {H.}~\bibnamefont
  {{Tanaka}}}\ and\ \bibinfo {author} {\bibfnamefont {W.~R.}\ \bibnamefont
  {{Ward}}},\ }\bibfield  {title} {\bibinfo {title} {{Three-dimensional
  Interaction between a Planet and an Isothermal Gaseous Disk. II. Eccentricity
  Waves and Bending Waves}},\ }\href {https://doi.org/10.1086/380992}
  {\bibfield  {journal} {\bibinfo  {journal} {\apj}\ }\textbf {\bibinfo
  {volume} {602}},\ \bibinfo {pages} {388} (\bibinfo {year}
  {2004})}\BibitemShut {NoStop}%
\bibitem [{\citenamefont {{Tsang}}(2014{\natexlab{a}})}]{Tsang14}%
  \BibitemOpen
  \bibfield  {author} {\bibinfo {author} {\bibfnamefont {D.}~\bibnamefont
  {{Tsang}}},\ }\bibfield  {title} {\bibinfo {title} {{Linear Corotation
  Torques in Non-barotropic Disks}},\ }\href
  {https://doi.org/10.1088/0004-637X/782/2/112} {\bibfield  {journal} {\bibinfo
   {journal} {\apj}\ }\textbf {\bibinfo {volume} {782}},\ \bibinfo {eid} {112}
  (\bibinfo {year} {2014}{\natexlab{a}})},\ \Eprint
  {https://arxiv.org/abs/1310.8626} {arXiv:1310.8626 [astro-ph.EP]}
  \BibitemShut {NoStop}%
\bibitem [{\citenamefont {{Miranda}}\ and\ \citenamefont
  {{Rafikov}}(2020{\natexlab{a}})}]{Miranda20}%
  \BibitemOpen
  \bibfield  {author} {\bibinfo {author} {\bibfnamefont {R.}~\bibnamefont
  {{Miranda}}}\ and\ \bibinfo {author} {\bibfnamefont {R.~R.}\ \bibnamefont
  {{Rafikov}}},\ }\bibfield  {title} {\bibinfo {title} {{Planet-Disk
  Interaction in Disks with Cooling: Basic Theory}},\ }\href
  {https://doi.org/10.3847/1538-4357/ab791a} {\bibfield  {journal} {\bibinfo
  {journal} {\apj}\ }\textbf {\bibinfo {volume} {892}},\ \bibinfo {eid} {65}
  (\bibinfo {year} {2020}{\natexlab{a}})},\ \Eprint
  {https://arxiv.org/abs/1911.01428} {arXiv:1911.01428 [astro-ph.EP]}
  \BibitemShut {NoStop}%
\bibitem [{\citenamefont {{Fairbairn}}\ and\ \citenamefont
  {{Rafikov}}(2022{\natexlab{a}})}]{Fairbairn22}%
  \BibitemOpen
  \bibfield  {author} {\bibinfo {author} {\bibfnamefont {C.~W.}\ \bibnamefont
  {{Fairbairn}}}\ and\ \bibinfo {author} {\bibfnamefont {R.~R.}\ \bibnamefont
  {{Rafikov}}},\ }\bibfield  {title} {\bibinfo {title} {{Density waves in
  protoplanetary discs excited by eccentric planets: linear theory}},\ }\href
  {https://doi.org/10.1093/mnras/stac2802} {\bibfield  {journal} {\bibinfo
  {journal} {\mnras}\ }\textbf {\bibinfo {volume} {517}},\ \bibinfo {pages}
  {2121} (\bibinfo {year} {2022}{\natexlab{a}})},\ \Eprint
  {https://arxiv.org/abs/2207.14637} {arXiv:2207.14637 [astro-ph.EP]}
  \BibitemShut {NoStop}%
\bibitem [{\citenamefont {{Tanaka}}\ and\ \citenamefont
  {{Okada}}(2024)}]{Tanaka2024}%
  \BibitemOpen
  \bibfield  {author} {\bibinfo {author} {\bibfnamefont {H.}~\bibnamefont
  {{Tanaka}}}\ and\ \bibinfo {author} {\bibfnamefont {K.}~\bibnamefont
  {{Okada}}},\ }\bibfield  {title} {\bibinfo {title} {{Three-dimensional
  Interaction between a Planet and an Isothermal Gaseous Disk. III. Locally
  Isothermal Cases}},\ }\href {https://doi.org/10.3847/1538-4357/ad410d}
  {\bibfield  {journal} {\bibinfo  {journal} {\apj}\ }\textbf {\bibinfo
  {volume} {968}},\ \bibinfo {eid} {28} (\bibinfo {year} {2024})},\ \Eprint
  {https://arxiv.org/abs/2404.12521} {arXiv:2404.12521 [astro-ph.EP]}
  \BibitemShut {NoStop}%
\bibitem [{\citenamefont {{Fairbairn}}\ and\ \citenamefont
  {{Dittmann}}(2025)}]{FairbairnDittmann:2025}%
  \BibitemOpen
  \bibfield  {author} {\bibinfo {author} {\bibfnamefont {C.~W.}\ \bibnamefont
  {{Fairbairn}}}\ and\ \bibinfo {author} {\bibfnamefont {A.~J.}\ \bibnamefont
  {{Dittmann}}},\ }\bibfield  {title} {\bibinfo {title} {{Pushing the limits of
  eccentricity in planet--disc interactions}},\ }\href
  {https://doi.org/10.1093/mnras/staf1399} {\bibfield  {journal} {\bibinfo
  {journal} {\mnras}\ }\textbf {\bibinfo {volume} {543}},\ \bibinfo {pages}
  {565} (\bibinfo {year} {2025})},\ \Eprint {https://arxiv.org/abs/2506.19917}
  {arXiv:2506.19917 [astro-ph.EP]} \BibitemShut {NoStop}%
\bibitem [{\citenamefont {{Kley}}\ and\ \citenamefont
  {{Nelson}}(2012)}]{KleyNelson:2012}%
  \BibitemOpen
  \bibfield  {author} {\bibinfo {author} {\bibfnamefont {W.}~\bibnamefont
  {{Kley}}}\ and\ \bibinfo {author} {\bibfnamefont {R.~P.}\ \bibnamefont
  {{Nelson}}},\ }\bibfield  {title} {\bibinfo {title} {{Planet-Disk Interaction
  and Orbital Evolution}},\ }\href
  {https://doi.org/10.1146/annurev-astro-081811-125523} {\bibfield  {journal}
  {\bibinfo  {journal} {\araa}\ }\textbf {\bibinfo {volume} {50}},\ \bibinfo
  {pages} {211} (\bibinfo {year} {2012})},\ \Eprint
  {https://arxiv.org/abs/1203.1184} {arXiv:1203.1184 [astro-ph.EP]}
  \BibitemShut {NoStop}%
\bibitem [{\citenamefont {{Copparoni}}\ \emph {et~al.}(2025)\citenamefont
  {{Copparoni}}, \citenamefont {{Barausse}}, \citenamefont {{Speri}},
  \citenamefont {{Sberna}},\ and\ \citenamefont
  {{Derdzinski}}}]{Copparoni+2025}%
  \BibitemOpen
  \bibfield  {author} {\bibinfo {author} {\bibfnamefont {L.}~\bibnamefont
  {{Copparoni}}}, \bibinfo {author} {\bibfnamefont {E.}~\bibnamefont
  {{Barausse}}}, \bibinfo {author} {\bibfnamefont {L.}~\bibnamefont {{Speri}}},
  \bibinfo {author} {\bibfnamefont {L.}~\bibnamefont {{Sberna}}},\ and\
  \bibinfo {author} {\bibfnamefont {A.}~\bibnamefont {{Derdzinski}}},\
  }\bibfield  {title} {\bibinfo {title} {{Implications of stochastic gas
  torques for asymmetric binaries in the LISA band}},\ }\href
  {https://doi.org/10.1103/PhysRevD.111.104079} {\bibfield  {journal} {\bibinfo
   {journal} {\prd}\ }\textbf {\bibinfo {volume} {111}},\ \bibinfo {eid}
  {104079} (\bibinfo {year} {2025})},\ \Eprint
  {https://arxiv.org/abs/2502.10087} {arXiv:2502.10087 [gr-qc]} \BibitemShut
  {NoStop}%
\bibitem [{\citenamefont {{Hirata}}(2011)}]{Hirata1}%
  \BibitemOpen
  \bibfield  {author} {\bibinfo {author} {\bibfnamefont {C.~M.}\ \bibnamefont
  {{Hirata}}},\ }\bibfield  {title} {\bibinfo {title} {{Lindblad resonance
  torques in relativistic discs - I. Basic equations}},\ }\href
  {https://doi.org/10.1111/j.1365-2966.2011.18617.x} {\bibfield  {journal}
  {\bibinfo  {journal} {\mnras}\ }\textbf {\bibinfo {volume} {414}},\ \bibinfo
  {pages} {3198} (\bibinfo {year} {2011})},\ \Eprint
  {https://arxiv.org/abs/1010.0758} {arXiv:1010.0758 [astro-ph.HE]}
  \BibitemShut {NoStop}%
\bibitem [{\citenamefont {Hirata}(2011)}]{Hirata2}%
  \BibitemOpen
  \bibfield  {author} {\bibinfo {author} {\bibfnamefont {C.~M.}\ \bibnamefont
  {Hirata}},\ }\bibfield  {title} {\bibinfo {title} {{Lindblad resonance
  torques in relativistic discs: II. Computation of resonance strengths}},\
  }\href {https://doi.org/10.1111/j.1365-2966.2011.18619.x} {\bibfield
  {journal} {\bibinfo  {journal} {Mon. Not. Roy. Astron. Soc.}\ }\textbf
  {\bibinfo {volume} {414}},\ \bibinfo {pages} {3212} (\bibinfo {year}
  {2011})},\ \Eprint {https://arxiv.org/abs/1010.0759} {arXiv:1010.0759
  [astro-ph.HE]} \BibitemShut {NoStop}%
\bibitem [{\citenamefont {Brito}\ and\ \citenamefont
  {Shah}(2023)}]{Brito:2023pyl}%
  \BibitemOpen
  \bibfield  {author} {\bibinfo {author} {\bibfnamefont {R.}~\bibnamefont
  {Brito}}\ and\ \bibinfo {author} {\bibfnamefont {S.}~\bibnamefont {Shah}},\
  }\bibfield  {title} {\bibinfo {title} {{Extreme mass-ratio inspirals into
  black holes surrounded by scalar clouds}},\ }\href
  {https://doi.org/10.1103/PhysRevD.108.084019} {\bibfield  {journal} {\bibinfo
   {journal} {Phys. Rev. D}\ }\textbf {\bibinfo {volume} {108}},\ \bibinfo
  {pages} {084019} (\bibinfo {year} {2023})},\ \Eprint
  {https://arxiv.org/abs/2307.16093} {arXiv:2307.16093 [gr-qc]} \BibitemShut
  {NoStop}%
\bibitem [{\citenamefont {{Datta}}\ and\ \citenamefont
  {{Maselli}}(2025)}]{Datta:2025ruh}%
  \BibitemOpen
  \bibfield  {author} {\bibinfo {author} {\bibfnamefont {S.}~\bibnamefont
  {{Datta}}}\ and\ \bibinfo {author} {\bibfnamefont {A.}~\bibnamefont
  {{Maselli}}},\ }\bibfield  {title} {\bibinfo {title} {{A multi-parameter
  expansion for the evolution of asymmetric binaries in astrophysical
  environments}},\ }\href {https://doi.org/10.48550/arXiv.2507.04471}
  {\bibfield  {journal} {\bibinfo  {journal} {arXiv e-prints}\ ,\ \bibinfo
  {eid} {arXiv:2507.04471}} (\bibinfo {year} {2025})},\ \Eprint
  {https://arxiv.org/abs/2507.04471} {arXiv:2507.04471 [gr-qc]} \BibitemShut
  {NoStop}%
\bibitem [{\citenamefont {{Ipser}}\ and\ \citenamefont
  {{Lindblom}}(1991)}]{Ipser91}%
  \BibitemOpen
  \bibfield  {author} {\bibinfo {author} {\bibfnamefont {J.~R.}\ \bibnamefont
  {{Ipser}}}\ and\ \bibinfo {author} {\bibfnamefont {L.}~\bibnamefont
  {{Lindblom}}},\ }\bibfield  {title} {\bibinfo {title} {{On the Adiabatic
  Pulsations of Accretion Disks and Rotating Stars}},\ }\href
  {https://doi.org/10.1086/170503} {\bibfield  {journal} {\bibinfo  {journal}
  {\apj}\ }\textbf {\bibinfo {volume} {379}},\ \bibinfo {pages} {285} (\bibinfo
  {year} {1991})}\BibitemShut {NoStop}%
\bibitem [{\citenamefont {{Ipser}}\ and\ \citenamefont
  {{Lindblom}}(1992)}]{Ipser92}%
  \BibitemOpen
  \bibfield  {author} {\bibinfo {author} {\bibfnamefont {J.~R.}\ \bibnamefont
  {{Ipser}}}\ and\ \bibinfo {author} {\bibfnamefont {L.}~\bibnamefont
  {{Lindblom}}},\ }\bibfield  {title} {\bibinfo {title} {{On the Pulsations of
  Relativistic Accretion Disks and Rotating Stars: The Cowling
  Approximation}},\ }\href {https://doi.org/10.1086/171214} {\bibfield
  {journal} {\bibinfo  {journal} {\apj}\ }\textbf {\bibinfo {volume} {389}},\
  \bibinfo {pages} {392} (\bibinfo {year} {1992})}\BibitemShut {NoStop}%
\bibitem [{\citenamefont {Goodman}\ and\ \citenamefont
  {Rafikov}(2001)}]{Goodman_2001}%
  \BibitemOpen
  \bibfield  {author} {\bibinfo {author} {\bibfnamefont {J.}~\bibnamefont
  {Goodman}}\ and\ \bibinfo {author} {\bibfnamefont {R.~R.}\ \bibnamefont
  {Rafikov}},\ }\bibfield  {title} {\bibinfo {title} {Planetary torques as the
  viscosity of protoplanetary disks},\ }\href {https://doi.org/10.1086/320572}
  {\bibfield  {journal} {\bibinfo  {journal} {The Astrophysical Journal}\
  }\textbf {\bibinfo {volume} {552}},\ \bibinfo {pages} {793} (\bibinfo {year}
  {2001})}\BibitemShut {NoStop}%
\bibitem [{\citenamefont {{Stewart}}\ and\ \citenamefont
  {{Walker}}(1974)}]{StewartWalker}%
  \BibitemOpen
  \bibfield  {author} {\bibinfo {author} {\bibfnamefont {J.~M.}\ \bibnamefont
  {{Stewart}}}\ and\ \bibinfo {author} {\bibfnamefont {M.}~\bibnamefont
  {{Walker}}},\ }\bibfield  {title} {\bibinfo {title} {{Perturbations of
  space-times in general relativity}},\ }\href
  {https://doi.org/10.1098/rspa.1974.0172} {\bibfield  {journal} {\bibinfo
  {journal} {Proceedings of the Royal Society of London Series A}\ }\textbf
  {\bibinfo {volume} {341}},\ \bibinfo {pages} {49} (\bibinfo {year}
  {1974})}\BibitemShut {NoStop}%
\bibitem [{\citenamefont {{Stone}}\ \emph {et~al.}(2024)\citenamefont
  {{Stone}}, \citenamefont {{Mullen}}, \citenamefont {{Fielding}},
  \citenamefont {{Grete}}, \citenamefont {{Guo}}, \citenamefont {{Kempski}},
  \citenamefont {{Most}}, \citenamefont {{White}},\ and\ \citenamefont
  {{Wong}}}]{AthenaK:2024}%
  \BibitemOpen
  \bibfield  {author} {\bibinfo {author} {\bibfnamefont {J.~M.}\ \bibnamefont
  {{Stone}}}, \bibinfo {author} {\bibfnamefont {P.~D.}\ \bibnamefont
  {{Mullen}}}, \bibinfo {author} {\bibfnamefont {D.}~\bibnamefont
  {{Fielding}}}, \bibinfo {author} {\bibfnamefont {P.}~\bibnamefont {{Grete}}},
  \bibinfo {author} {\bibfnamefont {M.}~\bibnamefont {{Guo}}}, \bibinfo
  {author} {\bibfnamefont {P.}~\bibnamefont {{Kempski}}}, \bibinfo {author}
  {\bibfnamefont {E.~R.}\ \bibnamefont {{Most}}}, \bibinfo {author}
  {\bibfnamefont {C.~J.}\ \bibnamefont {{White}}},\ and\ \bibinfo {author}
  {\bibfnamefont {G.~N.}\ \bibnamefont {{Wong}}},\ }\bibfield  {title}
  {\bibinfo {title} {{AthenaK: A Performance-Portable Version of the Athena++
  AMR Framework}},\ }\href {https://doi.org/10.48550/arXiv.2409.16053}
  {\bibfield  {journal} {\bibinfo  {journal} {arXiv e-prints}\ ,\ \bibinfo
  {eid} {arXiv:2409.16053}} (\bibinfo {year} {2024})},\ \Eprint
  {https://arxiv.org/abs/2409.16053} {arXiv:2409.16053 [astro-ph.IM]}
  \BibitemShut {NoStop}%
\bibitem [{\citenamefont {{Fairbairn}}(2025)}]{Fairbairn25}%
  \BibitemOpen
  \bibfield  {author} {\bibinfo {author} {\bibfnamefont {C.~W.}\ \bibnamefont
  {{Fairbairn}}},\ }\bibfield  {title} {\bibinfo {title} {{Linear Bending Wave
  Propagation in Laminar and Turbulent Disks}},\ }\href
  {https://doi.org/10.3847/1538-4357/ad9c73} {\bibfield  {journal} {\bibinfo
  {journal} {\apj}\ }\textbf {\bibinfo {volume} {979}},\ \bibinfo {eid} {156}
  (\bibinfo {year} {2025})},\ \Eprint {https://arxiv.org/abs/2412.06955}
  {arXiv:2412.06955 [astro-ph.SR]} \BibitemShut {NoStop}%
\bibitem [{\citenamefont {{Ogilvie}}(2001)}]{Ogilvie01}%
  \BibitemOpen
  \bibfield  {author} {\bibinfo {author} {\bibfnamefont {G.~I.}\ \bibnamefont
  {{Ogilvie}}},\ }\bibfield  {title} {\bibinfo {title} {{Non-linear fluid
  dynamics of eccentric discs}},\ }\href
  {https://doi.org/10.1046/j.1365-8711.2001.04416.x} {\bibfield  {journal}
  {\bibinfo  {journal} {\mnras}\ }\textbf {\bibinfo {volume} {325}},\ \bibinfo
  {pages} {231} (\bibinfo {year} {2001})},\ \Eprint
  {https://arxiv.org/abs/astro-ph/0102245} {arXiv:astro-ph/0102245 [astro-ph]}
  \BibitemShut {NoStop}%
\bibitem [{\citenamefont {{Ogilvie}}\ and\ \citenamefont
  {{Lubow}}(2002)}]{Ogilvie02}%
  \BibitemOpen
  \bibfield  {author} {\bibinfo {author} {\bibfnamefont {G.~I.}\ \bibnamefont
  {{Ogilvie}}}\ and\ \bibinfo {author} {\bibfnamefont {S.~H.}\ \bibnamefont
  {{Lubow}}},\ }\bibfield  {title} {\bibinfo {title} {{On the wake generated by
  a planet in a disc}},\ }\href
  {https://doi.org/10.1046/j.1365-8711.2002.05148.x} {\bibfield  {journal}
  {\bibinfo  {journal} {\mnras}\ }\textbf {\bibinfo {volume} {330}},\ \bibinfo
  {pages} {950} (\bibinfo {year} {2002})},\ \Eprint
  {https://arxiv.org/abs/astro-ph/0111265} {arXiv:astro-ph/0111265 [astro-ph]}
  \BibitemShut {NoStop}%
\bibitem [{\citenamefont {{Miranda}}\ and\ \citenamefont
  {{Rafikov}}(2019{\natexlab{a}})}]{Miranda19}%
  \BibitemOpen
  \bibfield  {author} {\bibinfo {author} {\bibfnamefont {R.}~\bibnamefont
  {{Miranda}}}\ and\ \bibinfo {author} {\bibfnamefont {R.~R.}\ \bibnamefont
  {{Rafikov}}},\ }\bibfield  {title} {\bibinfo {title} {{Multiple Spiral Arms
  in Protoplanetary Disks: Linear Theory}},\ }\href
  {https://doi.org/10.3847/1538-4357/ab0f9e} {\bibfield  {journal} {\bibinfo
  {journal} {\apj}\ }\textbf {\bibinfo {volume} {875}},\ \bibinfo {eid} {37}
  (\bibinfo {year} {2019}{\natexlab{a}})},\ \Eprint
  {https://arxiv.org/abs/1811.09628} {arXiv:1811.09628 [astro-ph.EP]}
  \BibitemShut {NoStop}%
\bibitem [{\citenamefont {{Zhang}}\ and\ \citenamefont
  {{Lai}}(2006)}]{Zhang2006}%
  \BibitemOpen
  \bibfield  {author} {\bibinfo {author} {\bibfnamefont {H.}~\bibnamefont
  {{Zhang}}}\ and\ \bibinfo {author} {\bibfnamefont {D.}~\bibnamefont
  {{Lai}}},\ }\bibfield  {title} {\bibinfo {title} {{Wave excitation in
  three-dimensional discs by external potential}},\ }\href
  {https://doi.org/10.1111/j.1365-2966.2006.10167.x} {\bibfield  {journal}
  {\bibinfo  {journal} {\mnras}\ }\textbf {\bibinfo {volume} {368}},\ \bibinfo
  {pages} {917} (\bibinfo {year} {2006})},\ \Eprint
  {https://arxiv.org/abs/astro-ph/0510069} {arXiv:astro-ph/0510069 [astro-ph]}
  \BibitemShut {NoStop}%
\bibitem [{\citenamefont {{Miranda}}\ and\ \citenamefont
  {{Rafikov}}(2019{\natexlab{b}})}]{Miranda19ALMA}%
  \BibitemOpen
  \bibfield  {author} {\bibinfo {author} {\bibfnamefont {R.}~\bibnamefont
  {{Miranda}}}\ and\ \bibinfo {author} {\bibfnamefont {R.~R.}\ \bibnamefont
  {{Rafikov}}},\ }\bibfield  {title} {\bibinfo {title} {{On the Planetary
  Interpretation of Multiple Gaps and Rings in Protoplanetary Disks Seen By
  ALMA}},\ }\href {https://doi.org/10.3847/2041-8213/ab22a7} {\bibfield
  {journal} {\bibinfo  {journal} {\apjl}\ }\textbf {\bibinfo {volume} {878}},\
  \bibinfo {eid} {L9} (\bibinfo {year} {2019}{\natexlab{b}})},\ \Eprint
  {https://arxiv.org/abs/1905.08259} {arXiv:1905.08259 [astro-ph.EP]}
  \BibitemShut {NoStop}%
\bibitem [{\citenamefont {{Fairbairn}}\ and\ \citenamefont
  {{Rafikov}}(2022{\natexlab{b}})}]{Fairbairn2022}%
  \BibitemOpen
  \bibfield  {author} {\bibinfo {author} {\bibfnamefont {C.~W.}\ \bibnamefont
  {{Fairbairn}}}\ and\ \bibinfo {author} {\bibfnamefont {R.~R.}\ \bibnamefont
  {{Rafikov}}},\ }\bibfield  {title} {\bibinfo {title} {{Density waves in
  protoplanetary discs excited by eccentric planets: linear theory}},\ }\href
  {https://doi.org/10.1093/mnras/stac2802} {\bibfield  {journal} {\bibinfo
  {journal} {\mnras}\ }\textbf {\bibinfo {volume} {517}},\ \bibinfo {pages}
  {2121} (\bibinfo {year} {2022}{\natexlab{b}})},\ \Eprint
  {https://arxiv.org/abs/2207.14637} {arXiv:2207.14637 [astro-ph.EP]}
  \BibitemShut {NoStop}%
\bibitem [{\citenamefont {Potter}(2021)}]{Potter_2021}%
  \BibitemOpen
  \bibfield  {author} {\bibinfo {author} {\bibfnamefont {W.~J.}\ \bibnamefont
  {Potter}},\ }\bibfield  {title} {\bibinfo {title} {A full relativistic thin
  disc – the physics of the plunging region and the value of the stress at
  the isco},\ }\href {https://doi.org/10.1093/mnras/stab636} {\bibfield
  {journal} {\bibinfo  {journal} {Monthly Notices of the Royal Astronomical
  Society}\ }\textbf {\bibinfo {volume} {503}},\ \bibinfo {pages} {5025–5045}
  (\bibinfo {year} {2021})}\BibitemShut {NoStop}%
\bibitem [{\citenamefont {{Miranda}}\ and\ \citenamefont
  {{Rafikov}}(2020{\natexlab{b}})}]{2020ApJ...892...65M}%
  \BibitemOpen
  \bibfield  {author} {\bibinfo {author} {\bibfnamefont {R.}~\bibnamefont
  {{Miranda}}}\ and\ \bibinfo {author} {\bibfnamefont {R.~R.}\ \bibnamefont
  {{Rafikov}}},\ }\bibfield  {title} {\bibinfo {title} {{Planet-Disk
  Interaction in Disks with Cooling: Basic Theory}},\ }\href
  {https://doi.org/10.3847/1538-4357/ab791a} {\bibfield  {journal} {\bibinfo
  {journal} {\apj}\ }\textbf {\bibinfo {volume} {892}},\ \bibinfo {eid} {65}
  (\bibinfo {year} {2020}{\natexlab{b}})},\ \Eprint
  {https://arxiv.org/abs/1911.01428} {arXiv:1911.01428 [astro-ph.EP]}
  \BibitemShut {NoStop}%
\bibitem [{\citenamefont {{Tsang}}(2014{\natexlab{b}})}]{2014ApJ...782..112T}%
  \BibitemOpen
  \bibfield  {author} {\bibinfo {author} {\bibfnamefont {D.}~\bibnamefont
  {{Tsang}}},\ }\bibfield  {title} {\bibinfo {title} {{Linear Corotation
  Torques in Non-barotropic Disks}},\ }\href
  {https://doi.org/10.1088/0004-637X/782/2/112} {\bibfield  {journal} {\bibinfo
   {journal} {\apj}\ }\textbf {\bibinfo {volume} {782}},\ \bibinfo {eid} {112}
  (\bibinfo {year} {2014}{\natexlab{b}})},\ \Eprint
  {https://arxiv.org/abs/1310.8626} {arXiv:1310.8626 [astro-ph.EP]}
  \BibitemShut {NoStop}%
\bibitem [{\citenamefont {Mino}\ \emph {et~al.}(1997)\citenamefont {Mino},
  \citenamefont {Sasaki},\ and\ \citenamefont {Tanaka}}]{Mino_1997}%
  \BibitemOpen
  \bibfield  {author} {\bibinfo {author} {\bibfnamefont {Y.}~\bibnamefont
  {Mino}}, \bibinfo {author} {\bibfnamefont {M.}~\bibnamefont {Sasaki}},\ and\
  \bibinfo {author} {\bibfnamefont {T.}~\bibnamefont {Tanaka}},\ }\bibfield
  {title} {\bibinfo {title} {Gravitational radiation reaction to a particle
  motion},\ }\href {https://doi.org/10.1103/physrevd.55.3457} {\bibfield
  {journal} {\bibinfo  {journal} {Physical Review D}\ }\textbf {\bibinfo
  {volume} {55}},\ \bibinfo {pages} {3457–3476} (\bibinfo {year}
  {1997})}\BibitemShut {NoStop}%
\bibitem [{\citenamefont {Pound}\ and\ \citenamefont
  {Wardell}(2021)}]{Pound_2021}%
  \BibitemOpen
  \bibfield  {author} {\bibinfo {author} {\bibfnamefont {A.}~\bibnamefont
  {Pound}}\ and\ \bibinfo {author} {\bibfnamefont {B.}~\bibnamefont
  {Wardell}},\ }\bibinfo {title} {Black hole perturbation theory and
  gravitational self-force},\ in\ \href
  {https://doi.org/10.1007/978-981-15-4702-7_38-1} {\emph {\bibinfo {booktitle}
  {Handbook of Gravitational Wave Astronomy}}}\ (\bibinfo  {publisher}
  {Springer Singapore},\ \bibinfo {year} {2021})\ p.\ \bibinfo {pages}
  {1–119}\BibitemShut {NoStop}%
\bibitem [{\citenamefont {Toomani}\ \emph {et~al.}(2022)\citenamefont
  {Toomani}, \citenamefont {Zimmerman}, \citenamefont {Spiers}, \citenamefont
  {Hollands}, \citenamefont {Pound},\ and\ \citenamefont
  {Green}}]{Toomani:2021jlo}%
  \BibitemOpen
  \bibfield  {author} {\bibinfo {author} {\bibfnamefont {V.}~\bibnamefont
  {Toomani}}, \bibinfo {author} {\bibfnamefont {P.}~\bibnamefont {Zimmerman}},
  \bibinfo {author} {\bibfnamefont {A.}~\bibnamefont {Spiers}}, \bibinfo
  {author} {\bibfnamefont {S.}~\bibnamefont {Hollands}}, \bibinfo {author}
  {\bibfnamefont {A.}~\bibnamefont {Pound}},\ and\ \bibinfo {author}
  {\bibfnamefont {S.~R.}\ \bibnamefont {Green}},\ }\bibfield  {title} {\bibinfo
  {title} {{New metric reconstruction scheme for gravitational self-force
  calculations}},\ }\href {https://doi.org/10.1088/1361-6382/ac37a5} {\bibfield
   {journal} {\bibinfo  {journal} {Class. Quant. Grav.}\ }\textbf {\bibinfo
  {volume} {39}},\ \bibinfo {pages} {015019} (\bibinfo {year} {2022})},\
  \Eprint {https://arxiv.org/abs/2108.04273} {arXiv:2108.04273 [gr-qc]}
  \BibitemShut {NoStop}%
\bibitem [{\citenamefont {Dolan}\ \emph {et~al.}(2022)\citenamefont {Dolan},
  \citenamefont {Kavanagh},\ and\ \citenamefont {Wardell}}]{Dolan_2022}%
  \BibitemOpen
  \bibfield  {author} {\bibinfo {author} {\bibfnamefont {S.~R.}\ \bibnamefont
  {Dolan}}, \bibinfo {author} {\bibfnamefont {C.}~\bibnamefont {Kavanagh}},\
  and\ \bibinfo {author} {\bibfnamefont {B.}~\bibnamefont {Wardell}},\
  }\bibfield  {title} {\bibinfo {title} {Gravitational perturbations of
  rotating black holes in lorenz gauge},\ }\bibfield  {journal} {\bibinfo
  {journal} {Physical Review Letters}\ }\textbf {\bibinfo {volume} {128}},\
  \href {https://doi.org/10.1103/physrevlett.128.151101}
  {10.1103/physrevlett.128.151101} (\bibinfo {year} {2022})\BibitemShut
  {NoStop}%
\bibitem [{\citenamefont {Dolan}\ \emph {et~al.}(2024)\citenamefont {Dolan},
  \citenamefont {Durkan}, \citenamefont {Kavanagh},\ and\ \citenamefont
  {Wardell}}]{Dolan_2024}%
  \BibitemOpen
  \bibfield  {author} {\bibinfo {author} {\bibfnamefont {S.~R.}\ \bibnamefont
  {Dolan}}, \bibinfo {author} {\bibfnamefont {L.}~\bibnamefont {Durkan}},
  \bibinfo {author} {\bibfnamefont {C.}~\bibnamefont {Kavanagh}},\ and\
  \bibinfo {author} {\bibfnamefont {B.}~\bibnamefont {Wardell}},\ }\bibfield
  {title} {\bibinfo {title} {Metric perturbations of kerr spacetime in lorenz
  gauge: circular equatorial orbits},\ }\bibfield  {journal} {\bibinfo
  {journal} {Classical and Quantum Gravity}\ }\textbf {\bibinfo {volume}
  {41}},\ \href {https://doi.org/10.1088/1361-6382/ad52e3}
  {10.1088/1361-6382/ad52e3} (\bibinfo {year} {2024})\BibitemShut {NoStop}%
\bibitem [{\citenamefont {Wardell}\ \emph {et~al.}(2025)\citenamefont
  {Wardell}, \citenamefont {Kavanagh},\ and\ \citenamefont
  {Dolan}}]{Wardell_2025}%
  \BibitemOpen
  \bibfield  {author} {\bibinfo {author} {\bibfnamefont {B.}~\bibnamefont
  {Wardell}}, \bibinfo {author} {\bibfnamefont {C.}~\bibnamefont {Kavanagh}},\
  and\ \bibinfo {author} {\bibfnamefont {S.~R.}\ \bibnamefont {Dolan}},\
  }\bibfield  {title} {\bibinfo {title} {Sourced metric perturbations of kerr
  spacetime in lorenz gauge},\ }\href
  {https://doi.org/10.1088/1361-6382/ae0918} {\bibfield  {journal} {\bibinfo
  {journal} {Classical and Quantum Gravity}\ }\textbf {\bibinfo {volume}
  {42}},\ \bibinfo {pages} {205007} (\bibinfo {year} {2025})}\BibitemShut
  {NoStop}%
\bibitem [{\citenamefont {{Teukolsky}}(1973)}]{Teukolsky73}%
  \BibitemOpen
  \bibfield  {author} {\bibinfo {author} {\bibfnamefont {S.~A.}\ \bibnamefont
  {{Teukolsky}}},\ }\bibfield  {title} {\bibinfo {title} {{Perturbations of a
  Rotating Black Hole. I. Fundamental Equations for Gravitational,
  Electromagnetic, and Neutrino-Field Perturbations}},\ }\href
  {https://doi.org/10.1086/152444} {\bibfield  {journal} {\bibinfo  {journal}
  {\apj}\ }\textbf {\bibinfo {volume} {185}},\ \bibinfo {pages} {635} (\bibinfo
  {year} {1973})}\BibitemShut {NoStop}%
\bibitem [{\citenamefont
  {Berndtson}(2009)}]{berndtson2009harmonicgaugeperturbationsschwarzschild}%
  \BibitemOpen
  \bibfield  {author} {\bibinfo {author} {\bibfnamefont {M.~V.}\ \bibnamefont
  {Berndtson}},\ }\href {https://arxiv.org/abs/0904.0033} {\bibinfo {title}
  {Harmonic gauge perturbations of the schwarzschild metric}} (\bibinfo {year}
  {2009}),\ \Eprint {https://arxiv.org/abs/0904.0033} {arXiv:0904.0033 [gr-qc]}
  \BibitemShut {NoStop}%
\bibitem [{\citenamefont {Brown}\ and\ \citenamefont
  {Ogilvie}(2024)}]{brown2024horseshoesspiralwavescapturing}%
  \BibitemOpen
  \bibfield  {author} {\bibinfo {author} {\bibfnamefont {J.~J.}\ \bibnamefont
  {Brown}}\ and\ \bibinfo {author} {\bibfnamefont {G.~I.}\ \bibnamefont
  {Ogilvie}},\ }\href {https://arxiv.org/abs/2409.02687} {\bibinfo {title}
  {Horseshoes and spiral waves: capturing the 3d flow induced by a low-mass
  planet analytically}} (\bibinfo {year} {2024}),\ \Eprint
  {https://arxiv.org/abs/2409.02687} {arXiv:2409.02687 [astro-ph.EP]}
  \BibitemShut {NoStop}%
\bibitem [{\citenamefont {Torok}\ and\ \citenamefont
  {Stuchlik}(2005)}]{Torok:2005ct}%
  \BibitemOpen
  \bibfield  {author} {\bibinfo {author} {\bibfnamefont {G.}~\bibnamefont
  {Torok}}\ and\ \bibinfo {author} {\bibfnamefont {Z.}~\bibnamefont
  {Stuchlik}},\ }\bibfield  {title} {\bibinfo {title} {{Radial and vertical
  epicyclic frequencies of Keplerian motion in the field of Kerr naked
  singularities - Comparison with the black hole case and possible instability
  of naked singularity accretion discs}},\ }\href
  {https://doi.org/10.1051/0004-6361:20052825} {\bibfield  {journal} {\bibinfo
  {journal} {Astron. Astrophys.}\ }\textbf {\bibinfo {volume} {437}},\ \bibinfo
  {pages} {775} (\bibinfo {year} {2005})},\ \Eprint
  {https://arxiv.org/abs/astro-ph/0502127} {arXiv:astro-ph/0502127}
  \BibitemShut {NoStop}%
\bibitem [{\citenamefont {Tsang}(2014)}]{Tsang_2014}%
  \BibitemOpen
  \bibfield  {author} {\bibinfo {author} {\bibfnamefont {D.}~\bibnamefont
  {Tsang}},\ }\bibfield  {title} {\bibinfo {title} {Linear corotation torques
  in non-barotropic disks},\ }\href
  {https://doi.org/10.1088/0004-637X/782/2/112} {\bibfield  {journal} {\bibinfo
   {journal} {The Astrophysical Journal}\ }\textbf {\bibinfo {volume} {782}},\
  \bibinfo {pages} {112} (\bibinfo {year} {2014})}\BibitemShut {NoStop}%
\bibitem [{\citenamefont {Fairbairn}(2025)}]{Fairbairn_2025}%
  \BibitemOpen
  \bibfield  {author} {\bibinfo {author} {\bibfnamefont {C.~W.}\ \bibnamefont
  {Fairbairn}},\ }\bibfield  {title} {\bibinfo {title} {Linear bending wave
  propagation in laminar and turbulent disks},\ }\href
  {https://doi.org/10.3847/1538-4357/ad9c73} {\bibfield  {journal} {\bibinfo
  {journal} {The Astrophysical Journal}\ }\textbf {\bibinfo {volume} {979}},\
  \bibinfo {pages} {156} (\bibinfo {year} {2025})}\BibitemShut {NoStop}%
\bibitem [{\citenamefont {Blanchet}\ and\ \citenamefont
  {Faye}(2000)}]{Blanchet_2000}%
  \BibitemOpen
  \bibfield  {author} {\bibinfo {author} {\bibfnamefont {L.}~\bibnamefont
  {Blanchet}}\ and\ \bibinfo {author} {\bibfnamefont {G.}~\bibnamefont
  {Faye}},\ }\bibfield  {title} {\bibinfo {title} {Hadamard regularization},\
  }\href {https://doi.org/10.1063/1.1308506} {\bibfield  {journal} {\bibinfo
  {journal} {Journal of Mathematical Physics}\ }\textbf {\bibinfo {volume}
  {41}},\ \bibinfo {pages} {7675–7714} (\bibinfo {year} {2000})}\BibitemShut
  {NoStop}%
\bibitem [{\citenamefont {{Rafikov}}(2002)}]{Rafikov_NLTidWave:2002}%
  \BibitemOpen
  \bibfield  {author} {\bibinfo {author} {\bibfnamefont {R.~R.}\ \bibnamefont
  {{Rafikov}}},\ }\bibfield  {title} {\bibinfo {title} {{Nonlinear Propagation
  of Planet-generated Tidal Waves}},\ }\href {https://doi.org/10.1086/339399}
  {\bibfield  {journal} {\bibinfo  {journal} {\apj}\ }\textbf {\bibinfo
  {volume} {569}},\ \bibinfo {pages} {997} (\bibinfo {year} {2002})},\ \Eprint
  {https://arxiv.org/abs/astro-ph/0110496} {arXiv:astro-ph/0110496 [astro-ph]}
  \BibitemShut {NoStop}%
\bibitem [{\citenamefont {Cardoso}\ \emph
  {et~al.}(2022{\natexlab{b}})\citenamefont {Cardoso}, \citenamefont
  {Destounis}, \citenamefont {Duque}, \citenamefont {Macedo},\ and\
  \citenamefont {Maselli}}]{Cardoso:2021wlq}%
  \BibitemOpen
  \bibfield  {author} {\bibinfo {author} {\bibfnamefont {V.}~\bibnamefont
  {Cardoso}}, \bibinfo {author} {\bibfnamefont {K.}~\bibnamefont {Destounis}},
  \bibinfo {author} {\bibfnamefont {F.}~\bibnamefont {Duque}}, \bibinfo
  {author} {\bibfnamefont {R.~P.}\ \bibnamefont {Macedo}},\ and\ \bibinfo
  {author} {\bibfnamefont {A.}~\bibnamefont {Maselli}},\ }\bibfield  {title}
  {\bibinfo {title} {{Black holes in galaxies: Environmental impact on
  gravitational-wave generation and propagation}},\ }\href
  {https://doi.org/10.1103/PhysRevD.105.L061501} {\bibfield  {journal}
  {\bibinfo  {journal} {Phys. Rev. D}\ }\textbf {\bibinfo {volume} {105}},\
  \bibinfo {pages} {L061501} (\bibinfo {year} {2022}{\natexlab{b}})},\ \Eprint
  {https://arxiv.org/abs/2109.00005} {arXiv:2109.00005 [gr-qc]} \BibitemShut
  {NoStop}%
\bibitem [{\citenamefont {Rahman}\ \emph {et~al.}(2024)\citenamefont {Rahman},
  \citenamefont {Kumar},\ and\ \citenamefont {Bhattacharyya}}]{Rahman:2023sof}%
  \BibitemOpen
  \bibfield  {author} {\bibinfo {author} {\bibfnamefont {M.}~\bibnamefont
  {Rahman}}, \bibinfo {author} {\bibfnamefont {S.}~\bibnamefont {Kumar}},\ and\
  \bibinfo {author} {\bibfnamefont {A.}~\bibnamefont {Bhattacharyya}},\
  }\bibfield  {title} {\bibinfo {title} {{Probing astrophysical environment
  with eccentric extreme mass-ratio inspirals}},\ }\href
  {https://doi.org/10.1088/1475-7516/2024/01/035} {\bibfield  {journal}
  {\bibinfo  {journal} {JCAP}\ }\textbf {\bibinfo {volume} {01}},\ \bibinfo
  {pages} {035}},\ \Eprint {https://arxiv.org/abs/2306.14971} {arXiv:2306.14971
  [gr-qc]} \BibitemShut {NoStop}%
\bibitem [{\citenamefont {{Derdzinski}}\ \emph {et~al.}(2025)\citenamefont
  {{Derdzinski}}, \citenamefont {{Dittmann}}, \citenamefont {{Franchini}},
  \citenamefont {{Lupi}}, \citenamefont {{Brucy}}, \citenamefont {{Capelo}},
  \citenamefont {{Masset}}, \citenamefont {{Mignon-Risse}}, \citenamefont
  {{Rizzo Smith}}, \citenamefont {{Santiago-Leandro}}, \citenamefont
  {{Toscani}}, \citenamefont {{Velasco-Romero}}, \citenamefont {{Wissing}},
  \citenamefont {{Garg}}, \citenamefont {{Mayer}}, \citenamefont
  {{Serafinelli}}, \citenamefont {{Souvaitzis}}, \citenamefont {{D'Orazio}},\
  and\ \citenamefont {{Menu}}}]{LISA_IMRI:2025}%
  \BibitemOpen
  \bibfield  {author} {\bibinfo {author} {\bibfnamefont {A.}~\bibnamefont
  {{Derdzinski}}}, \bibinfo {author} {\bibfnamefont {A.~J.}\ \bibnamefont
  {{Dittmann}}}, \bibinfo {author} {\bibfnamefont {A.}~\bibnamefont
  {{Franchini}}}, \bibinfo {author} {\bibfnamefont {A.}~\bibnamefont {{Lupi}}},
  \bibinfo {author} {\bibfnamefont {N.}~\bibnamefont {{Brucy}}}, \bibinfo
  {author} {\bibfnamefont {P.~R.}\ \bibnamefont {{Capelo}}}, \bibinfo {author}
  {\bibfnamefont {F.~S.}\ \bibnamefont {{Masset}}}, \bibinfo {author}
  {\bibfnamefont {R.}~\bibnamefont {{Mignon-Risse}}}, \bibinfo {author}
  {\bibfnamefont {M.}~\bibnamefont {{Rizzo Smith}}}, \bibinfo {author}
  {\bibfnamefont {E.}~\bibnamefont {{Santiago-Leandro}}}, \bibinfo {author}
  {\bibfnamefont {M.}~\bibnamefont {{Toscani}}}, \bibinfo {author}
  {\bibfnamefont {D.~A.}\ \bibnamefont {{Velasco-Romero}}}, \bibinfo {author}
  {\bibfnamefont {R.}~\bibnamefont {{Wissing}}}, \bibinfo {author}
  {\bibfnamefont {M.}~\bibnamefont {{Garg}}}, \bibinfo {author} {\bibfnamefont
  {L.}~\bibnamefont {{Mayer}}}, \bibinfo {author} {\bibfnamefont
  {R.}~\bibnamefont {{Serafinelli}}}, \bibinfo {author} {\bibfnamefont
  {L.}~\bibnamefont {{Souvaitzis}}}, \bibinfo {author} {\bibfnamefont {D.~J.}\
  \bibnamefont {{D'Orazio}}},\ and\ \bibinfo {author} {\bibfnamefont
  {J.}~\bibnamefont {{Menu}}},\ }\bibfield  {title} {\bibinfo {title} {{The
  LISA Astrophysics ``Disc-IMRI'' Code Comparison Project:
  Intermediate-Mass-Ratio Binaries in AGN-Like Discs}},\ }\href
  {https://doi.org/10.48550/arXiv.2512.10893} {\bibfield  {journal} {\bibinfo
  {journal} {arXiv e-prints}\ ,\ \bibinfo {eid} {arXiv:2512.10893}} (\bibinfo
  {year} {2025})},\ \Eprint {https://arxiv.org/abs/2512.10893}
  {arXiv:2512.10893 [astro-ph.HE]} \BibitemShut {NoStop}%
\bibitem [{\citenamefont {{Andrade}}\ \emph {et~al.}(2021)\citenamefont
  {{Andrade}}, \citenamefont {{Salo}}, \citenamefont {{Aurrekoetxea}},
  \citenamefont {{Bamber}}, \citenamefont {{Clough}}, \citenamefont {{Croft}},
  \citenamefont {{de Jong}}, \citenamefont {{Drew}}, \citenamefont {{Duran}},
  \citenamefont {{Ferreira}}, \citenamefont {{Figueras}}, \citenamefont
  {{Finkel}}, \citenamefont {{Fran{\c{c}}a}}, \citenamefont {{Ge}},
  \citenamefont {{Gu}}, \citenamefont {{Helfer}}, \citenamefont
  {{J{\"a}ykk{\"a}}}, \citenamefont {{Joana}}, \citenamefont {{Kunesch}},
  \citenamefont {{Kornet}}, \citenamefont {{Lim}}, \citenamefont {{Muia}},
  \citenamefont {{Nazari}}, \citenamefont {{Radia}}, \citenamefont {{Ripley}},
  \citenamefont {{Shellard}}, \citenamefont {{Sperhake}}, \citenamefont
  {{Traykova}}, \citenamefont {{Tunyasuvunakool}}, \citenamefont {{Wang}},
  \citenamefont {{Widdicombe}},\ and\ \citenamefont
  {{Wong}}}]{2021JOSS....6.3703A}%
  \BibitemOpen
  \bibfield  {author} {\bibinfo {author} {\bibfnamefont {T.}~\bibnamefont
  {{Andrade}}}, \bibinfo {author} {\bibfnamefont {L.}~\bibnamefont {{Salo}}},
  \bibinfo {author} {\bibfnamefont {J.}~\bibnamefont {{Aurrekoetxea}}},
  \bibinfo {author} {\bibfnamefont {J.}~\bibnamefont {{Bamber}}}, \bibinfo
  {author} {\bibfnamefont {K.}~\bibnamefont {{Clough}}}, \bibinfo {author}
  {\bibfnamefont {R.}~\bibnamefont {{Croft}}}, \bibinfo {author} {\bibfnamefont
  {E.}~\bibnamefont {{de Jong}}}, \bibinfo {author} {\bibfnamefont
  {A.}~\bibnamefont {{Drew}}}, \bibinfo {author} {\bibfnamefont
  {A.}~\bibnamefont {{Duran}}}, \bibinfo {author} {\bibfnamefont
  {P.}~\bibnamefont {{Ferreira}}}, \bibinfo {author} {\bibfnamefont
  {P.}~\bibnamefont {{Figueras}}}, \bibinfo {author} {\bibfnamefont
  {H.}~\bibnamefont {{Finkel}}}, \bibinfo {author} {\bibfnamefont
  {T.}~\bibnamefont {{Fran{\c{c}}a}}}, \bibinfo {author} {\bibfnamefont
  {B.-X.}\ \bibnamefont {{Ge}}}, \bibinfo {author} {\bibfnamefont
  {C.}~\bibnamefont {{Gu}}}, \bibinfo {author} {\bibfnamefont {T.}~\bibnamefont
  {{Helfer}}}, \bibinfo {author} {\bibfnamefont {J.}~\bibnamefont
  {{J{\"a}ykk{\"a}}}}, \bibinfo {author} {\bibfnamefont {C.}~\bibnamefont
  {{Joana}}}, \bibinfo {author} {\bibfnamefont {M.}~\bibnamefont {{Kunesch}}},
  \bibinfo {author} {\bibfnamefont {K.}~\bibnamefont {{Kornet}}}, \bibinfo
  {author} {\bibfnamefont {E.}~\bibnamefont {{Lim}}}, \bibinfo {author}
  {\bibfnamefont {F.}~\bibnamefont {{Muia}}}, \bibinfo {author} {\bibfnamefont
  {Z.}~\bibnamefont {{Nazari}}}, \bibinfo {author} {\bibfnamefont
  {M.}~\bibnamefont {{Radia}}}, \bibinfo {author} {\bibfnamefont
  {J.}~\bibnamefont {{Ripley}}}, \bibinfo {author} {\bibfnamefont
  {P.}~\bibnamefont {{Shellard}}}, \bibinfo {author} {\bibfnamefont
  {U.}~\bibnamefont {{Sperhake}}}, \bibinfo {author} {\bibfnamefont
  {D.}~\bibnamefont {{Traykova}}}, \bibinfo {author} {\bibfnamefont
  {S.}~\bibnamefont {{Tunyasuvunakool}}}, \bibinfo {author} {\bibfnamefont
  {Z.}~\bibnamefont {{Wang}}}, \bibinfo {author} {\bibfnamefont
  {J.}~\bibnamefont {{Widdicombe}}},\ and\ \bibinfo {author} {\bibfnamefont
  {K.}~\bibnamefont {{Wong}}},\ }\bibfield  {title} {\bibinfo {title}
  {{GRChombo: An adaptable numerical relativity code for fundamental
  physics}},\ }\href {https://doi.org/10.21105/joss.03703} {\bibfield
  {journal} {\bibinfo  {journal} {The Journal of Open Source Software}\
  }\textbf {\bibinfo {volume} {6}},\ \bibinfo {eid} {3703} (\bibinfo {year}
  {2021})},\ \Eprint {https://arxiv.org/abs/2201.03458} {arXiv:2201.03458
  [gr-qc]} \BibitemShut {NoStop}%
\bibitem [{\citenamefont {Clough}(2021)}]{Clough:2021qlv}%
  \BibitemOpen
  \bibfield  {author} {\bibinfo {author} {\bibfnamefont {K.}~\bibnamefont
  {Clough}},\ }\bibfield  {title} {\bibinfo {title} {{Continuity equations for
  general matter: applications in numerical relativity}},\ }\href
  {https://doi.org/10.1088/1361-6382/ac10ee} {\bibfield  {journal} {\bibinfo
  {journal} {Class. Quant. Grav.}\ }\textbf {\bibinfo {volume} {38}},\ \bibinfo
  {pages} {167001} (\bibinfo {year} {2021})},\ \Eprint
  {https://arxiv.org/abs/2104.13420} {arXiv:2104.13420 [gr-qc]} \BibitemShut
  {NoStop}%
\bibitem [{\citenamefont {Miranda}\ \emph {et~al.}(2016)\citenamefont
  {Miranda}, \citenamefont {Muñoz},\ and\ \citenamefont {Lai}}]{Miranda_2016}%
  \BibitemOpen
  \bibfield  {author} {\bibinfo {author} {\bibfnamefont {R.}~\bibnamefont
  {Miranda}}, \bibinfo {author} {\bibfnamefont {D.~J.}\ \bibnamefont
  {Muñoz}},\ and\ \bibinfo {author} {\bibfnamefont {D.}~\bibnamefont {Lai}},\
  }\bibfield  {title} {\bibinfo {title} {Viscous hydrodynamics simulations of
  circumbinary accretion discs: variability, quasi-steady state and angular
  momentum transfer},\ }\href {https://doi.org/10.1093/mnras/stw3189}
  {\bibfield  {journal} {\bibinfo  {journal} {Monthly Notices of the Royal
  Astronomical Society}\ }\textbf {\bibinfo {volume} {466}},\ \bibinfo {pages}
  {1170–1191} (\bibinfo {year} {2016})}\BibitemShut {NoStop}%
\bibitem [{\citenamefont {Mathews}\ and\ \citenamefont
  {Pound}(2025)}]{Mathews:2025nyb}%
  \BibitemOpen
  \bibfield  {author} {\bibinfo {author} {\bibfnamefont {J.}~\bibnamefont
  {Mathews}}\ and\ \bibinfo {author} {\bibfnamefont {A.}~\bibnamefont
  {Pound}},\ }\bibfield  {title} {\bibinfo {title} {{Postadiabatic
  waveform-generation framework for asymmetric precessing binaries}},\ }\href
  {https://doi.org/10.1103/rbkb-qnxv} {\bibfield  {journal} {\bibinfo
  {journal} {Phys. Rev. D}\ }\textbf {\bibinfo {volume} {112}},\ \bibinfo
  {pages} {104078} (\bibinfo {year} {2025})},\ \Eprint
  {https://arxiv.org/abs/2501.01413} {arXiv:2501.01413 [gr-qc]} \BibitemShut
  {NoStop}%
\bibitem [{\citenamefont {{Lubow}}(1990)}]{1990ApJ...362..395L}%
  \BibitemOpen
  \bibfield  {author} {\bibinfo {author} {\bibfnamefont {S.~H.}\ \bibnamefont
  {{Lubow}}},\ }\bibfield  {title} {\bibinfo {title} {{On Mass Transport in
  Nonviscous, Non--Self-gravitating Fluid Disks}},\ }\href
  {https://doi.org/10.1086/169277} {\bibfield  {journal} {\bibinfo  {journal}
  {\apj}\ }\textbf {\bibinfo {volume} {362}},\ \bibinfo {pages} {395} (\bibinfo
  {year} {1990})}\BibitemShut {NoStop}%
\bibitem [{\citenamefont {{Muto}}\ \emph {et~al.}(2010)\citenamefont {{Muto}},
  \citenamefont {{Suzuki}},\ and\ \citenamefont
  {{Inutsuka}}}]{2010ApJ...724..448M}%
  \BibitemOpen
  \bibfield  {author} {\bibinfo {author} {\bibfnamefont {T.}~\bibnamefont
  {{Muto}}}, \bibinfo {author} {\bibfnamefont {T.~K.}\ \bibnamefont
  {{Suzuki}}},\ and\ \bibinfo {author} {\bibfnamefont {S.-i.}\ \bibnamefont
  {{Inutsuka}}},\ }\bibfield  {title} {\bibinfo {title} {{Two-dimensional Study
  of the Propagation of Planetary Wake and the Indication of Gap Opening in an
  Inviscid Protoplanetary Disk}},\ }\href
  {https://doi.org/10.1088/0004-637X/724/1/448} {\bibfield  {journal} {\bibinfo
   {journal} {\apj}\ }\textbf {\bibinfo {volume} {724}},\ \bibinfo {pages}
  {448} (\bibinfo {year} {2010})},\ \Eprint {https://arxiv.org/abs/1009.4963}
  {arXiv:1009.4963 [astro-ph.EP]} \BibitemShut {NoStop}%
\bibitem [{\citenamefont {Rafikov}\ \emph {et~al.}(2025)\citenamefont
  {Rafikov}, \citenamefont {Cimerman}, \citenamefont {Fairbairn},\ and\
  \citenamefont {Dittmann}}]{rafikov2025indirectforcesdiscplanetinteraction}%
  \BibitemOpen
  \bibfield  {author} {\bibinfo {author} {\bibfnamefont {R.~R.}\ \bibnamefont
  {Rafikov}}, \bibinfo {author} {\bibfnamefont {N.~P.}\ \bibnamefont
  {Cimerman}}, \bibinfo {author} {\bibfnamefont {C.~W.}\ \bibnamefont
  {Fairbairn}},\ and\ \bibinfo {author} {\bibfnamefont {A.~J.}\ \bibnamefont
  {Dittmann}},\ }\href {https://arxiv.org/abs/2511.10745} {\bibinfo {title}
  {Indirect forces in disc-planet interaction}} (\bibinfo {year} {2025}),\
  \Eprint {https://arxiv.org/abs/2511.10745} {arXiv:2511.10745 [astro-ph.EP]}
  \BibitemShut {NoStop}%
\bibitem [{\citenamefont {{Cimerman}}\ \emph {et~al.}(2024)\citenamefont
  {{Cimerman}}, \citenamefont {{Rafikov}},\ and\ \citenamefont
  {{Miranda}}}]{TrqWiggles:2024}%
  \BibitemOpen
  \bibfield  {author} {\bibinfo {author} {\bibfnamefont {N.~P.}\ \bibnamefont
  {{Cimerman}}}, \bibinfo {author} {\bibfnamefont {R.~R.}\ \bibnamefont
  {{Rafikov}}},\ and\ \bibinfo {author} {\bibfnamefont {R.}~\bibnamefont
  {{Miranda}}},\ }\bibfield  {title} {\bibinfo {title} {{Torque wiggles - a
  robust feature of the global disc-planet interaction}},\ }\href
  {https://doi.org/10.1093/mnras/stae467} {\bibfield  {journal} {\bibinfo
  {journal} {\mnras}\ }\textbf {\bibinfo {volume} {529}},\ \bibinfo {pages}
  {425} (\bibinfo {year} {2024})},\ \Eprint {https://arxiv.org/abs/2306.07341}
  {arXiv:2306.07341 [astro-ph.EP]} \BibitemShut {NoStop}%
\bibitem [{\citenamefont {Bemfica}\ \emph {et~al.}(2022)\citenamefont
  {Bemfica}, \citenamefont {Disconzi},\ and\ \citenamefont
  {Noronha}}]{PhysRevX.12.021044}%
  \BibitemOpen
  \bibfield  {author} {\bibinfo {author} {\bibfnamefont {F.~S.}\ \bibnamefont
  {Bemfica}}, \bibinfo {author} {\bibfnamefont {M.~M.}\ \bibnamefont
  {Disconzi}},\ and\ \bibinfo {author} {\bibfnamefont {J.}~\bibnamefont
  {Noronha}},\ }\bibfield  {title} {\bibinfo {title} {First-order
  general-relativistic viscous fluid dynamics},\ }\href
  {https://doi.org/10.1103/PhysRevX.12.021044} {\bibfield  {journal} {\bibinfo
  {journal} {Phys. Rev. X}\ }\textbf {\bibinfo {volume} {12}},\ \bibinfo
  {pages} {021044} (\bibinfo {year} {2022})}\BibitemShut {NoStop}%
\bibitem [{\citenamefont {Kovtun}(2019)}]{Kovtun:2019hdm}%
  \BibitemOpen
  \bibfield  {author} {\bibinfo {author} {\bibfnamefont {P.}~\bibnamefont
  {Kovtun}},\ }\bibfield  {title} {\bibinfo {title} {{First-order relativistic
  hydrodynamics is stable}},\ }\href {https://doi.org/10.1007/JHEP10(2019)034}
  {\bibfield  {journal} {\bibinfo  {journal} {JHEP}\ }\textbf {\bibinfo
  {volume} {10}},\ \bibinfo {pages} {034}},\ \Eprint
  {https://arxiv.org/abs/1907.08191} {arXiv:1907.08191 [hep-th]} \BibitemShut
  {NoStop}%
\bibitem [{\citenamefont {Bemfica}\ \emph {et~al.}(2018)\citenamefont
  {Bemfica}, \citenamefont {Disconzi},\ and\ \citenamefont
  {Noronha}}]{PhysRevD.98.104064}%
  \BibitemOpen
  \bibfield  {author} {\bibinfo {author} {\bibfnamefont {F.~S.}\ \bibnamefont
  {Bemfica}}, \bibinfo {author} {\bibfnamefont {M.~M.}\ \bibnamefont
  {Disconzi}},\ and\ \bibinfo {author} {\bibfnamefont {J.}~\bibnamefont
  {Noronha}},\ }\bibfield  {title} {\bibinfo {title} {Causality and existence
  of solutions of relativistic viscous fluid dynamics with gravity},\ }\href
  {https://doi.org/10.1103/PhysRevD.98.104064} {\bibfield  {journal} {\bibinfo
  {journal} {Phys. Rev. D}\ }\textbf {\bibinfo {volume} {98}},\ \bibinfo
  {pages} {104064} (\bibinfo {year} {2018})}\BibitemShut {NoStop}%
\bibitem [{\citenamefont {Bemfica}\ \emph {et~al.}(2019)\citenamefont
  {Bemfica}, \citenamefont {Disconzi},\ and\ \citenamefont
  {Noronha}}]{PhysRevD.100.104020}%
  \BibitemOpen
  \bibfield  {author} {\bibinfo {author} {\bibfnamefont {F.~S.}\ \bibnamefont
  {Bemfica}}, \bibinfo {author} {\bibfnamefont {M.~M.}\ \bibnamefont
  {Disconzi}},\ and\ \bibinfo {author} {\bibfnamefont {J.}~\bibnamefont
  {Noronha}},\ }\bibfield  {title} {\bibinfo {title} {Nonlinear causality of
  general first-order relativistic viscous hydrodynamics},\ }\href
  {https://doi.org/10.1103/PhysRevD.100.104020} {\bibfield  {journal} {\bibinfo
   {journal} {Phys. Rev. D}\ }\textbf {\bibinfo {volume} {100}},\ \bibinfo
  {pages} {104020} (\bibinfo {year} {2019})}\BibitemShut {NoStop}%
\bibitem [{\citenamefont {Boyanov}\ \emph {et~al.}(2025)\citenamefont
  {Boyanov}, \citenamefont {Cardoso}, \citenamefont {Kokkotas},\ and\
  \citenamefont {Redondo-Yuste}}]{Boyanov:2024jge}%
  \BibitemOpen
  \bibfield  {author} {\bibinfo {author} {\bibfnamefont {V.}~\bibnamefont
  {Boyanov}}, \bibinfo {author} {\bibfnamefont {V.}~\bibnamefont {Cardoso}},
  \bibinfo {author} {\bibfnamefont {K.~D.}\ \bibnamefont {Kokkotas}},\ and\
  \bibinfo {author} {\bibfnamefont {J.}~\bibnamefont {Redondo-Yuste}},\
  }\bibfield  {title} {\bibinfo {title} {{Dynamical Response of Viscous Objects
  to Gravitational Waves}},\ }\href {https://doi.org/10.1103/smlr-v7b2}
  {\bibfield  {journal} {\bibinfo  {journal} {Phys. Rev. Lett.}\ }\textbf
  {\bibinfo {volume} {135}},\ \bibinfo {pages} {151402} (\bibinfo {year}
  {2025})},\ \Eprint {https://arxiv.org/abs/2411.16861} {arXiv:2411.16861
  [gr-qc]} \BibitemShut {NoStop}%
\bibitem [{\citenamefont {Redondo-Yuste}(2025)}]{Redondo-Yuste:2024vdb}%
  \BibitemOpen
  \bibfield  {author} {\bibinfo {author} {\bibfnamefont {J.}~\bibnamefont
  {Redondo-Yuste}},\ }\bibfield  {title} {\bibinfo {title} {{Perturbations of
  relativistic dissipative stars}},\ }\href
  {https://doi.org/10.1088/1361-6382/adbfef} {\bibfield  {journal} {\bibinfo
  {journal} {Class. Quant. Grav.}\ }\textbf {\bibinfo {volume} {42}},\ \bibinfo
  {pages} {075012} (\bibinfo {year} {2025})},\ \Eprint
  {https://arxiv.org/abs/2411.16841} {arXiv:2411.16841 [gr-qc]} \BibitemShut
  {NoStop}%
\bibitem [{\citenamefont {Redondo-Yuste}\ and\ \citenamefont
  {Cardoso}(2025)}]{Redondo-Yuste:2025ktt}%
  \BibitemOpen
  \bibfield  {author} {\bibinfo {author} {\bibfnamefont {J.}~\bibnamefont
  {Redondo-Yuste}}\ and\ \bibinfo {author} {\bibfnamefont {V.}~\bibnamefont
  {Cardoso}},\ }\bibfield  {title} {\bibinfo {title} {{Superradiant
  amplification by rotating viscous compact objects}},\ }\href
  {https://doi.org/10.1103/gvd3-lqkv} {\bibfield  {journal} {\bibinfo
  {journal} {Phys. Rev. D}\ }\textbf {\bibinfo {volume} {112}},\ \bibinfo
  {pages} {L061501} (\bibinfo {year} {2025})},\ \Eprint
  {https://arxiv.org/abs/2506.13850} {arXiv:2506.13850 [gr-qc]} \BibitemShut
  {NoStop}%
\bibitem [{\citenamefont {Caballero}\ and\ \citenamefont
  {Yunes}(2025)}]{Caballero:2025omv}%
  \BibitemOpen
  \bibfield  {author} {\bibinfo {author} {\bibfnamefont {D.~A.}\ \bibnamefont
  {Caballero}}\ and\ \bibinfo {author} {\bibfnamefont {N.}~\bibnamefont
  {Yunes}},\ }\bibfield  {title} {\bibinfo {title} {{Neutron star radial
  perturbations for causal, viscous, relativistic fluids}},\ }\href
  {https://doi.org/10.1103/cl4s-n7nr} {\bibfield  {journal} {\bibinfo
  {journal} {Phys. Rev. D}\ }\textbf {\bibinfo {volume} {112}},\ \bibinfo
  {pages} {063050} (\bibinfo {year} {2025})},\ \Eprint
  {https://arxiv.org/abs/2506.09149} {arXiv:2506.09149 [gr-qc]} \BibitemShut
  {NoStop}%
\bibitem [{BHP()}]{BHPToolkit}%
  \BibitemOpen
  \href@noop {} {\bibinfo {title} {{Black Hole Perturbation Toolkit}}},\
  \bibinfo {howpublished}
  {(\href{http://bhptoolkit.org/}{bhptoolkit.org})}\BibitemShut {NoStop}%
\bibitem [{\citenamefont {Brizuela}\ \emph {et~al.}(2009)\citenamefont
  {Brizuela}, \citenamefont {Martin-Garcia},\ and\ \citenamefont
  {Mena~Marugan}}]{Brizuela:2008ra}%
  \BibitemOpen
  \bibfield  {author} {\bibinfo {author} {\bibfnamefont {D.}~\bibnamefont
  {Brizuela}}, \bibinfo {author} {\bibfnamefont {J.~M.}\ \bibnamefont
  {Martin-Garcia}},\ and\ \bibinfo {author} {\bibfnamefont {G.~A.}\
  \bibnamefont {Mena~Marugan}},\ }\bibfield  {title} {\bibinfo {title} {{xPert:
  Computer algebra for metric perturbation theory}},\ }\href
  {https://doi.org/10.1007/s10714-009-0773-2} {\bibfield  {journal} {\bibinfo
  {journal} {Gen. Rel. Grav.}\ }\textbf {\bibinfo {volume} {41}},\ \bibinfo
  {pages} {2415} (\bibinfo {year} {2009})},\ \Eprint
  {https://arxiv.org/abs/0807.0824} {arXiv:0807.0824 [gr-qc]} \BibitemShut
  {NoStop}%
\end{thebibliography}%

\end{document}